\documentclass[12pt]{article}
\usepackage{setspace}

\usepackage{amsmath, amssymb, color, epsfig, amsthm, natbib}
\usepackage{graphicx,rotating}
\usepackage{fancyhdr}
\usepackage{algpseudocode}
\makeatletter

\newtheorem{theorem}{Theorem}
\newtheorem{lemma}{Lemma}
\newtheorem{proposition}{Proposition}

\theoremstyle{definition}

\newtheorem{remark}{Remark}

\newtheorem{assumption}{Assumption}

\newcommand{\Date}[1]{\def\@Date{#1}}
\def\today{\number\day~\ifcase\month\or
 January\or February\or March\or April\or May\or June\or
 July\or August\or September\or October\or November\or December\fi~\number\year}

\def\T{{ \mathrm{\scriptscriptstyle T} }}

\def \b1{{\bf 1}}
\def \bA{{\bf A}}
\def \bD{{\bf D}}
\def \bY{{\bf Y}}
\def \bv{{\bf v}}

\def \bh{{\bf h}}
\def \bW{{\bf W}}
\def \bH{{\bf H}}

\def \bV{{\bf V}}
\def \bG{{\bf G}}
\def \bzero{\boldsymbol{0}}

\def \bSigma{\boldsymbol{\Sigma}}
\def \bvartheta{\boldsymbol{\vartheta}}
\def \btheta{\boldsymbol{\theta}}
\def \bDelta{\boldsymbol{\Delta}}

\begin{document}
\begin{center}
{\bf\Large Estimation of Subgraph Densities in Noisy Networks
% Signal-plus-noise Model\\
}
\end{center}
\begin{center}
Jinyuan Chang\footnote{School of Statistics, Southwestern University of Finance and Economics, Chengdu, China}, ~~~Eric D. Kolaczyk\footnote{Corresponding author. Department of Mathematics and Statistics, Boston University, Boston, MA, USA. Email: kolaczyk@math.bu.edu. Postal address: Department of Mathematics and Statistics,
Boston University, 111 Cummington Mall, Boston, MA 02215, USA}, ~~~Qiwei Yao\footnote{Department of Statistics, London School of Economics and Political Science, London,
UK}
\end{center}

%\begin{center}
%Southwestern University of Finance and Economics, Boston University, London School of Economics and Political Science
%\end{center}

\begin{abstract}
While it is common practice in applied network analysis to report various standard network summary statistics, these numbers are rarely accompanied by uncertainty quantification.  Yet any error inherent in the measurements underlying the construction of the network, or in the network construction procedure itself, necessarily must propagate to any summary statistics reported.  Here we study the problem of estimating the density of an arbitrary subgraph, given a noisy version of some underlying network as data.  Under a simple model of network error, we show that consistent estimation of such densities is impossible when the rates of error are unknown and only a single network is observed.  Accordingly, we develop method-of-moment estimators of network subgraph densities and error rates for the case where a minimal number of network replicates are available.  These estimators are shown to be asymptotically normal as the number of vertices increases to infinity.  We also provide confidence intervals for quantifying the uncertainty in these estimates based on the asymptotic normality. To construct the confidence intervals, a new and non-standard bootstrap method is proposed to compute asymptotic variances, which is infeasible otherwise. We  illustrate the proposed methods in the context of gene coexpression networks.
\end{abstract}


%\newpage

\section{Introduction}

An applied analysis in network science typically includes the following
three steps: (i) gather basic measurements relevant to the interactions
among elements in a system of interest, (ii) construct a network-based
representation of that system, with nodes serving as elements and links
indicating interactions between pairs of elements, and (iii) report
various numerical summaries of network structure (e.g., density,
centralities, etc.).  Necessarily, uncertainty at the level of the basic
measurements in the first step will propagate to the network constructed
in the second step and thus to the summaries reported in the third
step.

The potential for measurement error arises in nearly every network analysis application.  Here, by ``measurement error" we will specifically mean true edges being observed as non-edges, and vice versa -- there are, of course, other notions of error that might be considered.  Such edge noise occurs in online social networks (e.g., Facebook), which are often based on the extraction and merging of lists of ``friends" from millions of individual accounts, where uniqueness of names is not assured.  Similarly, it can be found in biological networks (e.g., of gene regulatory relationships), which are often based on notions of association (e.g., correlation, partial correlation, etc.) among experimental measurements of gene activity levels that are determined by some form of statistical inference.  Finally, maps of the logical internet traditionally have been synthesized from the results of surveys in which paths along which information flows are learned experimentally through a large set of packet probes (e.g., via traceroute).  See~Chapter 3.5 of \cite{kolaczyk2009statistical} for several detailed examples of applied network analyses associated with such data.

That there is measurement error associated with these and other common
types of network constructions is typically well-understood by
practitioners.  And in many settings the general issue has received
substantial attention, such as, for example, in the context of
protein-protein interaction networks (e.g., \cite{hart2006complete}) or
social networks (e.g,~\cite{almquist2012random}).  But, to
our best knowledge, there has been little attention to date given toward
formal development of statistical methods accounting for propagation of
network error. Exceptions include statistical methodology for predicting
network topology or attributes with models that explicitly include a
component for network noise
(e.g.,~\cite{jiang2011network,jiang2012latent}), the ``denoising" of noisy
networks (e.g.~\cite{chatterjee2015matrix}), and the adaptation of
methods for vertex classification using networks observed with
errors~\citep{priebe2015statistical}.

Motivating our own work is that of~\cite{balachandran2017propagation}.
Working with the analogue of a ``signal plus noise"  model for networks, these authors
characterize the asymptotic distribution of the empirical edge density (i.e., formally, the density of observed edges)
 in noisy networks, in the context of what they call low-rate measurement error.   \cite{GanKolaczyk2018} offered a
refinement.  The edge density is an important prototype, as it is a fundamental characteristic of networks.  Its calculation generally is one of the first steps in an applied network analysis, analogous to computing a sample mean in
analyzing traditional % univariate
data.  Additionally, the edge density is understood to be a key driver of various other network characteristics -- for example, placing limits on the frequency of higher-order subgraphs (e.g.,
\cite{turan.1941}).  We note that the work in these two papers is entirely probabilistic in nature, focused on approximation error using Stein's method.  Here our focus is statistical in nature.

In particular, here we study the problem of estimating subgraph densities, with the
edge density serving as a critical initial case.
% Following~\cite{balachandran2017propagation},
We adopt a simple model for noisy networks that, conditional on some true underlying
network, assumes we observe a version of that network corrupted by an
independent random noise that effectively flips the status of (non)edges.
 If it is known the rates at which edges are instead observed as
non-edges, and non-edges as edges, then it is
straightforward to construct a moment-based estimator of the density of a
given subgraph of interest from a single noisy network.
%  We describe this estimator along with a
% bootstrap-based procedure for constructing confidence intervals.
However, in the more realistic setting in which one or both of these
error rates are unknown and must themselves be estimated, the problem of
identifiability arises.  The problem in this case is analogous to
estimation under a two-component mixture model.  We show that consistent
estimation of any subgraph density is in fact impossible under this setting.

The primary contribution in this paper is our development of
method-of-moments estimators for network subgraph densities and the
underlying rates of error when replicates of the observed network are
available.  Beginning with the fundamental case of edge density, we
provide estimators that are asymptotically normal  (as the number of vertices increases to infinity) when one or both of the
error rates are unknown, using a minimum of two or three replicates,
respectively.  The asymptotic normality in turn facilitates interval
estimation for network edge density.
% We offer a bootstrap-based procedure as well for this purpose.
We then extend the method-of-moments
estimator to the context of an arbitrary higher-order subgraph density, and illustrate
with the cases of two-star and triangle densities, as well as the clustering
coefficient (or transitivity). To construct their confidence intervals,
a new and non-standard bootstrap method is proposed in order to
compute asymptotic variances, which is infeasible otherwise.
% {\color{red} Bootstrap
% interval estimators are proposed for the case of two-star and triangle
% densities. Interestingly, these estimators require the development of a
% new algorithm to generate a graph with pre-specified counts of edges,
% two-star edges and triangles.  The algorithm is based on the notion of
% rewiring~\citep{mahadevan2006}, and is of some independent interest.
% Asymptotic expansions indicate that the approach can be extended to
% interval estimation for subgraph densities involving more than three
% nodes.}
Numerical simulation suggests that high accuracy is possible for
networks of even modest size.  We illustrate the practical use of our
estimators in the context of gene coexpression networks, where a small
number of replicates of the basic underlying measurements (e.g.,
microarray expression) are frequently available.

It is difficult to overstate how ubiquitous is the use of subgraph densities in empirical network analysis.  As a result, our work here is relevant to a broad and diverse cross-section of literature in humanities, social, and natural sciences, as touched by the applied network analysis literature.  Certain subgraph densities (i.e., the edge density and the two-star and triangle densities, through the clustering coefficient) are reported as commonly in network analysis as one reports, say, the mean, median, and standard deviation in standard data analysis.  In fact, they feature in what  at least one author has termed ``the network analysis `five-number summary' '' (\cite{luke2015user}).   Prolific use of subgraph densities is also found in the so-called ``triad census" that is standard in social network analysis (e.g., \cite{wasserman1994social}) and in the context of ``motif analysis" (\cite{milo2002network}), the latter being fundamental to both computational biology (e.g., \cite{stone2019network}) and computational neuroscience (e.g., \cite{sporns2016modular}).

To date researchers doing empirical network analysis have necessarily had to report these and other types of subgraph densities simply as descriptive summaries, lacking a statistically principled framework for assessing and correcting for bias and for quantifying uncertainty due to network noise.  Our work here not only provides such a framework but also demonstrates, in the context of a typical exercise in computational biology, that the nature and impact of network noise on the standard practice of reporting subgraph densities is almost surely more nuanced and pronounced than the general practitioner likely imagines.  In addition, of independent interest specifically to statisticians within our work are (i) the impossibility theorem described in Theorem~\ref{tm:1}, and (ii) the nonstandard bootstrap algorithm following Theorem~\ref{tm:bootstrap}.

The rest of the paper is organized as follows. Section 2 introduces
the problem to be tackled. Section 3 deals with the estimation of
error rates and the inference of edge density. It also reveals the
innate difficulty associated with estimation when the error rates are unknown.
Section 4 addresses the inference of subgraph densities in general.
Numerical illustration is reported in Section 5. Some further discussion of our work is stated in Section 6.
All technical proofs are relegated to supplementary materials.

\section{Notation, assumptions, and problem statement}\label{se:known}

\subsection{Noisy networks}

Let $G=(V,E)$ be a graph, with vertices $V=\{1,\ldots,p\}$ and edges $E\subseteq V^2$. We observe a noisy version of $G$, say, $G^{obs}=(V,E^{obs})$, where we implicitly assume that the vertex set $V$ is known. Denote the $p\times p$ adjacency matrix of $G$ by $\bA= (A_{i,j})_{p\times p}$, and that of $G^{obs}$ by $\bY=(Y_{i,j})_{p\times p}$. Hence $A_{i,j}=1$ if there is a true edge between the $i$-th vertex and the $j$-th vertex, and 0 otherwise, while $Y_{i,j}=1 $ if an edge is observed between the $i$-th vertex and the $j$-th vertex, and 0 otherwise. We assume throughout that $G$ and $G^{obs}$ are simple, i.e., that they possess neither multi-edges nor self-loops.  An implication of the latter is that $A_{i,i} = Y_{i,i} \equiv 0$.  Note that for the sake of exposition, we assume $G$ to be undirected.  Then $A_{i,j}=A_{j,i}$ and $Y_{i,j}=Y_{j,i}$ for any $i\neq j$.  Extension to directed graphs is straightforward and discussed briefly in Section~\ref{sec:disc}.

Following~\cite{balachandran2017propagation},
we specify the errors in the noisy network $G^{obs}$ as follows:
\begin{equation} \label{error2}
\mathbb{P}(Y_{i,j}=1\, |\, A_{i,j}=0) = \alpha \quad {\rm and} \quad
\mathbb{P}(Y_{i,j}=0\, |\, A_{i,j}=1) = \beta
\end{equation}
for any $1 \le i < j \le
p$. Note that $\alpha$ and $\beta$ may be interpreted, respectively, as Type I and II
error rates.
We assume that both $\alpha$ and $\beta$ remain constant over different edges.
For some applications, $\alpha$ is known as, for example,
the nominal significance level of statistical tests for the null hypothesis
that there is no edge between one vertex and another. If one applies the
same test method over different vertex pairs, and assumes (approximately)
equal strength of ``signal" across the network, then the power of the test
$1-\beta$, though unknown, also remains (approximately) the same.

Inspired by the conventional treatment of regression analysis in which
inference is conditionally on regressors (i.e. treating them as constants)
and with additive noise, we treat $A_{i,j}$ as constants and assume
\begin{equation}\label{eq:model}
Y_{i,j}=A_{i,j}I(\varepsilon_{i,j}=0)+I(\varepsilon_{i,j}=1)
\end{equation}
for any $1\le i< j \le p$, where $I(\cdot)$ denotes
the indicator function, and $\{\varepsilon_{i,j}\}$ are specified in
Assumption 1 below.
\begin{assumption} The $\varepsilon_{i,j}$, for all $1\le i< j \le p$,
are independent random variables with
$
\mathbb{P}(\varepsilon_{i,j}=1)=\alpha$,
$\mathbb{P}(\varepsilon_{i,j}=0)=1-\alpha-\beta$ and
$\mathbb{P}(\varepsilon_{i,j}=-1) = \beta$,
where $\alpha, \beta \ge 0$ and $\alpha + \beta<1$.
\end{assumption}

Now (\ref{error2}) follows from (\ref{eq:model}) and Assumption 1 immediately.
The independence condition in Assumption 1 is not strictly necessary. See
Remark~\ref{remark1} % and \ref{remark4}
in Section \ref{se:oneunknown}.

\subsection{Subgraph density}

A standard quantity of general interest in practice is the density of certain subgraphs in $G$.  Subgraphs of common interest include (i) edges, (ii) two-stars (also called triples) and other higher-order $k$-stars, (iii) triangles and other higher-order cliques, (iv) chains, and (v) cycles.  Subgraph density is simply the total number of times a given subgraph, say $H$, is found in $G$ (where, note, overlap among copies of $H$ is allowed), divided by the maximum number of copies possible in a graph of the same number of vertices as $G$.  There are different ways to express this notion formally.  Intuitively, for example, the count $f_H(G)$ of the number of distinct copies of a subgraph $H$ in $G$ is represented as
\begin{equation}
f_H(G) = \frac{1}{|{\rm Iso}(H)|} \sum_{H'\subseteq K_{p},H'\cong H} I{(H'\subseteq G)}\,,
\label{eq:subg.cnt}
\end{equation}
where $K_{p}$ is the complete graph on $p$ vertices and $H\subseteq G$ indicates that $H$ is a subgraph of $G$ (i.e., $V(H)\subseteq V(G)$ and $E(H)\subseteq E(G)$).  The value $|{\rm Iso}(H)|$ is a normalization factor for the number of isomorphisms of $H$.
  Normalizing $f_H(G)$, in turn, by the total number of copies of $H$ possible in the complete graph $K_p$ then yields the density of subgraph $H$ in $G$.

For our purposes, it is more convenient to adopt an alternative expression for subgraph density -- albeit one that is notationally more cumbersome.
Consider an arbitrary subgraph $H=(V_H, E_H)$ of interest, of order $|V_H|\ge 2$.  We characterize such subgraphs in terms of an index set $\mathcal{V} = \mathcal{V}_H$ of the following generic form
\begin{equation}\label{eq:V}
\begin{split}
\mathcal{V}=\{(i_1,i_1',\ldots,i_k,i_k'):&~i_\ell\neq i_\ell'~\mbox{for each}~\ell=1,\ldots,k,~|\{i_{\ell_1},i_{\ell_1}'\}\cap\{i_{\ell_2},i_{\ell_2}'\}|\leq 1\\
&~\mbox{for any}~\ell_1\neq\ell_2,~\mbox{and}~i_1,i_1',\ldots,i_k,i_k'~\textrm{also}\\
&~\mbox{satisfying other restrictions imposed by $H$}\}\,,
\end{split}
\end{equation}
and $k$ prescribed values $\tau_1,\ldots,\tau_k\in\{0,1\}$.  We then represent the
subgraph density
% normalized count (up to isomorphisms) of
for any subgraph $H$ in $G$ as
\begin{equation}\label{eq:C}
C_{\mathcal{V}}(\tau_1,\ldots,\tau_k)=\frac{1}{|\mathcal{V}|}\sum_{\bv=(i_1,i_1',\ldots,i_k,i_k')\in\mathcal{V}}A_{i_1,i_1'}^{\tau_1}(1-A_{i_1,i_1'})^{1-\tau_1}\cdots A_{i_k,i_k'}^{\tau_k}(1-A_{i_k,i_k'})^{1-\tau_k}\,.
\end{equation}
Here we adopt the convention  $0^0=1$.

The quantity $C_{\mathcal{V}}(\tau_1,\ldots,\tau_k)$ defined in
(\ref{eq:C}) is quite general. For example, if we let $k=1$ and $\tau_1=1$,
it reduces to the edge density defined in (\ref{b2}) below, which is arguably the
most important single-number summary for networks.
If we select
$\tau_1=\cdots=\tau_k=1$ and
$\mathcal{V}=\{(i_1,i_1',\ldots,i_k,i_k'):i_{\ell}'
=i_{\ell+1}~\textrm{for each}~\ell=1,\ldots,k-1, i_1\neq
i_2\neq\cdots\neq i_k\neq i_{k}'\}$, then
\[
C_{\mathcal{V}}(\tau_1,\ldots,\tau_k)=\frac{1}{p\cdots(p-k)}\sum_{i_1\neq\cdots\neq i_{k+1}}A_{i_1,i_2}\cdots A_{i_k,i_{k+1}}\,,
\]
which is the density of $k$ connected edges in $G$ passing through $k+1$ different nodes -- that is, the density of paths of length $k$.  If in addition we impose the constraint that the path must start and end with the same vertex, we select $\tau_1=\cdots=\tau_k=1$ and $\mathcal{V}=\{(i_1,i_1',\ldots,i_k,i_k'):i_{\ell}'=i_{\ell+1}~\textrm{for each}~\ell=1,\ldots,k-1, i_k'=i_1, i_1\neq i_2\neq\cdots\neq i_k\}$, yielding
\[
C_{\mathcal{V}}(\tau_1,\ldots,\tau_k)=\frac{1}{p\cdots(p-k+1)}\sum_{i_1\neq\cdots\neq i_{k}}A_{i_1,i_2}\cdots A_{i_k,i_{1}}\,,
\]
which is the density of cycles of length $k$ in $G$.   An important
special case of the latter is when $k=3$, which yields the density of
closed triples in $G$ (generally interpreted as three times the density
of triangles).  Similarly, if the summands $A_{i_1,i_2}A_{i_2,i_3}
A_{i_3,i_1}$ associated with the triangle density are instead replaced by
 $A_{i_1,i_2}A_{i_2,i_3}(1- A_{i_3,i_1})$, we obtain the density of
(open) connected triples or two-stars.  In turn, the ratio of the first
of these two quantities to its sum with the second defines the clustering
coefficient (also called the transitivity) of $G$ -- arguably the second
most important summary statistic in practice after the edge density.

In practice, given a noisy network, researchers currently report the empirical subgraph densities (i.e., $C_{\mathcal{V}}$ applied to $G^{obs}$ with adjacency matrix $\bY$) and assume that they are reflective of the corresponding true subgraph densities (i.e., $C_{\mathcal{V}}$ applied to $G$ with adjacency matrix $\bA$).  The work of~\cite{balachandran2017propagation} shows that, under conditions similar to those assumed here, there is in general no reason to expect that these empirical (or ``plug in") estimates are even consistent.  Our goal in this paper is to produce principled and accurate estimates of subgraph densities.  In what follows, we treat the estimation of edge density as a special base case, which helps inform the exposition of our results for general subgraph density estimation.

\section{Inference for edge density}

In this section, we consider inference of the edge density with unknown error rates
$\alpha$ and $\beta$. The edge is the simplest subgraph. The count of the number of edges or, upon normalization, the so-called edge density (aka network density) is defined as follows:
\begin{equation} \label{b2}
\delta = {2 \over p(p-1)} \sum_{i < j} A_{i,j}\,.
\end{equation}
It is both useful, from the perspective of our mathematical development,
and fundamental, from the perspective of network theory and applications,
to focus first on the edge density $\delta$ as the estimand of interest.
It reveals the innate difficulty associated with estimation under
unknown error rates. See Section \ref{se:oneknown}.
The inference for general
subgraphs will be presented in Section~\ref{sec4}.

\subsection{Difficulty of estimating subgraph densities}\label{se:oneknown}

\begin{figure}
\centerline{
\includegraphics[width=4.5in]{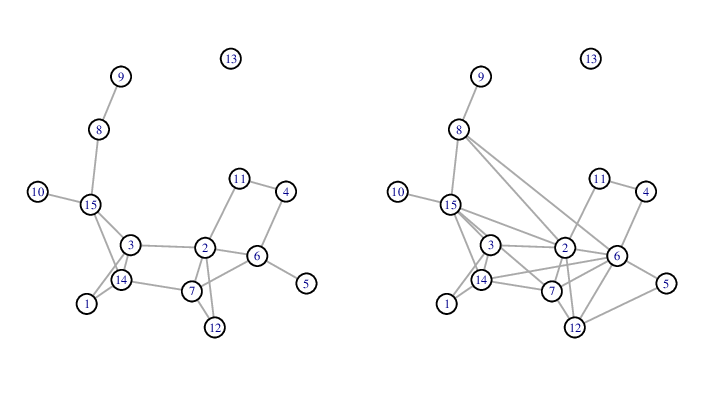}}
\caption{{\sl Left Panel} -- True network $G$, with $p=15$ nodes and density $\delta=0.181$.
{\sl Right Panel} -- Noisy network $G^{obs}$, with $\alpha =0.05, \beta=0.15$ and $\bar Y = 0.229$.  The goal is to estimate $\delta$ based on the noisy network.}
\label{fig1}
\end{figure}

Consider estimation of the network edge density $\delta$ in (\ref{b2}).
Figure~\ref{fig1} presents a simple visual illustration of our task.  The
network on the left  with $p=15$ nodes is defined by a deterministic
adjacency matrix $\bA$ with 19 edges, and hence the network density
$\delta =2\times 19/(15\times 14)=0.181$. The noisy network on the right defined by
the adjacency matrix $\bY$ was observed with 24 edges, where $\bY = (Y_{i,j})_{15\times 15}$ is
generated from $\bA$ by (\ref{eq:model}) with $\alpha =5\%$ and $\beta = 15\%$. Our
task is to estimate $\delta$ based on $\bY$.

A natural estimator for $\delta$ is given by
\[
\bar{Y}=\frac{2}{p(p-1)}\sum_{i < j}Y_{i,j}\,.
\]
In the illustration of Figure~\ref{fig1}, this value is $\bar Y = 0.229$,
in comparison to the true value $\delta=0.181$.  Let
$\mathcal{S}=\{(i,j):A_{i,j}=1,~i < j\}$ and
$\mathcal{S}^c=\{(i,j):A_{i,j}=0,~i < j\}$. From (\ref{b2}), we know
$\bar{Y}$ is a biased estimator for $\delta$. More specifically,
we have
\begin{equation}\label{eq:bary}
\begin{split}
\mathbb{E}(\bar{Y})=&~\frac{2}{p(p-1)}\sum_{(i,j)\in\mathcal{S}}\mathbb{E}(Y_{i,j})+\frac{2}{p(p-1)}\sum_{(i,j)\in\mathcal{S}^c}\mathbb{E}(Y_{i,j})\\
=&~\delta(1-\beta)+(1-\delta)\alpha\,.
\end{split}
\end{equation}
But if $\alpha$ and $\beta$ are known, (\ref{eq:bary}) suggests estimating $\delta$ instead by
\begin{equation} \label{b3}
\widetilde{Y}=\frac{\bar{Y}-\alpha}{1-\alpha-\beta}\,.
\end{equation}
Equation (\ref{b3}) defines a consistent estimator for $\delta$.

In practice, however, values for  $\alpha$ and $\beta$ typically are not
readily obtainable, and one or both must be estimated.  This makes the
problem of estimating $\delta$ decidedly more difficult.  In fact, it is
essentially impossible to estimate any subgraph count $f_H(G)$ from a
single noisy observation~$G^{obs}$.

Formally, let $\mathcal{M}=\{(\alpha,\beta,\bA):0\leq\alpha\leq1,
0\leq\beta\leq1, A_{i,j}=0~\textrm{or}~1, A_{i,j}=A_{j,i}\}$ be the class
of all models defined under (\ref{eq:model}) and Assumption~1.  For any
model $M=(\alpha,\beta,\bA)\in\mathcal{M}$, we define its dual model as
$M^*=(1-\beta,1-\alpha,\bA^*)$, where $\bA^*=(A_{i,j}^*)_{p\times p}$
satisfies $A_{i,j}^*=1-A_{i,j}$ for any $i\neq j$. Denote by $F_M$ and
$F_{M^*}$ the joint distributions of $\bY$ when $\bY$ follows models $M$
and $M^*$, respectively. Finally, for a given subgraph density $f$ of
interest, define
\[
d_f=\sup_{M\in\mathcal{M}}|f(M)-f(M^*)|\,,
\]
where $f(M)$ and $f(M^*)$ are the associated subgraph densities based on
model $M$ and its dual model $M^*$, respectively.  We then have the
following  result.
Note that Theorem 1 holds for any subgraph density $f$. When $f$ is the
edge density, $d_f=1$.

\begin{theorem}\label{tm:1}
Write $\mathcal{E}$ for the class of all measurable functionals of the
data $\bY$. Let Assumption {\rm1} hold. If $d_f>0$, then it holds that
\[
\inf_{\hat{f}\in\mathcal{E}}\sup_{\mathcal{M}}\mathbb{P}\bigg(|\hat{f}-f|\geq\frac{d_f}{2}\bigg)\geq \frac{1}{2}\,.
\]
\end{theorem}

Theorem \ref{tm:1} indicates that it is in general impossible to produce
a consistent estimate of a subgraph density $f$ based on only one noisy
version of the adjacency matrix $\bA$.
% In our earlier illustration -- that of estimating the edge density
% $\delta$ -- we have $d_f=1$, for example.

To build intuition for the difficulty of this problem, consider again equation (\ref{eq:bary}), which indicates that $\bar{Y}$ is an unbiased estimate of
\begin{equation*}\label{eq:u1}
u_1\equiv (1-\delta)\alpha+\delta(1-\beta) \, ,
\end{equation*}
rather than of $\delta$.  This observation suggests use of the (asymptotically) unbiased estimating equation
\begin{equation}\label{eq:est1}
\hat{u}_1=(1-\delta)\alpha+\delta(1-\beta)\,,
\end{equation}
where $\hat{u}_1=\bar{Y}$.  It is obvious that $\alpha, \beta$ and $\delta$ cannot all be uniquely identified from this single equation.

Fortunately, in certain key areas of application we may observe more than one noisy version of the target network $G$.  For example, in computational biology, the common use of replicates at the most basic level of measurement (e.g., microarray expression) often allows for the construction of replicate networks (e.g., coexpression networks), as we demonstrate in Section~\ref{sec:illus}.  Similarly, in the context of computational neuroscience, it has become common now to obtain imaging measurements (e.g., fMRI) on multiple individuals within a given subpopulation (e.g., healthy females of a given age) and to create networks (e.g., functional connectivity networks) for each individual.  In the remainder of this section, we demonstrate how to estimate the edge density of the adjacency matrix $\bA$ consistently using just two or three replicates.  We then develop generalizations of these results for the case of arbitrary subgraphs in Section~\ref{sec4}.

\subsection{Estimation of unknown error rates}
\label{sec:edge.den}
\subsubsection{One of $\alpha$ or $\beta$ known}\label{se:oneunknown}

In some settings, one of either $\alpha$ or $\beta$ may be known.  For example, if the edges in $\bY$ are inferred through formal hypothesis testing, then $\alpha$ would be the user-specified rate of Type I error.  In this case, there are only two unknown parameters that need to be estimated, and we demonstrate how two replicates are sufficient to do so.

Suppose that $\bY$ is defined as above, and that $\bY_*=(Y_{i,j,*})_{p\times p}$ is an independent and identically distributed replicate of $\bY$.  Both are then noisy versions of the same adjacency matrix $\bA$, observed with the same error rates $\alpha$ and $\beta$. It follows from (\ref{eq:model}) that for $(i,j)$ with $A_{i,j}=1$,
\[
 Y_{i,j,*}-Y_{i,j}=\left\{ \begin{aligned}
         -1\,,~~&~\textrm{with probability}~\beta(1-\beta)\,,\\
         0\,,~~&~\textrm{with probability}~1-2\beta(1-\beta)\,,\\
         1\,,~~&~\textrm{with probability}~\beta(1-\beta)\,,\\
                          \end{aligned} \right.
 \]
and for $(i,j)$ with $A_{i,j}=0$,
 \[
Y_{i,j,*}-Y_{i,j}=\left\{ \begin{aligned}
         -1\,,~~&~\textrm{with probability}~\alpha(1-\alpha)\,,\\
         0\,,~~&~\textrm{with probability}~1-2\alpha(1-\alpha)\,,\\
         1\,,~~&~\textrm{with probability}~\alpha(1-\alpha)\,.\\
                          \end{aligned} \right.
 \]
Similar to (\ref{eq:bary}), we have
\begin{equation*}\label{eq:2}
\begin{split}
\mathbb{E}\bigg\{\frac{2}{p(p-1)}\sum_{i< j}|Y_{i,j,*}-Y_{i,j}|\bigg\}%=&~\frac{1}{p(p-1)}\sum_{(i,j)\in\mathcal{S}}|Y_{ij,*}-Y_{ij}|+\frac{1}{p(p-1)}\sum_{(i,j)\in\mathcal{S}^c}|Y_{ij,*}-Y_{ij}|\\
%=&~\delta\cdot\frac{1}\sum_{(i,j)\in\mathcal{S}}\mathbb{I}(Y_{ij,*}-Y_{ij}=\pm1)+(1-\delta)\cdot\frac{1}{|\mathcal{S}^c|}\sum_{(i,j)\in\mathcal{S}^c}\mathbb{I}(Y_{ij,*}-Y_{ij}=\pm1)\\
=&~2\{(1-\delta)\alpha(1-\alpha)+\delta\beta(1-\beta)\}\,.
\end{split}
\end{equation*}
Let
\begin{equation}\label{eq:u2}
u_2\equiv (1-\delta)\alpha(1-\alpha)+\delta\beta(1-\beta)\,,
\end{equation}
for which the method of moment estimate is
\begin{equation*}\label{eq:u2est}
\hat{u}_2=\frac{1}{p(p-1)}\sum_{i< j}|Y_{i,j,*}-Y_{i,j}|\,.
\end{equation*}
Therefore, we have a second estimating equation:
\begin{equation}\label{eq:est2}
\hat{u}_2=(1-\delta)\alpha(1-\alpha)+\delta\beta(1-\beta)\,.
\end{equation}

Combining (\ref{eq:est1}) and (\ref{eq:est2}), when $\alpha$ is known, the estimators for $\beta$ and $\delta$
are
\begin{equation}\label{eq:MLE1}
\left\{ \begin{aligned}
         \hat{\beta}=&~\frac{\hat{u}_2-\alpha+\hat{u}_1\alpha}{\hat{u}_1-\alpha}\, , \\
\hat{\delta}=&~\frac{(\hat{u}_1-\alpha)^2}{\hat{u}_1-\hat{u}_2-2\hat{u}_1\alpha+\alpha^2} \, ,
                          \end{aligned} \right.
\end{equation}
and when $\beta$ is known, the estimators for  $\alpha $ and $\delta$ are
\begin{equation}\label{eq:MLE2}
\left\{ \begin{aligned}
         \hat{\alpha}=&~\frac{\hat{u}_1\beta-\hat{u}_2}{\hat{u}_1+\beta-1}\, , \\
         \hat{\delta}=&~\frac{\hat{u}_1^2-\hat{u}_1+\hat{u}_2}{\hat{u}_1+\hat{u}_2-2\hat{u}_1\beta-(1-\beta)^2} \, .
                          \end{aligned} \right.
\end{equation}
The following proposition gives the convergence rates for the proposed estimators.
\begin{proposition}\label{tm:2}
Let $N=p(p-1)/2$. Under Assumption {\rm 1}, if $N_1=p(p-1)\delta\rightarrow\infty$ and $N_2=p(p-1)(1-\delta)\rightarrow\infty$, it holds that {\rm(i)} $\hat{\beta}=\beta+O_p(N^{-1/2})$ and $\hat{\delta}=\delta+O_p(N^{-1/2})$, provided that $\alpha$ is known and $\delta(1-\alpha-\beta)^2\geq c$ for some positive constant $c$, {\rm(ii)}
$\hat{\alpha}=\alpha+O_p(N^{-1/2})$ and $\hat{\delta}=\delta+O_p(N^{-1/2})$, provided that $\beta$ is known and $(1-\delta)(1-\alpha-\beta)^2\geq c$ for some positive constant $c$.
\end{proposition}

\begin{remark}\label{remark1}
Since our estimation of the unknown parameters is based on moment
estimation, the independent noise dictated by Assumption~1 is not
strictly necessary. As is shown in the proof of Proposition 1, the
convergence rate for the moment estimation of the unknown parameters is
determined by the convergence rates of $\hat{u}_1-u_1$ and
$\hat{u}_2-u_2$. For any $i< j$, let
$e_{i,j}=I(\varepsilon_{i,j}=0,1)-(1-\beta)$ for $(i,j)\in\mathcal{S}$
and $e_{i,j}=I(\varepsilon_{i,j}=1)-\alpha$ for $(i,j)\in\mathcal{S}^c$.
Recall $\mathbb{P}(\varepsilon_{i,j}=1)=\alpha$,
$\mathbb{P}(\varepsilon_{i,j}=0)=1-\alpha-\beta$ and
$\mathbb{P}(\varepsilon_{i,j}=-1)=\beta$. Then $\mathbb{E}(e_{i,j})=0$
for any $i< j$. If $\textrm{Var}(N^{-1/2}\sum_{i< j}e_{i,j})\leq C$ for
some positive constant $C$, then $\hat u_1=u_1+O_p(N^{-1/2})$ without the
independence assumption. When Assumption 1 is satisfied,
$\textrm{Var}(N^{-1/2}\sum_{i<
j}e_{i,j})=\delta\beta(1-\beta)+(1-\delta)\alpha(1-\alpha)$. Analogously,
$\hat u_2=u_2+O_p(N^{-1/2})$ still holds when some dependency among
$\varepsilon_{i,j}$ $(i< j)$ is present. Hence, the results of
Proposition 1 still hold when there is some dependency among $\varepsilon_{i,j}$ $(i< j)$.
\end{remark}

\begin{remark} \label{remark2}
It is not strictly necessary that $\bY_*$ derive from the same underlying
adjacency matrix $\bA$ as $\bY$. More specifically, let
$\bA_*=(A_{i,j,*})_{p\times p}$ be the adjacency matrix underlying the observation
$\bY_*$, and let $\mathcal{B}_1=\{(i,j):A_{i,j}=A_{i,j,*}, i< j\}$. The average
of $|Y_{i,j,*}-Y_{i,j}|$ over $\mathcal{B}_1$ provides an unbiased
estimator for the parameter $u_2$ defined in (\ref{eq:u2}), while the
original estimator $\hat{u}_2$ defined in (\ref{eq:est2}) is no longer
unbiased if $|\mathcal{B}_1|<p(p-1)/2$. As long as
$\theta_1=2|\mathcal{B}_1|/\{p(p-1)\}$ is sufficiently close to 1, e.g. $|1 - \theta_1|=o(p^{-1})$, the
bias term in $\hat{u}_2$ will be asymptotically negligible, which means
the estimators (\ref{eq:MLE1}) and (\ref{eq:MLE2}) will still be consistent.
\end{remark}

\begin{theorem}\label{tm:twounknown}
Let $N=p(p-1)/2$. Under Assumptions {\rm 1}, if $N_1=p(p-1)\delta\rightarrow\infty$ and $N_2=p(p-1)(1-\delta)\rightarrow\infty$, it holds that {\rm(i)} $\sqrt{N}(\hat{\beta}-\beta,\hat{\delta}-\delta)^\T\rightarrow_d \mathcal{N}(\bzero,\bSigma_{1,\alpha})$ with $\bSigma_{1,\alpha}$ defined as {\rm(\ref{eq:sigma1alpha})} in the Appendix, provided that $\alpha$ is known and $\delta(1-\alpha-\beta)^2\geq c$ for some positive constant $c$, {\rm(ii)}
$\sqrt{N}(\hat{\alpha}-\alpha,\hat{\delta}-\delta)^\T\rightarrow_d
\mathcal{N}(\bzero,\bSigma_{1,\beta})$ with $\bSigma_{1,\beta}$ defined as
{\rm(\ref{eq:sigma1beta})} in the Appendix, provided that $\beta$
is known and $(1-\delta)(1-\alpha-\beta)^2\geq c$ for some positive
constant $c$.

\end{theorem}

We can construct approximate confidence intervals for $\delta$ based on the
asymptotic normality stated in Theorem \ref{tm:twounknown}. Let $\sigma^2$ denote
the asymptotic variance of $\sqrt{N}(\hat \delta-\delta)$. Then $\sigma$ depends on
unknown parameters $\delta$ and $\beta$ or $\alpha$. Replacing those
unknown parameters by their estimates, we obtain an estimated asymptotic
variance denoted by $\hat \sigma^2$. Then an approximate 95\% confidence
interval for $\delta$ is
\begin{equation} \label{CIdelta}
\big(\hat \delta - 1.96 \hat \sigma N^{-1/2}, \; \hat \delta + 1.96 \hat
\sigma N^{-1/2}\big)\,.
\end{equation}

\subsubsection{Both $\alpha$ and $\beta$ unknown}\label{se:twounknown}

When both $\alpha$ and $\beta$ are unknown, together with $\delta$ there
are three unknown parameters to be estimated.  We show that three replicates
are sufficient for asymptotically consistent estimation in this setting.

Let $\bY, \bY_*$ and $\bY_{**}$ be independent and identically distributed replicates from (\ref{eq:model}). %For any
%$(i,j)$, it holds that
%\[
%\begin{split}
%(Y_{ij,**}-Y_{ij,*})-(Y_{ij,*}-Y_{ij})=&~A_{ij}\{\mathbb{I}(\varepsilon_{ij,**}=0)-2\mathbb{I}(\varepsilon_{ij,*}=0)+\mathbb{I}(\varepsilon_{ij}=0)\}\\
%&+\{\mathbb{I}(\varepsilon_{ij,**}=1)-2\mathbb{I}(\varepsilon_{ij,*}=1)+\mathbb{I}(\varepsilon_{ij}=1)\}.
%\end{split}
%\]
Hence, for $(i,j)$ with $A_{i,j}=1$,
 \[
 Y_{i,j,**}-2Y_{i,j,*}+Y_{i,j}=\left\{ \begin{aligned}
         -2\,,~~&~\textrm{with probability}~\beta^2(1-\beta)\,, \\
         -1\,,~~&~\textrm{with probability}~2\beta(1-\beta)^2\,,\\
         0\,,~~&~\textrm{with probability}~\beta^3+(1-\beta)^3\,,\\
         1\,,~~&~\textrm{with probability}~2\beta^2(1-\beta)\,,\\
         2\,,~~&~\textrm{with probability}~\beta(1-\beta)^2\,,
                          \end{aligned} \right.
 \]
and for $(i,j)$ with $A_{i,j}=0$,
 \[
 Y_{i,j,**}-2Y_{i,j,*}+Y_{i,j}=\left\{ \begin{aligned}
         -2\,,~~&~\textrm{with probability}~\alpha(1-\alpha)^2\,, \\
         -1\,,~~&~\textrm{with probability}~2\alpha^2(1-\alpha)\,,\\
         0\,,~~&~\textrm{with probability}~\alpha^3+(1-\alpha)^3\,,\\
         1\,,~~&~\textrm{with probability}~2\alpha(1-\alpha)^2\,,\\
         2\,,~~&~\textrm{with probability}~\alpha^2(1-\alpha)\,.
                          \end{aligned} \right.
 \]

Arguing in a fashion analogous to that used in producing the parameters $u_1$ and $u_2$, we emerge with the parameter
\begin{equation} \label{eq:u3}
u_3\equiv
 (1-\delta)\alpha(1-\alpha)^2+\delta\beta^2(1-\beta)\,,
\end{equation}
with corresponding method of moment estimator
\begin{equation*}\label{eq:est3}
\hat{u}_3=\frac{2}{3p(p-1)}\sum_{i< j}I(Y_{i,j,**}-2Y_{i,j,*}+Y_{i,j}=1~\textrm{or}-2)\,,
\end{equation*}
from which we obtain a third estimating equation:
\begin{equation}\label{eq:estequ2}
\hat u_3 =(1-\delta)\alpha(1-\alpha)^2+\delta\beta^2(1-\beta)\,.
\end{equation}

Combining (\ref{eq:est1}), (\ref{eq:est2}), and (\ref{eq:estequ2}), we
have a nonlinear system of three equations with three unknowns. This nonlinear system can be solved by some simple numerical iterations. For example, it follows from (\ref{eq:u3}) that
\begin{equation}\label{iter1}
\hat \alpha =\frac{\hat u_3 - \delta \beta^2(1-\beta)}{ (1- \delta)(1-\alpha)^2}\,.
\end{equation}
Starting with an initial value $\alpha_0$, we compute the estimates for
$\beta, \, \delta$
and $\alpha$ recursively using (\ref{eq:MLE1}) and (\ref{iter1}) until the absolute difference between
two successive values for $ \hat\alpha$ is smaller than a prescribed small number.
% Like all iterative algorithms, it is always recommended to use multiple initial
% values.
Analogous to Proposition \ref{tm:2} and Theorem \ref{tm:twounknown}, we have the following result.
\begin{theorem}\label{tm:3}
Let $N=p(p-1)/2$. Under Assumptions {\rm 1}, if
$N_1=p(p-1)\delta\rightarrow\infty$ and
$N_2=p(p-1)(1-\delta)\rightarrow\infty$, it holds that
$\hat{\alpha}=\alpha+O_p(N^{-1/2})$, $\hat{\beta}=\beta+O_p(N^{-1/2})$
and $\hat{\delta}=\delta+O_p(N^{-1/2})$, provided that
$\delta(1-\delta)(1-\alpha-\beta)^4\geq c$ for some positive constant
$c$. More specifically, we have
$\sqrt{N}(\hat{\alpha}-\alpha,\hat{\beta}-\beta,\hat{\delta}-\delta)^\T\rightarrow_d
\mathcal{N}(\bzero,\bSigma_2)$ with $\bSigma_2$ defined as {\rm(\ref{eq:sigma2})} in the Appendix.
\end{theorem}

% The bootstrap procedure descripted in Section \ref{se:oneunknown} can be easily
% adapted to handle the current setting when both $\alpha$ and $\beta$ are unknown.
% To this end, we generate, in addition to $\bY^\star, \bY^\star_*$ defined
%  in (\ref{eq:data1}),
% a third bootstrap network
%  \begin{equation}\label{eq:data3}
%   Y_{i,j,**}^\star= A_{i,j}^\star
% I(\varepsilon_{i,j,**}^\star=0)+I(\varepsilon_{i,j,**}^\star=1), \quad i< j,
%   \end{equation}
% where $\varepsilon_{i,j,**}^\star$ are independent and of the same distribution
% as $\varepsilon_{i,j}^\star$, and $\varepsilon_{i,j,**}^\star=\varepsilon_{j,i,**}^\star$. Then the bootstrap estimators $(\hat
% \alpha^\star, \hat \beta^\star, \hat \delta^\star)$ are obtained in the same manner
% as $(\hat
% \alpha, \hat \beta, \hat \delta)$ above with $(\hat u_1, \hat u_2, \hat u_3)$ replaced
% by $(\hat u_1^\star, \hat u_2^\star, \hat u_3^\star)$, where $\hat u_1^\star, \hat u_2^\star$ are defined as in
% (\ref{eq:new3}), and
% \[
% \hat u_3^\star= \frac{2}{6N}\sum_{i < j}I(Y_{i,j,**}^{\star}-2Y_{i,j,*}^{\star}+Y_{i,j}^{\star}=1~\textrm{or}-2)\,.
% \]
% Repeating the above bootstrap sampling $B$ times, we obtain $B$ bootstrap estimates arranged in ascending order: $\hat \delta^\star_{[1]} \le \cdots \le \hat \delta^\star_{[B]}$.
% Then an approximate 95\% bootstrap confidence interval for $\delta$ admits
% the same form as in (\ref{bootInt}).

\section{Inference for higher-order subgraph densities}
\label{sec4}

Now we address the inference of higher-order subgraph densities
$C_{\mathcal{V}}(\tau_1,\ldots,\tau_k)$ defined in (\ref{eq:C}) with $k\geq2$.
We continue to use method of moments
estimation, but with the error rates $\alpha$ and/or $\beta$ replaced
by their estimators obtained in Section~\ref{sec:edge.den}. The resulting
estimators admit a uniform representation; see (\ref{hatCv}) below. However,
interval estimation for $C_{\mathcal{V}}(\tau_1,\ldots,\tau_k)$ requires the evaluation of an asymptotic variance that is a function of the individual (unknown) network edges $A_{i,j}$.
Accordingly, we propose a new and non-standard bootstrap method
to overcome this
obstacle. To highlight the key ideas, we first proceed in Section \ref{se:allknown} with both $\alpha$ and $\beta$
assumed to be known. The development with unknown $\alpha$ and $\beta$ is then presented in Section \ref{sec:higher.order}.

\subsection{Inference for subgraph densities with known error rates}\label{se:allknown}

In this subsection, we assume that both $\alpha$ and $\beta$ are known.
All inference will be based on one observed network $\bY=(Y_{i,j})_{p\times p}$ only.
It follows from (\ref{eq:model}) and Assumption 1 that
\[
A_{i,j} = \frac{\mathbb{E}(Y_{i,j} -\alpha)}{1-\alpha - \beta} \qquad {\rm and} \qquad
1- A_{i,j} = \frac{\mathbb{E}(1- \beta- Y_{i,j})}{1-\alpha - \beta}\,.
\]
Hence (\ref{eq:C}) admits a more compact representation
\begin{equation*}\label{eq:ex1}
C_{\mathcal{V}}=: C_{\mathcal{V}}(\tau_1,\ldots,\tau_k)
=\frac{1}{(1-\alpha-\beta)^k}\cdot\frac{1}{|\mathcal{V}|}\sum_{\bv=(i_1,i_1',\ldots,i_k,i_k')\in\mathcal{V}}\prod_{\ell=1}^k\mathbb{E}\big\{\varphi_\ell\big(Y_{i_\ell,i_\ell'}\big)\big\}\,,
\end{equation*}
where
% For given $\tau_1,\ldots,\tau_k\in\{0,1\}$, we define
\begin{equation}\label{eq:phiell}
\varphi_\ell(x)=(x-\alpha)^{\tau_\ell}(1-\beta-x)^{1-\tau_\ell}.
\end{equation}
% for $x\in\{0,1\}$ and $\ell=1,\ldots,k$. To simplify the notation and without causing confusion, we write $C_{\mathcal{V}}(\tau_1,\ldots,\tau_k)$ as $C_{\mathcal{V}}$.
% It follows from (\ref{eq:model}) that $
% A_{i_\ell,i_\ell'}^{\tau_\ell}(1-A_{i_\ell,i_\ell'})^{1-\tau_\ell}=(1-\alpha-\beta)^{-1}\mathbb{E}\{\varphi_\ell(Y_{i_\ell,i_\ell'})\}
% $,
% which implies that
% Due to
Note that
$|\{i_{\ell_1},i_{\ell_1}'\}\cap\{i_{\ell_2},i_{\ell_2}'\}|\leq 1$ for
any $\ell_1\neq \ell_2$, Assumption 1 implies that the
$\{Y_{i_\ell,i_\ell'}\}_{\ell=1}^k$ are independent of each other.
Therefore, a natural method of moments estimator for $C_{\mathcal{V}}$
can be defined as
\begin{equation} \label{Ctilda}
\widetilde{C}_{\mathcal{V}}=\frac{\widetilde{T}_{\mathcal{V}}}{(1-{\alpha}-{\beta})^k}\, ,
\end{equation}
where
% by (\ref{eq:ex1}), we have
% \begin{equation*}
% \begin{split}
% C_{\mathcal{V}}=&~\frac{1}{(1-\alpha-\beta)^k}\cdot\frac{1}{|\mathcal{V}|}\sum_{\bv=(i_1,i_1',\ldots,i_k,i_k')\in\mathcal{V}}\mathbb{E}\bigg\{\prod_{\ell=1}^k\varphi_\ell\big(Y_{i_\ell,i_\ell'}\big)\bigg\}\\
% =&:\frac{T_{\mathcal{V}}}{(1-\alpha-\beta)^k}\,.
% \end{split}
% \end{equation*}
% With known $(\alpha,\beta)$, we need only estimate $T_{\mathcal{V}}$. Based on the principle of method-of-moments estimation, we can estimate $T_{\mathcal{V}}$ as
\begin{equation*}
\widetilde{T}_{\mathcal{V}}=\frac{1}{|\mathcal{V}|}\sum_{\bv=(i_1,i_1',\ldots,i_k,i_k')\in\mathcal{V}}\prod_{\ell=1}^k\varphi_\ell\big(Y_{i_\ell,i_\ell'}\big)\,.
\end{equation*}

To state the asymptotic properties of $\widetilde{C}_{\mathcal{V}}$, we need to introduce some notation.
For any $\bv=(i_1,i_1',\ldots,i_k,i_k')\in\mathcal{V}$ with $\mathcal{V}$ given in (\ref{eq:V}) and $1\leq\ell_1<\cdots<\ell_s\leq k$ with $1\leq s\leq k-1$, we define
\[
\begin{split}
\mathcal{G}_{\ell_1,\ldots,\ell_s}(\bv)=\{(\theta_1,\theta'_1,\ldots,\theta_s,\theta'_s):&~(i_1,i_1',\ldots,i_{\ell_1-1},i_{\ell_1-1}',\theta_1,\theta'_1,\\
&~i_{\ell_1+1},i_{\ell_1+1}',\ldots,i_{\ell_2-1},i_{\ell_2-1}',\theta_2,\theta_2',\\
&~\ldots,i_{\ell_s-1},i_{\ell_s-1}',\theta_{s},\theta_{s}',i_{\ell_s+1},i_{\ell_s+1}',\ldots,i_k,i_k')\in\mathcal{V}\}\,.
\end{split}
\]
In turn, we define the quantity
\begin{equation}\label{eq:bound}
\aleph_{\mathcal{V}}(s)=\max_{\bv\in\mathcal{V}}\max_{1\leq \ell_1<\cdots<\ell_s\leq k}|\mathcal{G}_{\ell_1,\ldots,\ell_s}(\bv)|
\end{equation}
and
\[
\aleph_{\mathcal{V}}=\max_{1\leq s\leq k-1}\aleph_{\mathcal{V}}(s)\,.
\]

\begin{proposition}\label{tm:pre}
Under Assumption {\rm 1}, if $|1-\alpha-\beta|\geq c$ for
some positive constant $c$, it holds that
$|\widetilde{C}_{\mathcal{V}}-C_{\mathcal{V}}|=O_p(\sqrt{\aleph_{\mathcal{V}}/|\mathcal{V}|})$ as $p\rightarrow\infty$.
% as $p\rightarrow\infty$.
\end{proposition}

Notice that $\aleph_{\mathcal{V}}(1)\leq\cdots\leq\aleph_{\mathcal{V}}(k-1)$, so that $\aleph_{\mathcal{V}}=\aleph_{\mathcal{V}}(k-1)$. If we select $\mathcal{V}=\{(i_1,i_1',\ldots,i_k,i_k'):i_{\ell}'=i_{\ell+1}~\textrm{for each}~\ell=1,\ldots,k-1, i_1\neq i_2\neq\cdots\neq i_k\neq i_{k}'\}$, which corresponds to counting paths of length $k\geq2$, then $|\mathcal{V}|=p\cdots(p-k)$ and $\aleph_{\mathcal{V}}(s)=(p-k+s-1)\cdots(p-k)$ for any $1\leq s\leq k-1$. Alternately, if we select $\mathcal{V}=\{(i_1,i_1',\ldots,i_k,i_k'):i_{\ell}'=i_{\ell+1}~\textrm{for each}~\ell=1,\ldots,k-1, i_k'=i_1, i_1\neq i_2\neq\cdots\neq i_k\}$, which corresponds to counting cycles of length $k\geq3$, then $|\mathcal{V}|=p\cdots(p-k+1)$, $\aleph_{\mathcal{V}}(1)=1$ and $\aleph_{\mathcal{V}}(s)=(p-k+s-1)\cdots(p-k+1)$ for any $2\leq s\leq k-1$.  As a result, in the case of counting paths or cycles of length $k$,
$
\aleph_{\mathcal{V}}/|\mathcal{V}|=\{p(p-1)\}^{-1}
$. Letting $N=p(p-1)/2$, we then have $|\widetilde{C}_{\mathcal{V}}-C_{\mathcal{V}}|=O_p(N^{-1/2})$.

To investigate the asymptotic distribution of $\widetilde{C}_{\mathcal{V}}-C_{\mathcal{V}}$, we require the following mild assumption.

\begin{assumption}
(i) $\aleph_{\mathcal{V}}(s)/\aleph_{\mathcal{V}}\rightarrow0$ for any $1\leq s\leq k-2$, and (ii)
\[
\max_{\bv\in\mathcal{V}}\max_{1\leq \ell_1<\cdots<\ell_{k-1}\leq k}|\mathcal{G}_{\ell_1,\ldots,\ell_{k-1}}(\bv)|\asymp \min_{\bv\in\mathcal{V}}\min_{1\leq \ell_1<\cdots<\ell_{k-1}\leq k}|\mathcal{G}_{\ell_1,\ldots,\ell_{k-1}}(\bv)|\,.\]
\end{assumption}

Let
\begin{equation}\label{eq:Sv}
S_{\mathcal{V}}=\frac{\sqrt{N}}{(1-\alpha-\beta)^k}\sum_{j=1}^k\frac{(-1)^{1-\tau_j}}{|\mathcal{V}|}\sum_{\bv=(i_1,i_1',\ldots,i_k,i_k')\in\mathcal{V}}\bigg[\mathring{Y}_{i_j,i_j'}\prod_{\ell\neq j}\mathbb{E}\big\{\varphi_\ell\big(Y_{i_\ell,i_\ell'}\big)\big\}\bigg]\,,
\end{equation}
where $\mathring{Y}_{i_j,i_j'}=Y_{i_j,i_j'}-\mathbb{E}(Y_{i_j,i_j'})$ and $\varphi_\ell(\cdot)$ is defined as {\rm(\ref{eq:phiell})}.
%We have the following asymptotic expansion of $\sqrt{N}(\widetilde{C}_{\mathcal{V}}-C_{\mathcal{V}})$.

\begin{proposition}\label{pre2}
Let $N=p(p-1)/2$, $\aleph_{\mathcal{V}}/|\mathcal{V}|\asymp N^{-1}$ and
$|1-\alpha-\beta|\geq c$ for some positive constant $c$. Under
Assumptions {\rm 1} and {\rm 2}, it holds that
$
\sqrt{N}(\widetilde{C}_{\mathcal{V}}-C_{\mathcal{V}})=S_{\mathcal{V}}+o_p(1)
$
for $S_{\mathcal{V}}$ defined as {\rm(\ref{eq:Sv})}.
\end{proposition}

Recall $Y_{i,j}=Y_{j,i}$ for any $i\neq j$, and \[
\max_{\bv\in\mathcal{V}}\max_{1\leq \ell_1<\cdots<\ell_{k-1}\leq k}|\mathcal{G}_{\ell_1,\ldots,\ell_{k-1}}(\bv)|\asymp \min_{\bv\in\mathcal{V}}\min_{1\leq \ell_1<\cdots<\ell_{k-1}\leq k}|\mathcal{G}_{\ell_1,\ldots,\ell_{k-1}}(\bv)|\,.\]
Notice that $\aleph_{\mathcal{V}}/|\mathcal{V}|\asymp N^{-1}$. Then it holds that
\begin{equation*}\label{eq:asyknown}
S_{\mathcal{V}}=\frac{1}{\sqrt{N}}\sum_{i<j}\mathring{Y}_{i,j}K_{i,j}
\end{equation*}
for some constants $K_{i,j}$. Since $\{Y_{i,j}\}_{i<j}$ are independent, it follows from the Central Limit Theorem that
\[
\sqrt{N}(\widetilde{C}_{\mathcal{V}}-C_{\mathcal{V}})\xrightarrow{d}\mathcal{N}(0,\sigma_{\mathcal{V}}^2)
\]
as $p\rightarrow\infty$, where the asymptotic variance $\sigma^2_{\mathcal{V}}$ satisfies
\begin{equation}\label{eq:asyv}
\begin{split}
\sigma^2_{\mathcal{V}}=\lim_{p\rightarrow\infty} \frac{1}{N}\sum_{i<j}K_{i,j}^2\textrm{Var}(Y_{i,j})\,.
\end{split}
\end{equation}

It is easy to see from (\ref{eq:model}) that $\textrm{Var}(Y_{i,j})=A_{i,j}(1-\alpha-\beta)(\beta-\alpha)+\alpha(1-\alpha)$. As
we do not know $A_{i,j}$,  it is impossible
to compute $\sigma^2_{\mathcal{V}}$ based on (\ref{eq:asyv}) (except for
some simple special cases such as when $K_{i,j}^2$ does not vary with respect to $i$ and $j$).
To overcome this difficulty, we propose a non-standard bootstrap procedure as follows:
we draw bootstrap samples $Y^\dag$ according to
\begin{equation}\label{eq:dag}
Y_{i,j}^\dag \equiv  Y_{j,i}^\dag = Y_{i,j}I(\eta_{i,j}=0)+I(\eta_{i,j}=1)\, \quad
{\rm for} \; i<j\,,
\end{equation}
where $\{ \eta_{i,j}\}$ are independent random variables,
$\mathbb{P}(\eta_{i,j}=0)=\gamma_1$, $\mathbb{P}(\eta_{i,j}=1)=\gamma_2$ and
$\mathbb{P}(\eta_{i,j}=-1)=1-\gamma_1-\gamma_2$, with
 $\gamma_1>0$, $\gamma_2>0$ and $\gamma_1+ \gamma_2< 1$ satisfying
\begin{equation}\label{eq:theta}
\left\{ \begin{aligned}
\gamma_1(1-\gamma_1-2\gamma_2)=&~\beta-\alpha\,,\\
\gamma_2(1-\gamma_2)=&~\alpha(1-\beta)\,.
 \end{aligned}\right.
 \end{equation}
Now let
\[
S_{\mathcal{V}}^\dag=\frac{\sqrt{N}}{(1-\alpha-\beta)^k}\sum_{j=1}^k\frac{(-1)^{1-\tau_j}}{|\mathcal{V}|}\sum_{\bv=(i_1,i_1',\ldots,i_k,i_k')\in\mathcal{V}}\bigg\{\mathring{Y}_{i_j,i_j'}^\dag\prod_{\ell\neq j}\varphi_\ell\big(Y_{i_\ell,i_\ell'}\big)\bigg\}
\]
with
$\mathring{Y}_{i_j,i_j'}^\dag=Y_{i_j,i_j'}^\dag-Y_{i_j,i_j'}\gamma_1-\gamma_2$
and  $\varphi_\ell(\cdot)$ defined as  in {\rm(\ref{eq:phiell})}. Theorem
\ref{tm:bootstrap} below shows that the distribution of
$\sqrt{N}(\widetilde{C}_{\mathcal{V}}-C_{\mathcal{V}})$ can be
approximated by the conditional distribution of $S_{\mathcal{V}}^\dag$
given $\bY=(Y_{i,j})_{p\times p}$.

Note that (\ref{eq:theta}) may admit more than one legitimate
solution for $(\gamma_1, \gamma_2)$; any one of them can serve for our purpose. Furthermore,
the bootstrap sample $(Y_{i,j}^\dag)_{p\times p}$ does not necessarily resemble the full behavior of the original sample $(Y_{i,j})_{p\times p}$. What matters here is the fact that it has the correct (conditional expected) variance:
\[
\mathbb{E}\{{\rm Var}(Y^{\dag}_{i,j}\,|\,\bY)\}={\rm Var}(Y_{i,j})\,.
\]
Note that Var$(Y_{i,j}^\dag\,|\,\bY)= Y_{i,j}(\beta-\alpha)+\alpha(1-\beta)$, which is guaranteed by
(\ref{eq:theta}).

\begin{theorem}\label{tm:bootstrap}
Under the conditions of Proposition {\rm \ref{pre2}}, it holds that
\begin{equation*}\label{eq:expandnew}
\sup_{z\in\mathbb{R}}\big|\mathbb{P}\big\{\sqrt{N}(\widetilde{C}_{\mathcal{V}}-C_{\mathcal{V}})>z\big\}-\mathbb{P}(S_{\mathcal{V}}^\dag>z\,\big|\,\bY)\big|\rightarrow0
\end{equation*}
as $p\rightarrow\infty$.
\end{theorem}

Theorem \ref{tm:bootstrap} can be extended to multiple cases easily, which is required
for constructing the joint confidence regions for several subgraph densities, or
their functions such as the clustering coefficient.
For given
$(\mathcal{V}_1,\tau_{1,1},\ldots,\tau_{1,k_1}),\ldots,(\mathcal{V}_m,
\tau_{m,1},\ldots,\tau_{m,k_m})$, we approximate the joint distribution of
$
\sqrt{N}(\widetilde{C}_{\mathcal{V}_1}-C_{\mathcal{V}_1},\ldots,\widetilde{C}_{\mathcal{V}_m}-C_{\mathcal{V}_m})^\T
$ by the following parametric bootstrap procedure:

\begin{algorithmic}[1]
  \Repeat
  \State  given data $\bY=(Y_{i,j})_{p\times p}$ draw bootstrap samples $\bY^\dag=(Y_{i,j}^\dag)_{p\times p}$ as in (\ref{eq:dag})
  \State  calculate bootstrap estimate $\bvartheta^\dag=(\vartheta_1^\dag,\ldots,\vartheta_m^\dag)^\T$, where
  \[
\vartheta_q^\dag=\frac{\sqrt{N}}{(1-\alpha-\beta)^{k_q}}\sum_{j=1}^{k_q}\frac{(-1)^{1-\tau_{q,j}}}{|\mathcal{V}_q|}\sum_{\bv=(i_1,i_1',\ldots,i_{k_q},i_{k_q}')\in\mathcal{V}_q}\bigg\{\mathring{Y}_{i_j,i_j'}^\dag\prod_{\ell\neq j}\varphi_{q,\ell}\big(Y_{i_\ell,i_\ell'}\big)\bigg\}
\]
for each $q=1,\ldots,m$
  with $\mathring{Y}_{i_j,i_j'}^\dag=Y_{i_j,i_j'}^\dag-Y_{i_j,i_j'}\gamma_1-\gamma_2$ and $\varphi_{q,\ell}(x)=(x-\alpha)^{\tau_{q,\ell}}(1-\beta-x)^{1-\tau_{q,\ell}}$ for any $x\in\{0,1\}$
  \Until $B$ replicates obtained, for a large integer $B$
  \State {\bf approximate} the joint distribution by the empirical distribution function of $\{\bvartheta_1^\dag,\ldots,\bvartheta_B^\dag\}$
 \end{algorithmic}

% \begin{remark}
% If we are interested in edge density $\delta$, we can select $k=1$, $\tau_1=1$ and $\mathcal{V}=\{(i_1,i_1'):i_1\neq i_1'\}$. Then
%$
%S_{\mathcal{V}}=2^{-1}(1-\alpha-\beta)^{-1}N^{-1/2}\sum_{i_1\neq i_1'}\mathring{Y}_{i_1,i_1'}
%$ and $S_{\mathcal{V}}^\dag=2^{-1}(1-\alpha-\beta)^{-1}N^{-1/2}\sum_{i_1\neq i_1'}\mathring{Y}_{i_1,i_1'}^\dag$.
% \end{remark}

 \begin{remark} \label{2star0}
For estimating two-star density, we let $k=3$, $\tau_1=\tau_2=1$, $\tau_3=0$ and $\mathcal{V}=\{(i_1,i_1',i_2,i_2',i_3,i_3'):i_{1}'=i_{2}, i_{2}'=i_3, i_{3}'=i_1, i_1\neq i_2\neq i_3\}$. Then
\begin{align*}
&S_{\mathcal{V}}=\frac{(1-\alpha-\beta)^{-1}\sqrt{N}}{p(p-1)(p-2)}\sum_{i_1\neq i_2\neq i_3}\{\mathring{Y}_{i_1,i_2}A_{i_2,i_3}(1-A_{i_3,i_1})\\
&~~~~~~~~~~~~~~~~~~~~~~~~~~+\mathring{Y}_{i_2,i_3}A_{i_1,i_2}(1-A_{i_3,i_1})-\mathring{Y}_{i_3,i_1}A_{i_1,i_2}A_{i_2,i_3}\}
 \end{align*}and
 \begin{align*}
 &S_{\mathcal{V}}^\dag=\frac{(1-\alpha-\beta)^{-3}\sqrt{N}}{p(p-1)(p-2)}\sum_{i_1\neq i_2\neq i_3}\{\mathring{Y}_{i_1,i_2}^\dag(Y_{i_2,i_3}-\alpha)(1-\beta-Y_{i_3,i_1})\\
 &~~~~~~~~~~~~~~~~~~~~~~~+\mathring{Y}_{i_2,i_3}^\dag(Y_{i_1,i_2}-\alpha)(1-\beta-Y_{i_3,i_1})-\mathring{Y}_{i_3,i_1}^\dag(Y_{i_1,i_2}-\alpha)(Y_{i_2,i_3}-\alpha)\}\,.
 \end{align*}
 \end{remark}

\begin{remark} \label{3angle0}
For estimating triangle density, we let $k=3$,
$\tau_1=\tau_2=\tau_3=1$ and
$\mathcal{V}=\{(i_1,i_1',i_2,i_2',i_3, i_3'):i_{1}'=i_{2}, i_2'=i_3,
i_3'=i_1, i_1\neq i_2\neq i_3\}$. Then
$$S_{\mathcal{V}}=\frac{(1-\alpha-\beta)^{-1}\sqrt{N}}{p(p-1)(p-2)}
\sum_{i_1\neq i_2\neq
i_3}(\mathring{Y}_{i_1,i_2}A_{i_2,i_3}A_{i_3,i_1}+
\mathring{Y}_{i_2,i_3}A_{i_1,i_2}A_{i_3,i_1}+
\mathring{Y}_{i_3,i_1}A_{i_1,i_2}A_{i_2,i_3})$$ and
\[
\begin{split}
S_{\mathcal{V}}^\dag=\frac{(1-\alpha-\beta)^{-3}\sqrt{N}}{p(p-1)(p-2)}
\sum_{i_1\neq i_2\neq i_3}&\big\{\mathring{Y}_{i_1,i_2}^\dag(Y_{i_2,i_3}-\alpha)
(Y_{i_3,i_1}-\alpha)+\mathring{Y}_{i_2,i_3}^\dag(Y_{i_1,i_2}-\alpha)
(Y_{i_3,i_1}-\alpha) \\
&~~+\mathring{Y}_{i_3,i_1}^\dag(Y_{i_1,i_2}-\alpha)
(Y_{i_2,i_3}-\alpha)\big\}\,.
\end{split}
\]
\end{remark}

\subsection{Estimation of subgraph densities with unknown error rates}
\label{sec:higher.order}

When the error rates
 $\alpha$ and $\beta$ are unknown, we simply use the estimator $\widetilde C_{\mathcal{V}}$ defined in (\ref{Ctilda}) with $\alpha$ and $\beta$ replaced by their estimators
derived in Section \ref{sec:edge.den}. Then its asymptotic properties are more
complex, and, consequently, the construction of confidence sets is more involved.
Note that we need at most three samples $\bY, \bY_{*}, \bY_{**}$ for estimating
$\alpha$ and $\beta$ in Section \ref{sec:edge.den}. Obviously an
improvement to the approach outlined below can be entertained by combining the three
estimators obtained from computing (\ref{Ctilda}), using one of the three available
samples each time. For simplicity, we do not pursue this idea further here.

Given estimators $(\tilde{\alpha},\tilde{\beta})$ for $(\alpha,\beta)$, we define
\begin{equation} \label{hatCv}
\widehat{C}_{\mathcal{V}}=\frac{\widehat{T}_{\mathcal{V}}}{(1-\tilde{\alpha}-\tilde{\beta})^k}
\end{equation}
as an estimator for $C_{\mathcal{V}}$, where
\begin{equation*}
\widehat{T}_{\mathcal{V}}=\frac{1}{|\mathcal{V}|}\sum_{\bv=(i_1,i_1',\ldots,i_k,i_k')\in\mathcal{V}}\prod_{\ell=1}^k\big(Y_{i_\ell,i_\ell'}-\tilde{\alpha}\big)^{\tau_\ell}\big(1-\tilde{\beta}-Y_{i_\ell,i_\ell'}\big)^{1-\tau_\ell}\,.
\end{equation*}
See also (\ref{Ctilda}).
Here we let $(\tilde{\alpha},\tilde{\beta})=(\alpha,\hat{\beta})$ for
$\hat{\beta}$ defined in (\ref{eq:MLE1}) if $\alpha$ is known,
$(\tilde{\alpha},\tilde{\beta})=(\hat{\alpha},\beta)$ for $\hat{\alpha}$
defined in (\ref{eq:MLE2}) if $\beta$ is known, and
$(\tilde{\alpha},\tilde{\beta})=(\hat{\alpha},\hat{\beta})$ for
$(\hat{\alpha},\hat{\beta})$ defined in Section \ref{se:twounknown} if
both $\alpha$ and $\beta$ are unknown. Let
\begin{equation}\label{eq:da}
\begin{split}
\Delta_{\alpha,\mathcal{V}}=&~\frac{kC_{\mathcal{V}}}{1-\alpha-\beta}-\frac{1}{(1-\alpha-\beta)^k}\sum_{j:\tau_j=1}\frac{1}{|\mathcal{V}|}\sum_{\bv=(i_1,i_1',\ldots,i_k,i_k')\in\mathcal{V}}\prod_{\ell\neq j}\mathbb{E}\big\{\varphi_\ell\big(Y_{i_\ell,i_\ell'}\big)\big\}
\end{split}
\end{equation}
and
\begin{equation}\label{eq:db}
\begin{split}
\Delta_{\beta,\mathcal{V}}=&~\frac{kC_{\mathcal{V}}}{1-\alpha-\beta}-\frac{1}{(1-\alpha-\beta)^k}\sum_{j:\tau_j=0}\frac{1}{|\mathcal{V}|}\sum_{\bv=(i_1,i_1',\ldots,i_k,i_k')\in\mathcal{V}}\prod_{\ell\neq j}\mathbb{E}\big\{\varphi_\ell\big(Y_{i_\ell,i_\ell'}\big)\big\}
\end{split}
\end{equation}
with $\varphi_\ell(\cdot)$ defined as in {\rm(\ref{eq:phiell})}.

\begin{proposition}\label{pn2}
Let $N=p(p-1)/2$, $\aleph_{\mathcal{V}}/|\mathcal{V}|\asymp N^{-1}$,
$\max\{|\tilde{\alpha}-\alpha|,|\tilde{\beta}-\beta|\}=O_p(N^{-1/2})$ and
$|1-\alpha-\beta|\geq c$ for some positive constant $c$.
 Under Assumption {\rm 1}, it holds that
$
|\widehat{C}_{\mathcal{V}}-C_{\mathcal{V}}|=O_p(N^{-1/2})$.
Furthermore, if Assumption {\rm 2} also holds, then
$
\sqrt{N}(\widehat{C}_{\mathcal{V}}-C_{\mathcal{V}})=S_{\mathcal{V}}+\Delta_{\alpha,\mathcal{V}}\sqrt{N}(\tilde{\alpha}-\alpha)+\Delta_{\beta,\mathcal{V}}\sqrt{N}(\tilde{\beta}-\beta)+o_p(1)$,
%\begin{equation*}\label{eq:expand}
%\begin{split}
%\sqrt{N}(\widehat{C}_{\mathcal{V}}-C_{\mathcal{V}})=&~S_{\mathcal{V}}+\Delta_{\alpha,\mathcal{V}}\sqrt{N}(\tilde{\alpha}-\alpha)+\Delta_{\beta,\mathcal{V}}\sqrt{N}(\tilde{\beta}-\beta)+o_p(1)\,,\end{split}
%\end{equation*}
where $S_{\mathcal{V}}$ is defined as {\rm(\ref{eq:Sv})}.
\end{proposition}

In comparison to Proposition \ref{pre2}, the leading term of $\sqrt{N}(\widehat{C}_{\mathcal{V}}-C_{\mathcal{V}})$ with unknown $\alpha$ or/and $\beta$ has an additional part
\begin{equation}\label{eq:Xiv}
\begin{split}
\Xi_{\mathcal{V}}:=\Delta_{\alpha,\mathcal{V}}\sqrt{N}(\tilde{\alpha}-\alpha)+\Delta_{\beta,\mathcal{V}}\sqrt{N}(\tilde{\beta}-\beta)\, ,
\end{split}
\end{equation}
which is a linear combination of $\sqrt{N}(\tilde{\alpha}-\alpha)$ and
$\sqrt{N}(\tilde{\beta}-\beta)$. Since $S_{\mathcal{V}}$ and
$\Xi_{\mathcal{V}}$ both converge to normal distributions,
$\sqrt{N}(\widehat{C}_{\mathcal{V}}-C_{\mathcal{V}})$ is  also asymptotically
normal. Let $\kappa_1=\alpha(1-\alpha)$, $\kappa_2=\beta(1-\beta)$ and $\kappa_3=1-\alpha-\beta$. Define \begin{equation}\label{eq:G}
\bG=\left(\begin{array}{ccc}
                     g_{\alpha,1} & g_{\alpha,2} & g_{\alpha,3} \\
                     g_{\beta,1} & g_{\beta,2} & g_{\beta,3} \\
                   \end{array}
                 \right)\,,
\end{equation}
where
$(g_{\alpha,1},g_{\alpha,2},g_{\alpha,3},g_{\beta,1},g_{\beta,2},g_{\beta,3})$ are
specified as follows.

\begin{itemize}
\item If only $\alpha$ is known, $g_{\alpha,1}=g_{\alpha,2}=g_{\alpha,3}=0$, $g_{\beta,1}=\frac{\kappa_1-\kappa_2}{\delta\kappa_3^2}$, $g_{\beta,2}=\frac{1}{\delta\kappa_3}$ and $g_{\beta,3}=0$.

\item If only $\beta$ is known, $g_{\alpha,1}=\frac{\kappa_1-\kappa_2}{(1-\delta)\kappa_3^2}$, $g_{\alpha,2}=\frac{1}{(1-\delta)\kappa_3}$, $g_{\alpha,3}=0$ and $g_{\beta,1}=g_{\beta,2}=g_{\beta,3}=0$.

\item If both $\alpha$ and $\beta$ are unknown,
$
    g_{\alpha,1}=\frac{(1-2\beta)\alpha+\beta^{2}}{(1-\delta)\kappa_3^2}$, $g_{\alpha,2}=\frac{\alpha-2\beta}{(1-\delta)\kappa_3^2}$ , $g_{\alpha,3}=\frac{1}{(1-\delta)\kappa_3^2}$,
    $g_{\beta,1}=-\frac{(1-2\alpha)\beta+\alpha^{2}}{\delta\kappa_3^2}$, $g_{\beta,2}=\frac{\beta-2\alpha+1}{\delta\kappa_3^2}$ and $g_{\beta,3}=-\frac{1}{\delta\kappa_3^2}$.
\end{itemize}
Let $\kappa_4=\beta-\alpha$. We define a three-dimensional vector $\bh_{\mathcal{V}}$ such that
\begin{equation}\label{eq:hv}
\begin{split}
\bh_{\mathcal{V}}^\T=&\,\big[6\kappa_4,3(\kappa_4^2-\kappa_1-\kappa_2),2\{\kappa_4 (-6 \alpha  \beta +3 \kappa _3^2-4 \kappa _3)+(1-\alpha ) (\beta -2 \alpha )\}\big]\\
&~~~~~~~~~~~~~~~~~~~~~~~\times\frac{1}{3}\sum_{j=1}^k(-1)^{1-\tau_j}C_{\mathcal{V}}(\tau_1,\ldots,\tau_{j-1},1,\tau_{j+1},\ldots,\tau_k)\\
&+\big\{6\kappa_1,3{\kappa_1(1-2\alpha)},2\kappa_1(1-\alpha)(1-3\alpha)\big\}\\
&~~~~~~~~~~~~~~~~~~~~~~~\times\frac{1}{3\kappa_3^k}\sum_{j=1}^k\frac{(-1)^{1-\tau_j}}{|\mathcal{V}|}\sum_{\bv=(i_1,i_1',\ldots,i_k,i_k')\in\mathcal{V}}\prod_{\ell\neq j}\mathbb{E}\big\{\varphi_\ell\big(Y_{i_\ell,i_\ell'}\big)\big\}\,.
\end{split}
\end{equation}
Now we can state the following theorem.

\begin{theorem}\label{tm:5}
Let $N=p(p-1)/2$, $\aleph_{\mathcal{V}}/|\mathcal{V}|\asymp N^{-1}$,
$|1-\alpha-\beta|\geq c$ for some positive constant $c$,
$N_1=p(p-1)\delta\rightarrow\infty$ and
$N_2=p(p-1)(1-\delta)\rightarrow\infty$. Under Assumptions {\rm 1} and {\rm 2},
it holds that
$
\sqrt{N}(\widehat{C}_{\mathcal{V}}-C_{\mathcal{V}})\rightarrow_d\mathcal{N}(0,\phi_{\mathcal{V}}^2)$ with
$
\phi_{\mathcal{V}}^2=\sigma_{\mathcal{V}}^2+(\Delta_{\alpha,\mathcal{V}},\Delta_{\beta,\mathcal{V}})\bG\bSigma\bG^\T(\Delta_{\alpha,\mathcal{V}},\Delta_{\beta,\mathcal{V}})^\T+\bh_{\mathcal{V}}^\T\bG^\T(\Delta_{\alpha,\mathcal{V}},\Delta_{\beta,\mathcal{V}})^\T$, where
$\sigma_{\mathcal{V}}^2$ and $\bSigma$ are defined as {\rm(\ref{eq:asyv})} and {\rm(\ref{eq:Sigma})} in the Appendix, respectively, provided that one of the following three conditions holds:
{\rm(i)} $\delta(1-\alpha-\beta)^2\geq c$ for some positive constant $c$ when only $\alpha$ is known,
{\rm(ii)} $(1-\delta)(1-\alpha-\beta)^2\geq c$ for some positive constant $c$ when only $\beta$ is known,
or
{\rm(iii)} $\delta(1-\delta)(1-\alpha-\beta)^4\geq c$ for some positive constant
$c$ when both of $\alpha$ and $\beta$ are unknown.
\end{theorem}

Recall that
$\sqrt{N}(\widehat{C}_{\mathcal{V}}-C_{\mathcal{V}})=S_{\mathcal{V}}+\Xi_{\mathcal{V}}
+o_p(1)$. The asymptotic variance $\phi_{\mathcal{V}}^2$ stated in
Theorem \ref{tm:5} actually can be divided into three parts. The first
term $\sigma^2_{\mathcal{V}}$ is the asymptotic variance of
$S_{\mathcal{V}}$. The second term
$(\Delta_{\alpha,\mathcal{V}},\Delta_{\beta,\mathcal{V}})\bG\bSigma\bG^\T
(\Delta_{\alpha,\mathcal{V}},\Delta_{\beta,\mathcal{V}})^\T$ is the
asymptotic variance of $\Xi_{\mathcal{V}}$. The third term
$\bh_{\mathcal{V}}^\T\bG^\T(\Delta_{\alpha,\mathcal{V}},\Delta_{\beta,\mathcal{V}})^\T$
is two times the asymptotic covariance between $S_{\mathcal{V}}$ and $\Xi_{\mathcal{V}}$.

\begin{remark}
By way of comparison with Theorem~\ref{tm:5} here, based on Theorem~10 of~\cite{balachandran2017propagation} and the discussion immediately following that theorem, we can conclude that in general the empirically observed subgraph counts will not even be consistent estimates of $C_{\mathcal{V}}$.
\end{remark}

\subsection{Joint inference of subgraph densities with unknown error rates}
\label{sec34}

Theorem \ref{tm:5} can be extended to the case of multiple subgraph densities, which is required for constructing the joint confidence regions for several subgraph densities or
a smooth function thereof.
Given $(\mathcal{V}_1,\tau_{1,1},\ldots,\tau_{1,k_1}),\ldots,
(\mathcal{V}_m,\tau_{m,1},\ldots,\tau_{m,k_m})$, it holds that the random vector
$
\sqrt{N}(\widehat{C}_{\mathcal{V}_1}-C_{\mathcal{V}_1},\ldots,\widehat{C}_{\mathcal{V}_m}-C_{\mathcal{V}_m})^\T
$ converges to a multivariate normal distribution $\mathcal{N}(\bzero,\bV)$. Let \[
\bvartheta=(S_{\mathcal{V}_1},\ldots,S_{\mathcal{V}_m})^\T~~~\textrm{and}~~~\btheta=(\Xi_{\mathcal{V}_1},\ldots,\Xi_{\mathcal{V}_m})^\T
 \]
where
$S_{\mathcal{V}_q}=S_{\mathcal{V}_q}(\tau_{q,1},\ldots,\tau_{q,k_q})$
and $\Xi_{\mathcal{V}_q}=\Xi_{\mathcal{V}_q}(\tau_{q,1},
\ldots,\tau_{q,k_q})$ are defined in the same manner as (\ref{eq:Sv}) and
(\ref{eq:Xiv}), respectively, but in which $(\mathcal{V},\tau_1,\ldots,\tau_k)$ is
replaced by $(\mathcal{V}_q,\tau_{q,1},\ldots,\tau_{q,k_q})$ now.
 It follows from Proposition \ref{pn2} that $\bV=\lim_{p\rightarrow\infty}\bV_p$ with
\begin{equation} \label{Vp}
\bV_p=\underbrace{\textrm{Var}(\bvartheta)}_{\bV_{1,p}}+\underbrace{\textrm{Var}(\btheta)}_{\bV_{2,p}}+\underbrace{\textrm{Cov}(\bvartheta,\btheta)+\textrm{Cov}(\btheta,\bvartheta)}_{\bV_{3,p}}\,.
\end{equation}
% {\color{red} I have changed $\bV_n$ to $\bV_p$ as $n$ is never defined -- QY}
The first term $\bV_{1,p}$ can be consistently estimated by the bootstrap
procedure presented in Section \ref{se:allknown} with $(\alpha,\beta)$
replaced by $(\tilde{\alpha},\tilde{\beta})$. To evaluate $\bV_{2,p}$ and
$\bV_{3,p}$, we put
\[
\bDelta=\left(
            \begin{array}{cc}
              \Delta_{\alpha,\mathcal{V}_1} & \Delta_{\beta,\mathcal{V}_1} \\
              \vdots & \vdots \\
              \Delta_{\alpha,\mathcal{V}_m} & \Delta_{\beta,\mathcal{V}_m} \\
            \end{array}
          \right)~~~\textrm{and}~~~\bH=\left(
                                         \begin{array}{c}
                                           \bh_{\mathcal{V}_1}^\T \\
                                           \vdots \\
                                           \bh_{\mathcal{V}_m}^\T \\
                                         \end{array}
                                       \right)\,,
\]
where $\Delta_{\alpha,\mathcal{V}_q}$, $\Delta_{\beta,\mathcal{V}_q}$ and
$\bh_{\mathcal{V}_q}^\T$ are defined in the same manner as (\ref{eq:da}),
(\ref{eq:db}) and (\ref{eq:hv}), respectively, with
$(\mathcal{V},\tau_1,\ldots,\tau_k)$ replaced by
$(\mathcal{V}_q,\tau_{q,1},\ldots,\tau_{q,k_q})$ now.
Then it holds that
\begin{equation}\label{eq:V2}
\bV_{2,p}=\bDelta\bG\bSigma\bG^\T\bDelta^\T+o(1)~~~\textrm{and}~~~\bV_{3,p}=\frac{1}{2}\big(\bH\bG^\T\bDelta^\T+\bDelta\bG\bH^\T\big)+o(1)\,,
\end{equation}
where $\bG$ and $\bSigma$ are defined as (\ref{eq:G}) and (\ref{eq:Sigma}) in the Appendix, respectively.

For given $q=1,\ldots,m$, $\tau_{q,1},\ldots,\tau_{q,k_q}\in\{0,1\}$ and $(\tilde{\alpha},\tilde{\beta})$, define $\tilde{\varphi}_{q,\ell}(x)=(x-\tilde{\alpha})^{\tau_{q,\ell}}(1-\tilde{\beta}-x)^{1-\tau_{q,\ell}}$ for $x\in\{0,1\}$. Let $\tilde{\kappa}_1=\tilde{\alpha}(1-\tilde{\alpha})$, $\tilde{\kappa}_2=\tilde{\beta}(1-\tilde{\beta})$ and $\tilde{\kappa}_3=1-\tilde{\alpha}-\tilde{\beta}$. Since
\[
\frac{1}{|\mathcal{V}_q|}\sum_{\bv=(i_1,i_1',\ldots,i_{k_q},i_{k_q}')\in\mathcal{V}_q}\prod_{\ell\neq j}\mathbb{E}\big\{\varphi_{q,\ell}\big(Y_{i_\ell,i_\ell'}\big)\big\}
\]
can be consistently estimated by
\[
\frac{1}{|\mathcal{V}_q|}\sum_{\bv=(i_1,i_1',\ldots,i_{k_q},i_{k_q}')\in\mathcal{V}_q}\prod_{\ell\neq j}\tilde{\varphi}_{q,\ell}\big(Y_{i_\ell,i_\ell'}\big)\,,
\]
then
\begin{equation*}\label{eq:hatda}
\begin{split}
\widehat{\Delta}_{\alpha,\mathcal{V}_q}=&~\frac{k_q}{\tilde{\kappa}_3}\widehat{C}_{\mathcal{V}_q}-\frac{1}{\tilde{\kappa}_3^{k_q}}\sum_{j:\tau_{q,j}=1}\frac{1}{|\mathcal{V}_q|}\sum_{\bv=(i_1,i_1',\ldots,i_{k_q},i_{k_q}')\in\mathcal{V}_q}\prod_{\ell\neq j}\tilde{\varphi}_{q,\ell}\big(Y_{i_\ell,i_\ell'}\big)
\end{split}
\end{equation*}
and
\begin{equation*}\label{eq:hatdb}
\begin{split}
\widehat{\Delta}_{\beta,\mathcal{V}_q}=&~\frac{k_q}{\tilde{\kappa}_3}\widehat{C}_{\mathcal{V}_q}-\frac{1}{\tilde{\kappa}_3^{k_q}}\sum_{j:\tau_{q,j}=0}\frac{1}{|\mathcal{V}_q|}\sum_{\bv=(i_1,i_1',\ldots,i_{k_q},i_{k_q}')\in\mathcal{V}_q}\prod_{\ell\neq j}\tilde{\varphi}_{q,\ell}\big(Y_{i_\ell,i_\ell'}\big)
\end{split}
\end{equation*}
are consistent estimates for ${\Delta}_{\alpha,\mathcal{V}_q}$ and
${\Delta}_{\beta,\mathcal{V}_q}$, respectively. Replacing
$\Delta_{\alpha,\mathcal{V}_q}$, ${\Delta}_{\beta,\mathcal{V}_q}$ and
$(\alpha,\beta)$ by $\widehat{\Delta}_{\alpha,\mathcal{V}_q}$,
$\widehat{\Delta}_{\beta,\mathcal{V}_q}$ and
$(\tilde{\alpha},\tilde{\beta})$, respectively, we can obtain
consistent estimates of $\bDelta$, $\bG$, $\bH$ and $\bSigma$, and, consequently,
consistent estimates of $\bV_{2,p}$ and $\bV_{3,p}$.
% Note that $\bG\bSigma\bG^\T$ is actually the asymptotic
% covariance of $\sqrt{N}(\tilde{\alpha}-\alpha,\tilde{\beta}-\beta)^\T$.
% Alternatively, we can use the parametric bootstrap procedures suggested
% in Section \ref{sec:edge.den} to estimate $\bG\bSigma\bG^\T$, which can
% avoid the complicated formula of $\bG\bSigma\bG^\T$.
For $i=1,2,3$, denote by $\widehat{\bV}_{i,p}$ the consistent estimate of
$\bV_{i,p}$.  Then the joint distribution of $
\sqrt{N}(\widehat{C}_{\mathcal{V}_1}-C_{\mathcal{V}_1},\ldots,\widehat{C}_{\mathcal{V}_m}-C_{\mathcal{V}_m})^\T
$ can be approximated by $\mathcal{N}(\bzero,\widehat{\bV}_p)$ with $\widehat{\bV}_p=\widehat{\bV}_{1,p}+\widehat{\bV}_{2,p}+\widehat{\bV}_{3,p}$.

%  \begin{remark}
%  If we are interested in edge density $\delta$, we can select $k=1$, $\tau_1=1$ and $\mathcal{V}=\{(i_1,i_1'):i_1\neq i_1'\}$. Then $\widehat{C}_{\mathcal{V}}=\tilde{\kappa}_3^{-1}\{p(p-1)\}^{-1}\sum_{i_1\neq i_2}(Y_{i_1,i_2}-\tilde{\alpha})$,
% $
% \widehat{\Delta}_{\alpha,\mathcal{V}}=\tilde{\kappa}_3^{-1}\widehat{C}_{\mathcal{V}}-\tilde{\kappa}_3^{-1}
% $, $\widehat{\Delta}_{\beta,\mathcal{V}}=\tilde{\kappa}_3^{-1}\widehat{C}_{\mathcal{V}}$ and
% \[
% \begin{split}
% \widehat{\bh}_{\mathcal{V}}^\T=&~\frac{\widehat{C}_{\mathcal{V}}}{3}\big[6\tilde{\kappa}_4,3(\tilde{\kappa}_4^2-\tilde{\kappa}_1-\tilde{\kappa}_2),2\{\tilde{\kappa}_4 (-6 \tilde{\alpha}\tilde{\beta} +3 \tilde{\kappa} _3^2-4 \tilde{\kappa}_3)+(1-\tilde{\alpha}) (\tilde{\beta}-2\tilde{\alpha})\}\big]\\
% &+\frac{1}{3\tilde{\kappa}_3}\big\{6\tilde{\kappa}_1,3{\tilde{\kappa}_1(1-2\tilde{\alpha})},2\tilde{\kappa}_1(1-\tilde{\alpha})(1-3\tilde{\alpha})\big\}\,.
% \end{split}
% \]
%  \end{remark}

 \begin{remark} \label{2star}
For estimating two-star density,  we let $k=3$, $\tau_1=\tau_2=1$, $\tau_3=0$ and $\mathcal{V}=\{(i_1,i_1',i_2,i_2',i_3,i_3'):i_{1}'=i_{2},i_{2}'=i_3,i_{3}'=i_1, i_1\neq i_2\neq i_3\}$. Then $\widehat{C}_{\mathcal{V}}=\tilde{\kappa}_3^{-3}\{p(p-1)(p-2)\}^{-1}\sum_{i_1\neq i_2\neq i_3}(Y_{i_1,i_2}-\tilde{\alpha})(Y_{i_2,i_3}-\tilde{\alpha})(1-\tilde{\beta}-Y_{i_3,i_1})$, $\widehat{\Delta}_{\alpha,\mathcal{V}}=3\tilde{\kappa}_3^{-1}\widehat{C}_{\mathcal{V}}-2\tilde{\kappa}_3^{-3}\{p(p-1)(p-2)\}^{-1}\sum_{i_1\neq i_2\neq i_3}({Y}_{i_1,i_2}-\tilde{\alpha})(1-\tilde{\beta}-Y_{i_3,i_1})$, $\widehat{\Delta}_{\beta,\mathcal{V}}=3\tilde{\kappa}_3^{-1}\widehat{C}_{\mathcal{V}}-\tilde{\kappa}_3^{-3}\{p(p-1)(p-2)\}^{-1}\sum_{i_1\neq i_2\neq i_3}(Y_{i_1,i_2}-\tilde{\alpha})(Y_{i_2,i_3}-\tilde{\alpha})$ and
\[
\begin{split}
\widehat{\bh}_{\mathcal{V}}^\T=&~\frac{1}{3}\bigg\{2\widehat{C}_{\mathcal{V}}-\frac{\tilde{\kappa}_3^{-3}}{p(p-1)(p-2)}\sum_{i_1\neq i_2\neq i_3}(Y_{i_1,i_2}-\tilde{\alpha})(Y_{i_2,i_3}-\tilde{\alpha})(Y_{i_3,i_1}-\tilde{\alpha})\bigg\}\\
&~~~~~~~~~~\times\big[6\tilde{\kappa}_4,3(\tilde{\kappa}_4^2-\tilde{\kappa}_1-\tilde{\kappa}_2),2\{\tilde{\kappa}_4 (-6 \tilde{\alpha}\tilde{\beta} +3\tilde{\kappa}_3^2-4\tilde{\kappa}_3)+(1-\tilde{\alpha}) (\tilde{\beta}-2\tilde{\alpha})\}\big]\\
&+\frac{\tilde{\kappa}_3^{-3}}{3p(p-1)(p-2)}\sum_{i_1\neq i_2\neq i_3}\{2(Y_{i_1,i_2}-\tilde{\alpha})(1-\tilde{\beta}-Y_{i_3,i_1})-(Y_{i_1,i_2}-\tilde{\alpha})(Y_{i_2,i_3}-\tilde{\alpha})\}\\
&~~~~~~~~~~\times\big\{6\tilde{\kappa}_1,3{\tilde{\kappa}_1(1-2\tilde{\alpha})},2\tilde{\kappa}_1(1-\tilde{\alpha})(1-3\tilde{\alpha})\big\}\,.
\end{split}
\]
 \end{remark}

\begin{remark} \label{3angle}
For estimating triangle density,
we let $k=3$,
$\tau_1=\tau_2=\tau_3=1$ and
$\mathcal{V}=\{(i_1,i_1',i_2,i_2',i_3, i_3'):i_{1}'=i_{2}, i_2'=i_3,
i_3'=i_1, i_1\neq i_2\neq i_3\}$. Then
$\widehat{C}_{\mathcal{V}}=\tilde{\kappa}_3^{-3}\{p(p-1)(p-2)\}^{-1}
\sum_{i_1\neq i_2\neq i_3}(Y_{i_1,i_2}-\tilde{\alpha})(Y_{i_2,i_3}-\tilde{\alpha})(Y_{i_3,i_1}-\tilde{\alpha})$,
$\widehat{\Delta}_{\alpha,\mathcal{V}}=3\tilde{\kappa}_3^{-1}\widehat{C}_{\mathcal{V}}-3\tilde{\kappa}_3^{-3}\{p(p-1)(p-2)\}^{-1}
\sum_{i_1\neq i_2\neq i_3}(Y_{i_1,i_2}-\tilde{\alpha})(Y_{i_2,i_3}-\tilde{\alpha})$,
% +(Y_{i_1,i_2}-\tilde{\alpha})(Y_{i_3,i_1}-\tilde{\alpha})+(Y_{i_2,i_3}-\tilde{\alpha})(Y_{i_3,i_1}-\tilde{\alpha})\}$,
$\widehat{\Delta}_{\beta,\mathcal{V}}=3\tilde{\kappa}_3^{-1}\widehat{C}_{\mathcal{V}}$ and
\[
\begin{split}
\widehat{\bh}_{\mathcal{V}}^\T=&~\widehat{C}_{\mathcal{V}}\big[6\tilde{\kappa}_4,3(\tilde{\kappa}_4^2-\tilde{\kappa}_1-\tilde{\kappa}_2),2\{\tilde{\kappa}_4 (-6 \tilde{\alpha}\tilde{\beta} +3\tilde{\kappa}_3^2-4\tilde{\kappa}_3)+(1-\tilde{\alpha})(\tilde{\beta} -2 \tilde{\alpha})\}\big]\\
&+\frac{1}{\tilde{\kappa}_3^3}\big\{6\tilde{\kappa}_1,3{\tilde{\kappa}_1(1-2\tilde{\alpha})},2\tilde{\kappa}_1(1-\tilde{\alpha})(1-3\tilde{\alpha})\big\}\\
&~~~~~~~\times\frac{1}{p(p-1)(p-2)}\sum_{i_1\neq i_2\neq i_3}(Y_{i_1,i_2}-\tilde{\alpha})(Y_{i_2,i_3}-\tilde{\alpha})\,.
% +(Y_{i_1,i_2}-\tilde{\alpha})(Y_{i_3,i_1}-\tilde{\alpha})\\
% &~~~~~~~~~~~~~~~~~~~~~~~~~~~~~~~~~~~~~~~~~~~~+(Y_{i_2,i_3}-\tilde{\alpha})(Y_{i_3,i_1}-\tilde{\alpha})\big\}\,.
\end{split}
\]
\end{remark}

\section{Numerical illustration}
\label{sec:illus}

\subsection{Simulations} \label{sec51}

We conduct some simulations to illustrate the finite sample properties of the
proposed estimation methods.
For given $\delta \in (0, 1)$ and integers $p$, $N_{2*}$ and
$N_{\triangle}$, we specify a $p\times p$ deterministic adjacency
matrix $\bA$ with $\lfloor\delta p(p-1)/2\rfloor$ edges randomly allocated
among vertex pairs subject to the condition that there are
exactly $N_{2*}$ (open and closed) two-stars (also called triplets), and $N_{\triangle}$
triangles. Hence the clustering
coefficient of the corresponding network is
\begin{equation} \label{CC}
\gamma = 3N_{\triangle}/N_{2*}\,.
\end{equation}
Generating such $\bA$ is accomplished by an adaptation of the
rewiring ideas of \cite{mahadevan2006}, which, to our best knowledge, is new.
Note that $\delta$ is the edge density, $2N_{2*}/\{p(p-1)(p-2)\}$
and $6N_{\triangle}/\{p(p-1)(p-2)\}$ are, respectively,
the two-star density and the triangle density.
We set $\alpha=0.05$, $\beta=0.05$ or $0.20$, $p= 30, 50, 100$ and 200.
We assume that both $\alpha$ and $\beta$ are unknown.
Therefore we need 3 noisy observations $\bY, \bY_{*}, \bY_{**}$ to
facilitate the estimation, which
are generated according to
(\ref{eq:model}).

We evaluate the point estimates for $\delta, \alpha $ and $\beta$
iteratively using (\ref{eq:MLE1}) and (\ref{iter1}).
More precisely we set an initial value $\alpha_0=0.2$, and obtain
$\hat\beta$ and $\hat\delta$ from (\ref{eq:MLE1}). Plugging $(\alpha,
\hat\beta, \hat\delta)$
into the right-hand side of (\ref{iter1}), we obtain $\hat\alpha$. We
repeat this exercise by setting $\alpha = \hat\alpha$, and
  terminate the recursion
when the absolute difference of two successive values of $\alpha$ is smaller than
$10^{-4}$. We also calculate the approximate confidence intervals for $\delta$
based on the asymptotic normality stated in Theorem \ref{tm:3}. More precisely,
the confidence interval is in the same form as (\ref{CIdelta}) with the asymptotic
variance determined by (\ref{eq:sigma2}) in the Appendix in which $\alpha, \beta, \delta$
are replaced by their respective estimates.

Having obtained estimates $\hat \alpha$ and $\hat \beta$, the point
estimates for the densities of two-star edges and triangles are
$\widehat C_{\mathcal{V}}$ defined in
(\ref{hatCv}); see also Remarks \ref{2star} and
\ref{3angle}. Then a plug-in estimate for clustering coefficient is obtained
based on (\ref{CC}). To compute their confidence intervals is more involved,
and is based on the procedure described in Section~\ref{sec34}. More precisely,
we calculate the joint asymptotic distribution of the normalized
estimators for two-star edge density and triangle density, which is a
two-dimensional normal  distribution with zero mean and variance-covariance
matrix $\bV_p= \bV_{1,p} + \bV_{2,p}+ \bV_{3,p}$, as given in the form
(\ref{Vp}). Note that $\bV_{2,p}$ and $\bV_{3,p}$
can be calculated directly; see (\ref{eq:V2}) and also Remarks
\ref{2star} and \ref{3angle}. To calculate $\bV_{1,p}$, we have to
apply the bootstrap algorithm presented in Section \ref{se:allknown}
with $\alpha = \hat \alpha$ and $\beta =\hat \beta$;
see also Remarks \ref{2star0} and \ref{3angle0}. We replicate
bootstrap sampling 500 times. Then a 95\% confidence
interval is $\widehat C_{\mathcal{V}} \pm 1.96 s$, where $s$ is the
square-root of, respectively,  the (1,1)-element or the (2,2)-element
of $2\bV_p/\{p(p-1)\}$ for two-star density or triangle
density. Consequently a confidence interval for clustering coefficient
is deduced based on (\ref{CC}).

% \begin{sidewaystable}
\begin{table}[tbh]
\caption{Mean absolute errors (MAE) of the point estimates for error
rates $\alpha, \beta$,
edge density $\delta$, two-star count $N_{2*}$, triangle count $N_{\triangle}$
and clustering coefficient $\gamma$
in the simulation with 500
replications for noisy network with $p$ nodes, and  $\alpha=0.05$.
 }
\label{tab1}

\vspace{1ex}

\resizebox{\textwidth}{!}{
\begin{tabular}{rccccc|cccccc}
$p$& $\beta$ & $\delta$& $N_{2*}$ &$ N_{\triangle}$& $\gamma$
&MAE$(\hat \alpha)$
&MAE$(\hat \beta)$
&MAE$(\hat \delta)$
&MAE$(\widehat N_{2*})$
&MAE$(\widehat N_{\triangle})$
&MAE$(\hat \gamma)$\\
\hline
30& 0.05 & 0.1 & 100 & 15 & 0.4500 &
0.0057& 0.0369 & 0.0103& 24.61 & 3.450& 0.1079 \\
& 0.20 & & & & &
0.0064& 0.0622& 0.0150& 34.66 & 5.781& 0.1897 \\
30& 0.05 & 0.2 & 430 & 40 & 0.2791 &
0.0058& 0.0228& 0.0103& 48.54& 6.248& 0.0279\\
& 0.20 & & & & &
0.0072& 0.0385& 0.0162& 74.44& 11.25 & 0.0538\\
50& 0.05 & 0.1 & 1260 & 50 & 0.1190 &
0.0034 & 0.0243& 0.0058& 105.0 & 11.45 & 0.0204\\
& 0.20 & & & & &
0.0037 & 0.0397 & 0.0086& 170.3 & 17.57& 0.0334\\
50& 0.05 & 0.2 & 2300 & 140& 0.1826&
0.0037& 0.0138& 0.0061& 145.5& 16.54& 0.0132\\
& 0.20 & & & & &
0.0048& 0.0275& 0.0111& 255.4& 27.98 & 0.0230\\
100& 0.05& 0.1& 5000& 150& 0.0900 &
0.0017& 0.0125& 0.0030& 299.1& 22.84 & 0.0107\\
& 0.20 & & & & &
0.0020& 0.0237& 0.0048& 481.3& 35.45& 0.0170\\
100&0.05& 0.2& 22000& 1800& 0.2455 &
0.0019& 0.0071& 0.0031& 630.7& 82.42& 0.0054\\
& 0.20 & & & & &
0.0024& 0.0157& 0.0058 & 1199& 154.2& 0.0096\\
200& 0.05&0.1& 40000& 1500 & 0.1125&
0.0008& 0.0065& 0.0016& 1235& 82.09& 0.0039\\
& 0.20 & & & & &
0.0010& 0.0126& 0.0027& 2179& 137.1& 0.0063\\
200& 0.05& 0.2& 155000& 10000& 0.1935&
0.0008& 0.0036& 0.0016& 2444& 258.1& 0.0023\\
 & 0.20 & & & & &
 0.0012& 0.0078& 0.0027& 4249& 431.4& 0.0036\\
\end{tabular}
}
\end{table}
% \end{sidewaystable}

To assess the performance of the estimation procedure, we replicate the
simulation 500 times for each setting. The results are reported in
Tables \ref{tab1} and \ref{tab2}.
As the densities for two-stars and triangles are very
small (i.e. smaller than $10^{-2}$), we report the estimates
for the counts $N_{2*}$ and $N_{\triangle}$ instead.
The mean absolute errors (MAE) for the point estimates for the error rates
$\alpha, \beta$, the edge density $\delta$, the two-star count $N_{2*}$,
the triangle count $N_{\triangle}$, and the clustering coefficient $\gamma$
are reported in Table \ref{tab1}. For example, MAE$(\hat \delta) =
{1 \over 500} \sum_{i=1}^{500} |\delta_i - \delta|$, where
$\delta_1, \ldots, \delta_{500}$ denote the estimated values in the 500
replications of simulation, and $\delta$ denotes the true value.
When $p$ increases, the estimation errors for $\alpha, \beta, \delta$ and $\gamma$ decrease.
Furthermore the errors with $\beta=0.2$ are always greater than those with
$\beta=0.05$. This is due to greater (Type II) errors occurring in the observations $Y_{i,j}$.
The estimation for the edge density $\delta$ is very accurate, and is more accurate
than that for the clustering coefficient $\gamma$ which is a higher-order quantity,
though $\gamma$ can be estimated accurately too especially when $p\ge 100$.
Also noticeable are greater errors
 in estimating $\beta$ than those in estimating $\alpha$. For sparser
networks (such as $\delta=0.1$ or 0.2), there are a comparatively smaller number of $A_{i,j}$
taking value 1, and, hence, the information on $\beta$ is less.  Note that
the estimation for $\beta$ improves when $\delta$ increases from 0.1 to 0.2.
The MAE for the two-star count and the triangle count depend on the
magnitudes of the counts themselves. Note that the relative MAE (i.e.
MAE$(\widehat N_{2*})/N_{2*}$ or MAE$(\widehat N_{\triangle})/N_{\triangle}$) are
small or very small. Indeed they decrease too when $p$ increases.

The estimated 95\% confidence intervals for $\delta, N_{2*}, N_{\triangle}$ and $\gamma$
are reported in Table \ref{tab2}. The estimated coverage probabilities are indeed
around 95\%. The interval estimation for the edge density $\delta$ is accurate as
the average interval lengths are small, varying from 0.0602 when $p=30$ to
0.0072 when $p=200$. Note that the true value of
$\delta$ is either 0.1 or 0.2. The confidence intervals for the clustering coefficient
$\gamma$ tend to be conservative with the coverage probabilities ranging
from 96.4\% to 99.9\%. Nevertheless, the average interval lengths are also
small, especially for large $p$. For example, when $p=200$ and
$\gamma=0.1935$, the average interval length is 0.0111 when $\beta=0.05$,
or 0.0211 when $\beta=0.2$.

\begin{table}[tbh]
\caption{The 95\% confidence intervals for
edge density $\delta$, two-star count $N_{2*}$, triangle count $N_{\triangle}$
and clustering coefficient $\gamma$
in the simulation with 500
replications for noisy networks with $p$ nodes, and  $\alpha=0.05$.
Reported in the table are the relative frequencies (RF) of the event
that a confidence interval covers the corresponding true value, and
also the average {\sl Length} of the intervals. }
\label{tab2}

\vspace{1ex}

\resizebox{\textwidth}{!}{
\begin{tabular}{rccccc|cc|cc|cc|cc}
\multicolumn{6}{c}{True value}
& \multicolumn{2}{c}{$\delta$}
& \multicolumn{2}{c}{$ N_{2*}$}
& \multicolumn{2}{c}{$N_{\triangle}$}
& \multicolumn{2}{c}{$\gamma$}\\
$p$& $\beta$ & $\delta$& $N_{2*}$ &$ N_{\triangle}$& $\gamma$
& RF & Length
& RF & Length
& RF & Length
& RF & Length\\
\hline
30& 0.05 & 0.1 & 100 & 15 & 0.4500 &
0.950 & 0.0520 & 0.950 & 130.5 & 0.978& 20.92& 0.982 & 0.6316\\
& 0.20 & & & & &
0.938& 0.0602 & 0.899 & 146.0 & 0.939 & 26.10 & 0.986 & 0.9709\\
30& 0.05 & 0.2 & 430 & 40 & 0.2791 &
0.954 & 0.0496 & 0.950 & 239.1 & 0.960& 33.60 & 0.982& 0.1633\\
& 0.20 & & & & &
0.929& 0.0747& 0.920& 349.9 & 0.941 & 52.57& 0.990& 0.2582\\
50& 0.05 & 0.1 & 1260 & 50 & 0.1190 &
0.952& 0.0301& 0.956& 544.1& 0.964& 62.04& 0.966& 0.1144\\
& 0.20 & & & & &
0.950& 0.0396 & 0.946& 765.7 & 0.947& 82.04& 0.990& 0.1519\\
50& 0.05 & 0.2 & 2300 & 140& 0.1826&
0.942 & 0.0295 & 0.946& 705.8& 0.966& 88.06& 0.976& 0.0770\\
& 0.20 & & & & &
0.950& 0.0530 & 0.940& 1256& 0.955& 152.5 & 0.991& 0.1313\\
100& 0.05& 0.1& 5000& 150& 0.0900 &
0.960& 0.0150& 0.966& 1521& 0.970& 129.1 & 0.972& 0.0637\\
& 0.20 & & & & &
0.954& 0.0253& 0.954& 2571& 0.990& 216.2& 0.999& 0.1070\\
100&0.05& 0.2& 22000& 1800& 0.2455 &
0.954& 0.0145& 0.954& 3011& 0.956& 404.1 & 0.968& 0.0281\\
& 0.20 & & & & &
0.948& 0.0288& 0.950& 6081& 0.956& 808.0 &0.978& 0.0541\\
200& 0.05&0.1& 40000& 1500 & 0.1125&
0.948& 0.0074& 0.948& 6014& 0.958& 435.1& 0.968& 0.0228\\
& 0.20 & & & & &
0.944& 0.0131& 0.940& 10559& 0.960& 768.7& 0.986& 0.0399\\
200& 0.05& 0.2& 155000& 10000& 0.1935&
0.942& 0.0072& 0.940& 11399& 0.938& 1197 & 0.964& 0.0111\\
& 0.20 & & & & &
0.970& 0.0142& 0.972& 22323& 0.966& 2329& 0.970& 0.0211\\
\end{tabular}
}
\end{table}

\subsection{Application: Gene expression networks}

It is a standard exercise in computational biology to construct and analyze networks from gene expression data.  For the purpose of illustration, we consider the data and network construction described in Section~7.3.1 of \cite{kolaczyk2014statistical}.  These data, originally published by~\cite{clr}, contain (log) gene expression levels  in the bacteria {\it Escherichia coli (E. coli)},
measured for 153 genes under each of $40$ different experimental conditions, with three replicates of each condition. For each set of replicates, we constructed a network among the 153 genes by applying a threshold to the Fisher transformation of the Pearson correlation coefficients calculated for the expression levels between all pairs of genes.  A Bonferonni correction was used to adjust for multiple testing, with the family-wise error rate controled at the $0.05$ level.  While there are numerous other approaches to construction of gene coexpression networks, this simple method is both immediately amenable to our illustration and not uncommon in practice.

The empirical edge density in each of the three resulting networks is quite stable, i.e., approximately $0.073, 0.075$, and $0.074$, respectively.  With $153\times 152/2 = 11,628$ hypothesis tests, the nominal value of $\alpha$ in this analysis is at most $4.3\times 10^{-6}$.  Taking this value as known, and calculating the estimates in (\ref{eq:MLE1}) for two of the networks, we obtain $\hat\beta =0.456$ and $\hat\delta = 0.135$.  The corresponding approximate $95\%$ confidence interval for $\delta$ is $(0.131,0.139)$.  Similar results are obtained for the other possible pairings of the three networks.  These numbers suggest that the true edge density $\delta$ differs substantially from those observed empirically.  However, it is well known that the nominal Type I error rates in this setting can be quite inaccurate (e.g., \cite{cosgrove2010choice}).  If we instead treat $\alpha$ as unknown, the estimators defined by (\ref{eq:est1}), (\ref{eq:est2}), and (\ref{eq:estequ2}) yield estimates $\hat\alpha = 0.024$, $\hat\beta=0.232$, and $\hat\delta = 0.067$.   These numbers suggest that the Type I error rate is orders of magnitude higher than nominally expected, and furthermore that the Type II error rate is nearly one in four.  On the other hand, the resulting method-of-moments estimate of the edge density $\delta$ suggests that the empirical edge densities observed in our networks over-estimate only slightly.

However, consider now estimation of higher-order quantities -- specifically, of the number of two-stars $N_{2*}$, the number of triangles $N_\triangle$, and the clustering coefficient $\gamma$.  For the three networks, the empirical values of these numbers are, respectively, $19112$, $3373$, and $0.53$ for the first network, $22952$, $4814$, and $0.63$ for the second network, and $21820$, $4349$, and $0.60$ for the third network.  Thus we see substantially more variability in these numbers across networks than we did for the empirical edge density.  Applying our proposed method-of-moments estimators to these data, we obtain estimates of approximately $\widehat N_{2*}=25248$, $\widehat N_\triangle = 7243$, and $\hat\gamma = 0.86$.  These are all substantially higher than their empirical counterparts, indicating a nontrivial upward adjustment for network noise, presumably driven in large part by the high estimated rate of Type II error.

Finally, applying our bootstrap-based methodology for construction of asymptotic confidence intervals, we obtain an approximate $95\%$ confidence interval for $\delta$ of $(0.06,0.074)$, which further reinforces the evidence that the true network edge density is less than that observed empirically.  At the same time, the corresponding confidence interval for the clustering coefficient $\gamma$ is $(0.81,0.91)$, suggesting that the true network clustering coefficient is roughly 1/3 larger than observed empirically.  Furthermore, the confidence intervals for $N_{2*}$ and $N_\triangle$ are $(21580,28915)$ and $(5879,8607)$, respectively, by which we see that the triangle count appears to be more adversely affected by noise than the two-star count.

Ultimately, we see that the ability to account for network noise appropriately in reporting these basic summary statistics can lead to distinctly different numbers and conclusions.  From a biological perspective, the fact that the empirically observed edge density is inferred to be fairly accurate, while the clustering coefficient is inferred to be noticeably larger than observed empirically, is suggestive.  Specifically, increasing clustering coefficient has been found to trend with increasing modularity in a variety of biological networks (\cite{ravasz2002hierarchical,pavlopoulos2011using}).  Modules (i.e., groups of highly connected nodes) in gene co-expression networks are understood to be reflective of groups of genes that cooperate in common biological functions.  Our results suggest that the presence of modularity in gene co-expression -- and, hence, the level of functional cooperation among genes -- may well be even more pronounced than currently believed.

\section{Discussion}
\label{sec:disc}

Here we have developed a general framework for estimation and uncertainty quantification of arbitrary subgraph densities in contexts wherein one has observations of noisy networks.  Our approach requires as few as two or three replicates of network observations, and employs method-of-moments techniques to derive estimators and establish their asymptotic consistency and normality.  Simulations demonstrate that substantial inferential accuracy is possible in networks of even modest size when nontrivial noise is present.  And our application to coexpression networks in the context of computational biology shows that the gains offered by our approach over presenting traditional empirical network summaries can be substantial.

The approach we develop here is relevant and broadly applicable to numerous contexts wherein it is possible to obtain some notion of a handful of network replicates.  For example, multiple observations of networks are encountered in genetics (e.g., \cite{bartlett2014}), computational neuroscience (e.g., \cite{biswal2010}), on-line social media (e.g., \cite{mukherjee2017}), and in the study of psychiatric disorders (e.g., \cite{nelson2017}).  Similarly, we note that most papers on dynamic networks assume that the networks observed over
different times are (conditionally) independent of each other as the connection probabilities
evolve over time. As a result, for connection probabilities that do not evolve too quickly, our
results are directly applicable within small windows of time (i.e., in light of Remark 2, following Theorem 1).  See \cite{pensky2019} and \cite{zhao2019}, and the references therein, for a variety of examples of relevant dynamic networks.

Our development here is general and supported by formal theoretical
arguments.  In practice, other approaches have been utilized to date for
uncertainty quantification in certain specific contexts, albeit -- to our
best knowledge -- without the formal justification developed here.  For
example, in the context of gene expression measurements (as in the
application described in Section~5.2), investigators will sometimes use
bootstrapping of the original experiments to resample many pseudo-data
sets and construct many networks, from which in turn they generate
bootstrap distributions of network summaries of interest (e.g., \cite{xulvi2009}).

We have pursued a frequentist approach to the problem of uncertainty quantification for network summary statistics.  If the replicates necessary for our approach are unavailable in a given setting, a Bayesian approach is a natural alternative.  For example, posterior-predictive checks for goodness-of-fit based on examination of a handful of network summary measures is common practice (e.g., \cite[Sec 5.3]{bloem2018}).  Note, however, that the Bayesian approach requires careful modeling of the generative process underlying $G$ and typically does not distinguish between signal and noise components.  Our analysis is conditional on $G$, and hence does not require that $G$ be modeled. It is effectively a `signal plus noise' model, with the signal taken to be fixed but unknown.  Related and more formal work has been done in the context of graphon modeling, with the goal of estimating network motif frequencies (e.g., \cite{latouche2016}).  However, again, one typically does not distinguish between signal and noise components in this setting.  Additionally, we note that the problem of practical graphon estimation itself is still a developing area of research.

Our work here sets the stage for extensions of various levels of
difficulty.  For example, while we have focused here on the case of
undirected graphs, the extension to directed graphs is straightforward.
For directed graphs, $A_{i,j}\ne A_{j,i}$ and $Y_{i,j}\ne Y_{j,i}$.
The representation (\ref{eq:model}) relies on independent $\varepsilon_{i,j}$ for
$1\le i\ne j\le p$. The statistics used for estimation should be changed
accordingly too.  For example in (\ref{b2}) the sum should be taken for
all $i\ne j$ instead of $i<j$, and the sum should be divided by $p(p-1)$
instead of $p(p-1)/2$. Then the current technical proofs for
undirected graphs are applicable identically to directed graphs.
On the other hand, whereas we have focused on
estimation solely in the case of subgraph densities, which rests on the
behavior of counting statistics, we anticipate that the estimation of
non-counting network summaries (e.g., summaries based on shortest path
lengths) from noisy network data is likely nontrivial, due to the fact
that the latter are based on extremes rather than counts.

\section*{Acknowledgement}
The authors are grateful to the editor, an associate editor and two referees for their helpful suggestions. Chang was supported in part by the Fundamental Research Funds for the Central Universities
of China, the National Natural Science Foundation of China (Grant No.\;11871401, 71991472), the funds of Fok Ying-Tong Education
Foundation for Young Teachers in the Higher Education Institutions of China, and the Center of Statistical Research
and the Joint Lab of Data Science and Business Intelligence at SWUFE. Kolaczyk was supported in part by the US Air Force Office of Scientific Research.

\section*{Appendix}

Here we derive expressions for the covariance matrices in Theorem~\ref{tm:twounknown}.
Let $\kappa_1=\alpha(1-\alpha)$, $\kappa_2=\beta(1-\beta)$ and $\kappa_3=1-\alpha-\beta$. Let
\[
\begin{split}
\bW_\alpha=\left(
  \begin{array}{cc}
    \frac{\kappa_2-\kappa_1}{\delta\kappa_3^2} &-\frac{1}{\delta\kappa_3}\\
\frac{2\beta-1}{\kappa_3^2}  & -\frac{1}{\kappa_3^2} \\
  \end{array}
\right)\,,~~
\bW_\beta=\left(
  \begin{array}{cc}
    \frac{\kappa_2-\kappa_1}{(1-\delta)\kappa_3^2} & -\frac{1}{(1-\delta)\kappa_3} \\
    \frac{2\alpha-1}{\kappa_3^2} & \frac{1}{\kappa_3^2} \\
  \end{array}
\right)
\end{split}
\]
and
\[
\bW=\left(
  \begin{array}{ccc}
    \frac{(1-2\beta)\alpha+\beta^{2}}{(1-\delta)\kappa_3^2} & \frac{\alpha-2\beta}{(1-\delta)\kappa_3^2} & \frac{1}{(1-\delta)\kappa_3^2}\\
    -\frac{(1-2\alpha)\beta+\alpha^{2}}{\delta\kappa_3^2} & \frac{\beta-2\alpha+1}{\delta\kappa_3^2} & -\frac{1}{\delta\kappa_3^2}\\
    \frac{3\kappa_3+6\alpha\beta-2}{\kappa_3^3} & \frac{3\kappa_3+6\beta-2}{\kappa_3^3} & -\frac{2}{\kappa_3^3}\\
  \end{array}
\right)\,.
\]
Define a matrix
\begin{equation}\label{eq:Sigma}
\bSigma=(\sigma_{ij})_{3\times3}
\end{equation}
with $\sigma_{11}=\delta\kappa_2+(1-\delta)\kappa_1$,
$\sigma_{22}
=\delta\kappa_2({1}/{2}-\kappa_2)+(1-\delta)\kappa_1({1}/{2}-\kappa_1)$,
$\sigma_{33}=\delta\beta\kappa_2({1}/{3}-\beta\kappa_2)+(1-\delta)\kappa_1(1-\alpha)\{{1}/{3}-\kappa_1(1-\alpha)\}$,
$\sigma_{12}=\sigma_{21}=\delta\kappa_2(\beta-{1}/{2})+(1-\delta)\kappa_1({1}/{2}-\alpha)$,
$\sigma_{13}=\sigma_{31}=\delta\kappa_2({\beta^2}/{3}-{2\kappa_2}/{3})+(1-\delta)\kappa_1\{{(1-\alpha)^2}/{3}-{2\kappa_1}/{3}\}$ and
$\sigma_{23}=\sigma_{32}=\delta\beta\kappa_2({1}/{3}-\kappa_2)+(1-\delta)(1-\alpha)\kappa_1({1}/{3}-\kappa_1)$. Denote by $\bSigma_1=(\sigma_{ij})_{2\times2}$ the $2\times 2$ submatrix of $\bSigma$. Based on such defined $\bSigma$ and $\bSigma_1$, let
\begin{equation}\label{eq:sigma1alpha}
\begin{split}
\bSigma_{1,\alpha}=&~\bW_\alpha\bSigma_1\bW_\alpha^\T\,,
\end{split}
\end{equation}
\begin{equation}\label{eq:sigma1beta}
\begin{split}
\bSigma_{1,\beta}=&~\bW_\beta\bSigma_1\bW_\beta^\T
\end{split}
\end{equation}
and
\begin{equation}\label{eq:sigma2}
\bSigma_2=\bW\bSigma\bW^\T\,.
\end{equation}

\newpage

\setcounter{page}{1}
%\pagestyle{fancy}
%\fancyhf{}
\rhead{\bfseries\thepage}
\lhead{\bfseries SUPPLEMENTARY MATERIAL}

\setcounter{page}{1}
\begin{center}
{\bf\Large Supplementary Material for ``Estimation of Subgraph Densities in Noisy Networks'' by Chang, Kolaczyk and Yao}
\end{center}

\appendix

%\section*{Appendix}
\renewcommand{\theequation}{S.\arabic{equation}}
\setcounter{equation}{0}

\subsection*{Proof of Theorem \ref{tm:1}}

Recalling the definition of $F_M$ and $F_{M^*}$ as the joint distributions of $\bY$ when $\bY$ follows models $M$ and $M^*$, respectively, denote by $F_{i,j,M}$ and $F_{i,j,M^*}$ the corresponding marginal distribution of $Y_{i,j}$.  From Assumption 1, we have
\[
\mathcal{H}^2(F_{M},F_{M^*})\leq\sum_{(i,j)\in\mathcal{S}}\mathcal{H}^2(F_{i,j,M},F_{i,j,M^*})+\sum_{(i,j)\in\mathcal{S}^c}\mathcal{H}^2(F_{i,j,M},F_{i,j,M^*}) \enskip ,
\]
where $\mathcal{S}=\textrm{supp}(\bA)$, $\mathcal{S}^c=\textrm{supp}(\bA^*)$, and $\mathcal{H}(\cdot,\cdot)$ denotes the Hellinger distance between two distributions. Since $F_{i,j,M}=F_{i,j,M^*}$ for any $i\neq j$ which implies $\mathcal{H}^2(F_{i,j,M},F_{i,j,M^*})=0$, then $\mathcal{H}^2(F_{M},F_{M^*})=0$.

Without lose of generality, we assume $d_f=|f(M)-f(M^*)|$ for some $M\in\mathcal{M}$ with $f(M)<f(M^*)$. For any $\hat{f}\in\mathcal{E}$, we consider the hypothesis testing problem $H_0: \bY\sim M$ versus $H_1: \bY\sim M^*$, and define the test function $\Psi=I\{\hat{f}>f(M)+d_f/2\}$, which means we reject $H_0$ if $\Psi=1$ and accept $H_0$ if $\Psi=0$. The testing affinity \citep{lecam1973convergence,lecam2012asymptotic} is defined as \[
\pi=\inf_{0\leq \phi\leq 1\atop \phi\textrm{-measurable}}\mathbb{E}_{H_0}(\phi)+\mathbb{E}_{H_1}(1-\phi),
\]
and it is the minimal sum of type I and type II errors of any test between $H_0$ and $H_1$. Recall $\mathcal{H}(F_{M},F_{M^*})=0$ and $\pi\geq 1-\mathcal{H}(F_M,F_{M^*})$, then $\pi=1$. Notice that
$
\mathbb{P}_{M}(|\hat{f}-f|\geq d_f/2)\geq\mathbb{P}_{M}(\hat{f}> f+d_f/2)=\textrm{type I error}
$ and $
\mathbb{P}_{M^*}(|\hat{f}-f|\geq d_f/2)\geq\mathbb{P}_{M^*}(\hat{f}\leq f-d_f/2)=\mathbb{P}_{M^*}\{\hat{f}\leq f(M)+d_f/2\}=\textrm{type II error}
$.  Thus
$
\max\{\mathbb{P}_{M}(|\hat{f}-f|\geq d_f/2),\mathbb{P}_{M^*}(|\hat{f}-f|\geq d_f/2)\}\geq1/2
$ which implies
\[
\sup_{\mathcal{M}}\mathbb{P}\bigg(|\hat{f}-f|\geq\frac{d_f}{2}\bigg)\geq \frac{1}{2}\,.
\]
Since the above result holds for any $\hat{f}\in\mathcal{E}$, the proof of Theorem \ref{tm:1} is complete. $\hfill\Box$

\subsection*{A useful lemma}

To prove Proposition \ref{tm:2} and Theorems \ref{tm:twounknown} and \ref{tm:3}, we need the following lemma.
\begin{lemma}\label{la:1}
Let $N=p(p-1)/2$, $\kappa_1=\alpha(1-\alpha)$ and
$\kappa_2=\beta(1-\beta)$. Under Assumption {\rm 1}, if $N_1=p(p-1)\delta\rightarrow\infty$ and $N_2=p(p-1)(1-\delta)\rightarrow\infty$, it holds that $
\sqrt{N}(\hat{u}_1-u_1,\hat{u}_2-u_2,\hat{u}_3-u_3)^\T\rightarrow_d \mathcal{N}(\bzero,\bSigma)
$ with
$
\bSigma=(\sigma_{ij})_{3\times3}
$, where $\sigma_{11}=\delta\kappa_2+(1-\delta)\kappa_1$,
$\sigma_{22}
=\delta\kappa_2({1}/{2}-\kappa_2)+(1-\delta)\kappa_1({1}/{2}-\kappa_1)$,
$\sigma_{33}=\delta\beta\kappa_2({1}/{3}-\beta\kappa_2)+(1-\delta)\kappa_1(1-\alpha)\{{1}/{3}-\kappa_1(1-\alpha)\}$,
$\sigma_{12}=\sigma_{21}=\delta\kappa_2(\beta-{1}/{2})+(1-\delta)\kappa_1({1}/{2}-\alpha)$,
$\sigma_{13}=\sigma_{31}=\delta\kappa_2({\beta^2}/{3}-{2\kappa_2}/{3})+(1-\delta)\kappa_1\{{(1-\alpha)^2}/{3}-{2\kappa_1}/{3}\}$ and
$\sigma_{23}=\sigma_{32}=\delta\beta\kappa_2({1}/{3}-\kappa_2)+(1-\delta)(1-\alpha)\kappa_1({1}/{3}-\kappa_1)$.
\end{lemma}

\noindent{\it Proof.} Let $\mathcal{S}=\{(i,j):A_{i,j}=1, i<j\}$ and $\mathcal{S}^c=\{(i,j):A_{i,j}=0, i<j\}$. By the definition of $\hat{u}_k$ and $u_k$ $(k=1,2,3)$, we have
\begin{equation*}
\begin{split}
\hat{u}_1-u_1%=&~\frac{1}{N}\sum_{(i,j)\in\mathcal{S}}\{Y_{ij}-(1-\beta)\}+\frac{1}{N}\sum_{(i,j)\in\mathcal{S}^c}(Y_{ij}-\alpha)\\
=&~\frac{1}{N}\sum_{(i,j)\in\mathcal{S}}\{Y_{i,j}-(1-\beta)\}+\frac{1}{N}\sum_{(i,j)\in\mathcal{S}^c}(Y_{i,j}-\alpha),\\
\hat{u}_2-u_2%=&~\frac{1}{2N}\sum_{(i,j)\in\mathcal{S}}(|Y_{ij,*}-Y_{ij}|-2\kappa_2)+\frac{1}{2N}\sum_{(i,j)\in\mathcal{S}^c}(|Y_{ij,*}-Y_{ij}|-2\kappa_1)\\
=&~\frac{1}{2N}\sum_{(i,j)\in\mathcal{S}}(|Y_{i,j,*}-Y_{i,j}|-2\kappa_2)+\frac{1}{2N}\sum_{(i,j)\in\mathcal{S}^c}(|Y_{i,j,*}-Y_{i,j}|-2\kappa_1)\\
\hat{u}_3-u_3%=&~\frac{1}{3N}\sum_{(i,j)\in\mathcal{S}}(\xi_{ij}-3\beta\kappa_2)+\frac{1}{3N}\sum_{(i,j)\in\mathcal{S}^c}\{\xi_{ij}-3\kappa_1(1-\alpha)\}\\
=&~\frac{1}{3N}\sum_{(i,j)\in\mathcal{S}}(\xi_{i,j}-3\beta\kappa_2)+\frac{1}{3N}\sum_{(i,j)\in\mathcal{S}^c}\{\xi_{i,j}-3\kappa_1(1-\alpha)\}\\
\end{split}
\end{equation*}
where $\xi_{i,j}=I(Y_{i,j,**}-2Y_{i,j,*}+Y_{i,j}=1~\textrm{or}-2)$. It
follows from Assumption 1 that
\[
\begin{split}
N\mathbb{E}\{(\hat{u}_1-u_1)^2\}%=&~\frac{1}{N}\sum_{(i,j)\in\mathcal{S}}E[\{Y_{ij}-(1-\beta)\}^2]+\frac{1}{N}\sum_{(i,j)\in\mathcal{S}^c}E\{(Y_{ij}-\alpha)^2\}\\
=&~\delta\kappa_2+(1-\delta)\kappa_1=\sigma_{11},\\
N\mathbb{E}\{(\hat{u}_2-u_2)^2\}%=&~\frac{1}{4N}\sum_{(i,j)\in\mathcal{S}}E\{(|Y_{ij,*}-Y_{ij}|-2\kappa_2)^2\}+\frac{1}{4N}\sum_{(i,j)\in\mathcal{S}^c}E\{(|Y_{ij,*}-Y_{ij}|-2\kappa_1)^2\}\\
=&~\delta\kappa_2\bigg(\frac{1}{2}-\kappa_2\bigg)+(1-\delta)\kappa_1\bigg(\frac{1}{2}-\kappa_1\bigg)=\sigma_{22},\\
N\mathbb{E}\{(\hat{u}_3-u_3)^2\}%=&~\frac{1}{9N}\sum_{(i,j)\in\mathcal{S}}E\{(\xi_{ij}-3\beta\kappa_2)^2\}+\frac{1}{9N}\sum_{(i,j)\in\mathcal{S}^c}E[\{\xi_{ij}-3\kappa_1(1-\alpha)\}^2]\\
=&~\delta\beta\kappa_2\bigg(\frac{1}{3}-\beta\kappa_2\bigg)+(1-\delta)\kappa_1(1-\alpha)\bigg\{\frac{1}{3}-\kappa_1(1-\alpha)\bigg\}=\sigma_{33},\\
N\mathbb{E}\{(\hat{u}_1-u_1)(\hat{u}_2-u_2)\}%=&~\frac{1}{2N}\sum_{(i,j)\in\mathcal{S}}E[\{Y_{ij}-(1-\beta)\}(|Y_{ij,*}-Y_{ij}|-2\kappa_2)]\\
%&+\frac{1}{2N}\sum_{(i,j)\in\mathcal{S}^c}E\{(Y_{ij}-\alpha)(|Y_{ij,*}-Y_{ij}|-2\kappa_1)\}\\
=&~\delta\kappa_2\bigg(\beta-\frac{1}{2}\bigg)+(1-\delta)\kappa_1\bigg(\frac{1}{2}-\alpha\bigg)=\sigma_{12},\\
N\mathbb{E}\{(\hat{u}_1-u_1)(\hat{u}_3-u_3)\}%=&~\frac{1}{3N}\sum_{(i,j)\in\mathcal{S}}E[\{Y_{ij}-(1-\beta)\}(\xi_{ij}-3\beta\kappa_2)]\\
%&+\frac{1}{3N}\sum_{(i,j)\in\mathcal{S}}E[(Y_{ij}-\alpha)\{\xi_{ij}-3\kappa_1(1-\alpha)\}]\\
=&~\delta\kappa_2\bigg(\frac{\beta^2}{3}-\frac{2\kappa_2}{3}\bigg)+(1-\delta)\kappa_1\bigg\{\frac{(1-\alpha)^2}{3}-\frac{2\kappa_1}{3}\bigg\}=\sigma_{13},\\
N\mathbb{E}\{(\hat{u}_2-u_2)(\hat{u}_3-u_3)\}%=&~\frac{1}{6N}\sum_{(i,j)\in\mathcal{S}}E\{(|Y_{ij,*}-Y_{ij}|-2\kappa_2)(\xi_{ij}-3\beta\kappa_2)\}\\
%&+\frac{1}{6N}\sum_{(i,j)\in\mathcal{S}}E[(|Y_{ij,*}-Y_{ij}|-2\kappa_1)\{\xi_{ij}-3\kappa_1(1-\alpha)\}]\\
=&~\delta\beta\kappa_2\bigg(\frac{1}{3}-\kappa_2\bigg)+(1-\delta)(1-\alpha)\kappa_1\bigg(\frac{1}{3}-\kappa_1\bigg)=\sigma_{23}.
\end{split}
\]
By the Lindberg-Feller Central Limit Theorem, we have Lemma \ref{la:1}. $\hfill\Box$

\subsection*{Proof of Proposition \ref{tm:2}}

Define $g_1(x,y,z)=(1-z)x+z(1-y)$ and $g_2(x,y,z)=(1-z)x(1-x)+zy(1-y)$ for any $(x,y,z)\in(0,1)^3$. When $\alpha$ is known, it holds that $g_1(\alpha,\hat{\beta},\hat{\delta})-g_1(\alpha,\beta,\delta)=\hat{u}_1-u_1$ and $g_2(\alpha,\hat{\beta},\hat{\delta})-g_2(\alpha,\beta,\delta)=\hat{u}_2-u_2$. Since the equations $g_1(\alpha,y,z)=u_1$ and $g_2(\alpha,y,z)=u_2$ have the unique solution $(y,z)=(\beta,\delta)$, and $(\hat{u}_1,\hat{u}_2)=(u_1,u_2)+o_p(1)$, we have consistency of $(\hat{\beta},\hat{\delta})$. By Taylor expansion, we have
$
\bD_\alpha(
           \hat{\beta}-\beta,
           \hat{\delta}-\delta)^\T=(
                   \hat{u}_1-u_1,
                   \hat{u}_2-u_2)^\T$ with
               \begin{equation}
               \bD_\alpha=\left(
  \begin{array}{cc}
    \frac{\partial g_1(x,y,z)}{\partial y} & \frac{\partial g_1(x,y,z)}{\partial z} \\
    \frac{\partial g_2(x,y,z)}{\partial y} & \frac{\partial g_2(x,y,z)}{\partial z} \\
  \end{array}
\right)\bigg|_{(x,y,z)=(\alpha,\beta^*,\delta^*)}
\end{equation}
where $(\beta^*,\delta^*)=\lambda\cdot(\beta,\delta)+(1-\lambda)\cdot(\hat{\beta},\hat{\delta})$ for some $\lambda\in(0,1)$.
Notice that $\textrm{det}(\bD_\alpha)=-\delta^*(1-\alpha-\beta^*)^2$.  Since $\delta(1-\alpha-\beta)^2\geq c$ for some positive constant $c$, with the continuity of the function $\delta(1-\alpha-\beta)^2$ with respect to $(\beta,\delta)$, we know $\textrm{det}(\bD_\alpha)\leq -{c}/2$ with probability approaching one. Therefore, $(\hat{\beta}-\beta,\hat{\delta}-\delta)^\T=\bD_\alpha^{-1}(\hat{u}_1-u_1,\hat{u}_2-u_2)^\T$. From Lemma \ref{la:1}, $(\hat{u}_1-u_1,\hat{u}_2-u_2)=O_p(N^{-1/2})$ which implies part (i) of Proposition \ref{tm:2}. Analogously, we have part (ii). $\hfill\Box$

\subsection*{Proof of Theorem \ref{tm:twounknown}}

It follows from Lemma \ref{la:1} that $\sqrt{N}(\hat{u}_1-u_1,\hat{u}_2-u_2)^\T\rightarrow_d \mathcal{N}(\bzero,\bSigma_1)$ where $\bSigma_1=(\sigma_{ij})_{2\times2}$
with $\sigma_{ij}$ specified in Lemma \ref{la:1}. We first consider the case with known $\alpha$. As we have shown in the proof of Proposition \ref{tm:2}, $(\hat{\beta}-\beta,\hat{\delta}-\delta)^\T=\bD_\alpha^{-1}(\hat{u}_1-u_1,\hat{u}_2-u_2)^\T$ with
\[
\bD_\alpha^{-1}=-\frac{1}{\delta^*(1-\alpha-\beta^*)^2}\left(
  \begin{array}{cc}
    \beta^*(1-\beta^*)-\alpha(1-\alpha) & -(1-\alpha-\beta^*) \\
    -\delta^*(1-2\beta^*) & -\delta^* \\
  \end{array}
\right).
\]
Therefore, $\sqrt{N}(\hat{\beta}-\beta,\hat{\delta}-\delta)^\T\rightarrow_d \mathcal{N}(\bzero,\bSigma_{1,\alpha})$ with
\begin{equation*}
\begin{split}
\bSigma_{1,\alpha}=&~\frac{1}{\delta^2\kappa_3^4}\left(
  \begin{array}{cc}
    \kappa_2-\kappa_1 &-\kappa_3\\
-\delta(1-2\beta)  & -\delta \\
  \end{array}
\right)\bSigma_1\left(
  \begin{array}{cc}
    \kappa_2-\kappa_1 & -\delta(1-2\beta)\\
  -\kappa_3  & -\delta \\
  \end{array}
\right)\\
\end{split}
\end{equation*}
where $\kappa_1=\alpha(1-\alpha)$, $\kappa_2=\beta(1-\beta)$ and $\kappa_3=1-\alpha-\beta$.
This completes part (i) of Theorem \ref{tm:twounknown}. For part (ii), notice that
\[
\bD_\beta=\left(
  \begin{array}{cc}
    \frac{\partial g_1(x,y,z)}{\partial x} & \frac{\partial g_1(x,y,z)}{\partial z} \\
    \frac{\partial g_2(x,y,z)}{\partial x} & \frac{\partial g_2(x,y,z)}{\partial z} \\
  \end{array}
\right)\bigg|_{(x,y,z)=(\alpha^*,\beta,\delta^*)},
\]
where $(\alpha^*,\delta^*)=\lambda\cdot(\alpha,\delta)+(1-\lambda)\cdot(\hat{\alpha},\hat{\delta})$ for some $\lambda\in(0,1)$. Then
\[
\bD_\beta^{-1}=-\frac{1}{(1-\delta^*)(1-\alpha^*-\beta)^2}\left(
  \begin{array}{cc}
    \beta(1-\beta)-\alpha^*(1-\alpha^*) & -(1-\alpha^*-\beta) \\
    -(1-\delta^*)(1-2\alpha^*) & 1-\delta^* \\
  \end{array}
\right).
\]
Since $(\hat{\alpha}-\alpha,\hat{\delta}-\delta)^\T=\bD_\beta^{-1}(\hat{u}_1-u_1,\hat{u}_2-u_2)^\T$, then $\sqrt{N}(\hat{\alpha}-\alpha,\hat{\delta}-\delta)^\T\rightarrow_d \mathcal{N}(\bzero,\bSigma_{1,\beta})$ with
\begin{equation*}
\begin{split}
\bSigma_{1,\beta}=&~\frac{1}{(1-\delta)^2\kappa_3^4}\left(
  \begin{array}{cc}
    \kappa_2-\kappa_1 & -\kappa_3 \\
    -(1-\delta)(1-2\alpha) & 1-\delta \\
  \end{array}
\right)\bSigma_1\left(
  \begin{array}{cc}
    \kappa_2-\kappa_1 & -(1-\delta)(1-2\alpha)\\
   -\kappa_3 & 1-\delta \\
  \end{array}
\right)\,.
\end{split}
\end{equation*}
Therefore, we have part (ii). $\hfill\Box$

\subsection*{Proof of Theorem \ref{tm:3}}

Define $g_3(x,y,z)=(1-z)x(1-x)^2+zy^2(1-y)$ for any $(x,y,z)\in(0,1)^3$. Recall $g_1(x,y,z)=(1-z)x+z(1-y)$ and $g_2(x,y,z)=(1-z)x(1-x)+zy(1-y)$. Following the same arguments in the proof of Proposition \ref{tm:2} for the consistency of $(\hat{\beta},\hat{\delta})$, we have the consistency of $(\hat{\alpha},\hat{\beta},\hat{\delta})$. By Taylor expansion, we have $\bD(\hat{\alpha}-\alpha,\hat{\beta}-\beta,\hat{\delta}-\delta)^\T=(\hat{u}_1-u_1,\hat{u}_2-u_2,\hat{u}_3-u_3)^\T$ with
\[
\bD=\left(
  \begin{array}{ccc}
    \frac{\partial g_1(x,y,z)}{\partial x} & \frac{\partial g_1(x,y,z)}{\partial y} & \frac{\partial g_1(x,y,z)}{\partial z}\\
    \frac{\partial g_2(x,y,z)}{\partial x} & \frac{\partial g_2(x,y,z)}{\partial y} & \frac{\partial g_2(x,y,z)}{\partial z}\\
    \frac{\partial g_3(x,y,z)}{\partial x} & \frac{\partial g_3(x,y,z)}{\partial y} & \frac{\partial g_3(x,y,z)}{\partial z}\\
  \end{array}
\right)\bigg|_{(x,y,z)=(\alpha^*,\beta^*,\delta^*)},
\]
where $(\alpha^*,\beta^*,\delta^*)=\lambda\cdot(\alpha,\beta,\delta)+(1-\lambda)\cdot(\hat{\alpha},\hat{\beta},\hat{\delta})$ for some $\lambda\in(0,1)$. Notice that $\det(\bD)=-(1-\delta^*)\delta^*(1-\alpha^*-\beta^*)^4$.  Since $(1-\delta)\delta(1-\alpha-\beta)^4\geq c$ for some positive constant $c$, with the continuity of the function $(1-\delta)\delta(1-\alpha-\beta)^4$ with respect to $(\alpha,\beta,\delta)$, we know $\textrm{det}(\bD)\leq -{c}/2$ with probability approaching one. Therefore, $(\hat{\alpha}-\alpha,\hat{\beta}-\beta,\hat{\delta}-\delta)^\T=\bD^{-1}(\hat{u}_1-u_1,\hat{u}_2-u_2,\hat{u}_3-u_3)^\T$. From Lemma \ref{la:1}, $(\hat{u}_1-u_1,\hat{u}_2-u_2,\hat{u}_3-u_3)=O_p(N^{-1/2})$ which implies $(\hat{\alpha}-\alpha,\hat{\beta}-\beta,\hat{\delta}-\delta)=O_p(N^{-1/2})$. Since
\[
\bD^{-1}=\left(
  \begin{array}{ccc}
    \frac{(1-2\beta^*)\alpha^*+\beta^{*2}}{(1-\delta^*)(1-\alpha^*-\beta^*)^2} & \frac{\alpha^*-2\beta^*}{(1-\delta^*)(1-\alpha^*-\beta^*)^2} & \frac{1}{(1-\delta^*)(1-\alpha^*-\beta^*)^2}\\
    -\frac{(1-2\alpha^*)\beta^*+\alpha^{*2}}{\delta^*(1-\alpha^*-\beta^*)^2} & \frac{\beta^*-2\alpha^*+1}{\delta^*(1-\alpha^*-\beta^*)^2} & -\frac{1}{\delta^*(1-\alpha^*-\beta^*)^2}\\
    -\frac{3(\alpha^*+\beta^*)-6\alpha^*\beta^*-1}{(1-\alpha^*-\beta^*)^3} & -\frac{3\alpha^*-3\beta^*-1}{(1-\alpha^*-\beta^*)^3} & -\frac{2}{(1-\alpha^*-\beta^*)^3}\\
  \end{array}
\right),
\]
then $\sqrt{N}(\hat{\alpha}-\alpha,\hat{\beta}-\beta,\hat{\delta}-\delta)^\T\rightarrow_d \mathcal{N}(\bzero,\bSigma_2)$ with
\begin{equation*}
\bSigma_2=\left(
  \begin{array}{ccc}
    \frac{(1-2\beta)\alpha+\beta^{2}}{(1-\delta)\kappa_3^2} & \frac{\alpha-2\beta}{(1-\delta)\kappa_3^2} & \frac{1}{(1-\delta)\kappa_3^2}\\
    -\frac{(1-2\alpha)\beta+\alpha^{2}}{\delta\kappa_3^2} & \frac{\beta-2\alpha+1}{\delta\kappa_3^2} & -\frac{1}{\delta\kappa_3^2}\\
    \frac{3\kappa_3+6\alpha\beta-2}{\kappa_3^3} & \frac{3\kappa_3+6\beta-2}{\kappa_3^3} & -\frac{2}{\kappa_3^3}\\
  \end{array}
\right)\bSigma\left(
  \begin{array}{ccc}
    \frac{(1-2\beta)\alpha+\beta^{2}}{(1-\delta)\kappa_3^2} & -\frac{(1-2\alpha)\beta+\alpha^{2}}{\delta\kappa_3^2} &\frac{3\kappa_3+6\alpha\beta-2}{\kappa_3^3}\\
  \frac{\alpha-2\beta}{(1-\delta)\kappa_3^2}  & \frac{\beta-2\alpha+1}{\delta\kappa_3^2} &\frac{3\kappa_3+6\beta-2}{\kappa_3^3}\\
  \frac{1}{(1-\delta)\kappa_3^2}  & -\frac{1}{\delta\kappa_3^2} & -\frac{2}{\kappa_3^3}\\
  \end{array}
\right),
\end{equation*}
where $\kappa_1=\alpha(1-\alpha)$, $\kappa_2=\beta(1-\beta)$, $\kappa_3=1-\alpha-\beta$, and $\bSigma$ is specified in Lemma \ref{la:1}. This completes the proof of Theorem \ref{tm:3}. $\hfill\Box$

\subsection*{Proof of Proposition \ref{tm:pre}}

Define
\begin{align*}
T_{\mathcal{V}}(\tau_1,\ldots,\tau_k)=\frac{1}{|\mathcal{V}|}\sum_{\bv=(i_1,i_1',\ldots,i_k,i_k')\in\mathcal{V}}\prod_{\ell=1}^k\mathbb{E}\big\{\varphi_\ell\big(Y_{i_\ell,i_\ell'}\big)\big\}\,.
\end{align*}
Since $|1-\alpha-\beta|\geq c$ for some positive constant $c$, the convergence rate of $|\tilde{C}_{\mathcal{V}}(\tau_1,\ldots,\tau_k)-C_{\mathcal{V}}(\tau_1,\ldots,\tau_k)|$ is the same as that of $|\tilde{T}_{\mathcal{V}}(\tau_1,\ldots,\tau_k)-T_{\mathcal{V}}(\tau_1,\ldots,\tau_k)|$. To simplify the notation, we write $\tilde{T}_{\mathcal{V}}(\tau_1,\ldots,\tau_k)$ and $T_{\mathcal{V}}(\tau_1,\ldots,\tau_k)$ as $\tilde{T}_{\mathcal{V}}$ and $T_{\mathcal{V}}$, respectively. Let $\mathring{\varphi}_\ell(Y_{i_\ell,i_\ell'})=\varphi_\ell(Y_{i_\ell,i_\ell'})-\mathbb{E}\{\varphi_\ell(Y_{i_\ell,i_\ell'})\}$.
Notice that
\begin{align*}
\tilde{T}_{\mathcal{V}}-{T}_{\mathcal{V}}=&~\frac{1}{|\mathcal{V}|}\sum_{\bv\in\mathcal{V}}\bigg[\prod_{\ell=1}^k\varphi_\ell\big(Y_{i_\ell,i_\ell'}\big)-\prod_{\ell=1}^k\mathbb{E}\big\{\varphi_\ell\big(Y_{i_\ell,i_\ell'}\big)\big\}\bigg]\\
=&~\frac{1}{|\mathcal{V}|}\sum_{\bv\in\mathcal{V}}\sum_{\xi_1+\cdots+\xi_k=1\atop\xi_1,\ldots,\xi_k\in\{0,1\}}^k\prod_{\ell=1}^k\mathring{\varphi}_\ell\big(Y_{i_\ell,i_\ell'}\big)^{\xi_\ell}\big[\mathbb{E}\big\{\varphi_\ell\big(Y_{i_\ell,i_\ell'}\big)\big\}\big]^{1-\xi_\ell}\\
=&~\sum_{\xi_1+\cdots+\xi_k=1\atop\xi_1,\ldots,\xi_k\in\{0,1\}}^k\frac{1}{|\mathcal{V}|}\sum_{\bv\in\mathcal{V}}\prod_{\ell=1}^k\mathring{\varphi}_\ell\big(Y_{i_\ell,i_\ell'}\big)^{\xi_\ell}\big[\mathbb{E}\big\{\varphi_\ell\big(Y_{i_\ell,i_\ell'}\big)\big\}\big]^{1-\xi_\ell}\,.
\end{align*}
By Cauchy-Schwarz inequality, we have
\[
\begin{split}
&\mathbb{E}\big(|\tilde{T}_{\mathcal{V}}-{T}_{\mathcal{V}}|^2\big)\leq J_k\sum_{\xi_1+\cdots+\xi_k=1\atop\xi_1,\ldots,\xi_k\in\{0,1\}}^k\mathbb{E}\bigg\{\bigg(\frac{1}{|\mathcal{V}|}\sum_{\bv\in\mathcal{V}}\prod_{\ell=1}^k\mathring{\varphi}_\ell\big(Y_{i_\ell,i_\ell'}\big)^{\xi_\ell}\big[\mathbb{E}\big\{\varphi_\ell\big(Y_{i_\ell,i_\ell'}\big)\big\}\big]^{1-\xi_\ell}\bigg)^2\bigg\}
\end{split}
\]
where $J_k=2^k-1$. For any given $\xi_1,\ldots,\xi_k\in\{0,1\}$, define
\[
\psi_{\xi_1,\ldots,\xi_k}(\bv)=\prod_{\ell=1}^k\mathring{\varphi}_\ell\big(Y_{i_\ell,i_\ell'}\big)^{\xi_\ell}\big[\mathbb{E}\big\{\varphi_\ell\big(Y_{i_\ell,i_\ell'}\big)\big\}\big]^{1-\xi_\ell}
\]
with $\bv=(i_1,i_1',\ldots,i_k,i_k')\in\mathcal{V}$. Therefore,
\begin{equation}\label{eq:bb1}
\mathbb{E}\big(|\tilde{T}_{\mathcal{V}}-{T}_{\mathcal{V}}|^2\big)\leq J_k\sum_{\xi_1+\cdots+\xi_k=1\atop\xi_1,\ldots,\xi_k\in\{0,1\}}^k\mathbb{E}\bigg\{\bigg(\frac{1}{|\mathcal{V}|}\sum_{\bv\in\mathcal{V}}\psi_{\xi_1,\ldots,\xi_k}(\bv)\bigg)^2\bigg\}\,.
\end{equation}

For $\aleph_{\mathcal{V}}(s)$ defined in (\ref{eq:bound}), we adopt the convention $\aleph_{\mathcal{V}}(0)=1$. If $\xi_1+\cdots+\xi_k=s$ with $1\leq s\leq k$, without lose of generality, we assume $\xi_1=\cdots=\xi_s=1$ and $\xi_{s+1}=\cdots=\xi_k=0$. Then
\[
\frac{1}{|\mathcal{V}|}\sum_{\bv\in\mathcal{V}}\psi_{1,\ldots,1,0,\ldots,0}(\bv)=\frac{1}{|\mathcal{V}|}\sum_{\bv\in\mathcal{V}}\bigg(\prod_{\ell=1}^s\mathring{\varphi}_\ell\big(Y_{i_\ell,i_\ell'}\big)\cdot\prod_{\ell=s+1}^k\mathbb{E}\big\{\varphi_\ell\big(Y_{i_\ell,i_\ell'}\big)\big\}\bigg)\,.
\]
For any $\bv=(i_1,i_1',\ldots,i_k,i_k')\in\mathcal{V}$ and $\tilde{\bv}=(\tilde{i}_1,\tilde{i}_1',\ldots,\tilde{i}_k,\tilde{i}_k')\in\mathcal{V}$, if
$
|\{\{i_1,i_1'\},\ldots,\{i_s,i_s'\}\}\cap\{\{\tilde{i}_1,\tilde{i}_1'\},\ldots,\{\tilde{i}_s,\tilde{i}_s'\}\}|<s
$,
then $\mathbb{E}\{\psi_{1,\ldots,1,0,\ldots,0}(\bv)\psi_{1,\ldots,1,0,\ldots,0}(\tilde{\bv})\}=0$. Recall that $|\psi_{\xi_1,\ldots,\xi_k}(\bv)|\leq q_{\max}^k$ for any $\xi_1,\ldots,\xi_k\in\{0,1\}$ and $\bv\in\mathcal{V}$, where $q_{\max} =\max\{1-\alpha,\alpha,1-\beta,\beta\}$. Thus,
\[
\mathbb{E}\bigg[\bigg\{\frac{1}{|\mathcal{V}|}\sum_{\bv\in\mathcal{V}}\psi_{1,\ldots,1,0\ldots,0}(\bv)\bigg\}^2\bigg]\leq\frac{2^s s!\aleph_{\mathcal{V}}(k-s)}{|\mathcal{V}|^2}\sum_{\bv\in\mathcal{V}}q_{\max}^{2k}=\frac{2^{s} s! q_{\max}^{2k}\aleph_{\mathcal{V}}(k-s)}{|\mathcal{V}|}\,.
\]
Similarly, we know
\begin{equation}\label{eq:b2}
\mathbb{E}\bigg[\bigg\{\frac{1}{|\mathcal{V}|}\sum_{\bv\in\mathcal{V}}\psi_{\xi_1,\ldots,\xi_k}(\bv)\bigg\}^2\bigg]\leq\frac{2^{s} s! q_{\max}^{2k}\aleph_{\mathcal{V}}(k-s)}{|\mathcal{V}|}
\end{equation}for any $\xi_1,\ldots,\xi_k\in\{0,1\}$ such that $\xi_1+\cdots+\xi_k=s$. Therefore, from (\ref{eq:bb1}), it holds that
\begin{equation}\label{eq:b1}
\mathbb{E}\big(|\tilde{T}_{\mathcal{V}}-{T}_{\mathcal{V}}|^2\big)\leq \frac{2^{k} k! q_{\max}^{2k}J_k^2}{|\mathcal{V}|}\max_{1\leq s\leq k}\aleph_{\mathcal{V}}(k-s)=\frac{2^{k} k! q_{\max}^{2k}J_k^2}{|\mathcal{V}|}\aleph_{\mathcal{V}}\,.
\end{equation}
It follows from Markov inequality that
\[
|\tilde{T}_{\mathcal{V}}-{T}_{\mathcal{V}}|=O_p\bigg(\sqrt{\frac{\aleph_{\mathcal{V}}}{|\mathcal{V}|}}\bigg)\,.
\] We complete the proof of Proposition \ref{tm:pre}. $\hfill\Box$

\subsection*{Proof of Proposition \ref{pre2}}

Notice that $\aleph_{\mathcal{V}}(s)/\aleph_{\mathcal{V}}\rightarrow0$ for each $1\leq s\leq k-2$. By the definition of $\varphi_\ell(\cdot)$, we have $\mathring{\varphi}_\ell(Y_{i_\ell,i_\ell'})=(-1)^{1-\tau_\ell}\mathring{Y}_{i_\ell,i_\ell'}$ with $\mathring{Y}_{i_\ell,i_\ell'}=Y_{i_\ell,i_\ell'}-\mathbb{E}(Y_{i_\ell,i_\ell'})$. Then we have
\begin{equation}\label{eq:expandor}
\begin{split}
\tilde{T}_{\mathcal{V}}-T_{\mathcal{V}}=&~\sum_{\xi_1+\cdots+\xi_k=1\atop\xi_1,\ldots,\xi_k\in\{0,1\}}\frac{1}{|\mathcal{V}|}\sum_{\bv\in\mathcal{V}}\prod_{\ell=1}^k\mathring{\varphi}_\ell\big(Y_{i_\ell,i_\ell'}\big)^{\xi_\ell}\big[\mathbb{E}\big\{\varphi_\ell\big(Y_{i_\ell,i_\ell'}\big)\big\}\big]^{1-\xi_\ell}+o_p\bigg(\sqrt{\frac{\aleph_{\mathcal{V}}}{|\mathcal{V}|}}\bigg)\\
=&~\sum_{j=1}^k\frac{1}{|\mathcal{V}|}\sum_{\bv\in\mathcal{V}}\bigg[\mathring{\varphi}_j\big(Y_{i_j,i_j'}\big)\prod_{\ell\neq j}\mathbb{E}\big\{\varphi_\ell\big(Y_{i_\ell,i_\ell'}\big)\big\}\bigg]+o_p\bigg(\sqrt{\frac{\aleph_{\mathcal{V}}}{|\mathcal{V}|}}\bigg)\\
=&~\sum_{j=1}^k\frac{(-1)^{1-\tau_j}}{|\mathcal{V}|}\sum_{\bv\in\mathcal{V}}\bigg[\mathring{Y}_{i_j,i_j'}\prod_{\ell\neq j}\mathbb{E}\big\{\varphi_\ell\big(Y_{i_\ell,i_\ell'}\big)\big\}\bigg]+o_p\bigg(\sqrt{\frac{\aleph_{\mathcal{V}}}{|\mathcal{V}|}}\bigg)\,.
\end{split}
\end{equation}
Notice that $\aleph_{\mathcal{V}}/|\mathcal{V}|=O_p(N^{-1})$ and $\sqrt{N}(\tilde{C}_{\mathcal{V}}-C_\mathcal{V})=(1-\alpha-\beta)^{-k}\sqrt{N}(\tilde{T}_{\mathcal{V}}-T_{\mathcal{V}})$. Then we complete the proof of Proposition \ref{pre2}. $\hfill\Box$

\subsection*{Proof of Theorem \ref{tm:bootstrap}}

Let
\[
\theta=\mathbb{E} \bigg\{\bigg(\sum_{j=1}^k\frac{(-1)^{1-\tau_j}}{|\mathcal{V}|}\sum_{\bv\in\mathcal{V}}\bigg[\mathring{Y}_{i_j,i_j'}\prod_{\ell\neq j}\mathbb{E}\big\{\varphi_\ell\big(Y_{i_\ell,i_\ell'}\big)\big\}\bigg]\bigg)^2\bigg\}\,.
\]
Based on the Berry-Essen Theorem, we have
\begin{equation}\label{eq:be1}
\sup_{z\in\mathbb{R}}\bigg|\mathbb{P}\big\{\sqrt{N}(\tilde{C}_{\mathcal{V}}-C_{\mathcal{V}})\leq z\big\}-\Phi\bigg\{\frac{(1-\alpha-\beta)^kz}{\sqrt{N\theta}}\bigg\}\bigg|\rightarrow0\,,
\end{equation}
where $\Phi(\cdot)$ denotes the cumulative distribution function of standard normal distribution.
It holds that
\begin{equation}\label{eq:d}
\begin{split}
\theta=\sum_{j_1,j_2=1}^k\frac{(-1)^{2-\tau_{j_1}-\tau_{j_2}}}{|\mathcal{V}|^2}\sum_{\bv,\tilde{\bv}\in\mathcal{V}}\bigg[\mathbb{E}\big(\mathring{Y}_{i_{j_1},i_{j_1}'}\mathring{Y}_{\tilde{i}_{j_2},\tilde{i}_{j_2}'}\big)\prod_{\ell\neq j_1}\mathbb{E}\big\{\varphi_\ell\big(Y_{i_\ell,i_\ell'}\big)\big\}\prod_{\ell\neq j_2}\mathbb{E}\big\{\varphi_\ell\big(Y_{\tilde{i}_\ell,\tilde{i}_\ell'}\big)\big\}\bigg]\,.
\end{split}
\end{equation}
Notice that $\mathbb{E}(\mathring{Y}_{i_{j_1},i_{j_1}'}\mathring{Y}_{\tilde{i}_{j_2},\tilde{i}_{j_2}'})=\{A_{i_{j_1},i_{j_1}'}(1-\alpha-\beta)+\alpha\}\{1-\alpha-A_{i_{j_1},i_{j_1}'}(1-\alpha-\beta)\}={\rm Var}(Y_{i_{j_1},i_{j_1}'})$ if $\{i_{j_1},i_{j_1}'\}=\{\tilde{i}_{j_2},\tilde{i}_{j_2}'\}$, and $\mathbb{E}\big(\mathring{Y}_{i_{j_1},i_{j_1}'}\mathring{Y}_{\tilde{i}_{j_2},\tilde{i}_{j_2}'}\big)=0$ if $\{i_{j_1},i_{j_1}'\}\neq\{\tilde{i}_{j_2},\tilde{i}_{j_2}'\}$. For any $j_1,j_2=1,\ldots,k$ and $\bv=(i_1,i_1',\ldots,i_k,i_k')\in\mathcal{V}$, define $\mathcal{V}_{j_1,j_2}(\bv)=\{\tilde{\bv}=(\tilde{i}_1,\tilde{i}_1',\ldots,\tilde{i}_k,\tilde{i}_k')\in\mathcal{V}:\{\tilde{i}_{j_2},\tilde{i}_{j_2}'\}=\{{i}_{j_1},{i}_{j_1}'\}\}$. Then
\[
\begin{split}
\theta=&~\sum_{j_1,j_2=1}^k\frac{(-1)^{2-\tau_{j_1}-\tau_{j_2}}}{|\mathcal{V}|^2}\sum_{\bv\in\mathcal{V}}\bigg[{\rm Var}\big(Y_{i_{j_1},i_{j_1}'}\big)\prod_{\ell\neq j_1}\mathbb{E}\big\{\varphi_\ell\big(Y_{i_\ell,i_\ell'}\big)\big\}\sum_{\tilde{\bv}\in\mathcal{V}_{j_1,j_2}(\bv)}\prod_{\ell\neq j_2}\mathbb{E}\big\{\varphi_\ell\big(Y_{\tilde{i}_\ell,\tilde{i}_\ell'}\big)\big\}\bigg]\,.
\end{split}
\]

Define
\[
Z=\frac{\sqrt{N}}{(1-\alpha-\beta)^k}\sum_{j=1}^k\frac{(-1)^{1-\tau_j}}{|\mathcal{V}|}\sum_{\bv\in\mathcal{V}}\bigg\{\mathring{Y}_{i_j,i_j'}^\dag\prod_{\ell\neq j}\varphi_\ell\big(Y_{i_\ell,i_\ell'}\big)\bigg\}\,.
\]
Given $\bY=(Y_{i,j})_{p\times p}$, we have $Z\rightarrow_d \mathcal{N}(0,\hat{\sigma}^2_{\mathcal{V}})$ with
\[
\begin{split}
\hat{\sigma}^2_{\mathcal{V}}=&~\frac{1}{(1-\alpha-\beta)^{2k}}\lim_{p\rightarrow\infty}N\mathbb{E}^*\bigg(\bigg[\sum_{j=1}^k\frac{(-1)^{1-\tau_j}}{|\mathcal{V}|}\sum_{\bv\in\mathcal{V}}\bigg\{\mathring{Y}_{i_j,i_j'}^\dag\prod_{\ell\neq j}\varphi_\ell\big(Y_{i_\ell,i_\ell'}\big)\bigg\}\bigg]^2\bigg)\\
=&:\frac{1}{(1-\alpha-\beta)^{2k}}\lim_{p\rightarrow\infty}N\theta^*\,,
\end{split}
\]
where $\mathbb{E}^*(\cdot)$ denotes the conditional expectation given $\bY$.
Based on the Berry-Essen Theorem, we have
\begin{equation}\label{eq:be2}
\sup_{z\in\mathbb{R}}\bigg|\mathbb{P}\big(Z\leq z\,|\,\bY\big)-\Phi\bigg\{\frac{(1-\alpha-\beta)^kz}{\sqrt{N\theta^*}}\bigg\}\bigg|\rightarrow0\,.
\end{equation}
Same as (\ref{eq:d}), we have
\[
\begin{split}
\theta^*=&~\sum_{j_1,j_2=1}^k\frac{(-1)^{2-\tau_{j_1}-\tau_{j_2}}}{|\mathcal{V}|^2}\sum_{\bv,\tilde{\bv}\in\mathcal{V}}\bigg[\mathbb{E}^*\big(\mathring{Y}_{i_{j_1},i_{j_1}'}^\dag\mathring{Y}_{\tilde{i}_{j_2},\tilde{i}_{j_2}'}^\dag\big)\prod_{\ell\neq j_1}\varphi_\ell\big(Y_{i_\ell,i_\ell'}\big)\prod_{\ell\neq j_2}\varphi_\ell\big(Y_{\tilde{i}_\ell,\tilde{i}_\ell'}\big)\bigg]\\
=&~\sum_{j_1,j_2=1}^k\frac{(-1)^{2-\tau_{j_1}-\tau_{j_2}}}{|\mathcal{V}|^2}\sum_{\bv\in\mathcal{V}}\bigg[{\rm Var}^*\big(Y_{i_{j_1},i_{j_1}'}^\dag\big)\prod_{\ell\neq j_1}\varphi_\ell\big(Y_{i_\ell,i_\ell'}\big)\sum_{\tilde{\bv}\in\mathcal{V}_{j_1,j_2}(\bv)}\prod_{\ell\neq j_2}\varphi_\ell\big(Y_{\tilde{i}_\ell,\tilde{i}_\ell'}\big)\bigg]\,,
\end{split}
\]
where ${\rm Var}^*(\cdot)$ denotes the conditional variance given $\bY$. It follows from (\ref{eq:be1}) and (\ref{eq:be2}) that
\[
\begin{split}
&\sup_{z\in\mathbb{R}}\big|\mathbb{P}\big\{\sqrt{N}(\tilde{C}_{\mathcal{V}}-C_{\mathcal{V}})\leq z\big\}-\mathbb{P}\big(Z\leq z\,|\,\bY\big)\big|\\
&~~~~~~~~\leq \sup_{z\in\mathbb{R}}\bigg|\Phi\bigg\{\frac{(1-\alpha-\beta)^kz}{\sqrt{N\theta}}\bigg\}-\Phi\bigg\{\frac{(1-\alpha-\beta)^kz}{\sqrt{N\theta^*}}\bigg\}\bigg|+o(1)\,.\\
\end{split}
\]In the sequel, we show $|\theta^*-\theta|=o_p(N^{-1})$. To do this, we only need to show
\[
\begin{split}
\Delta_{j_1,j_2}:=&~\frac{1}{|\mathcal{V}|^2}\sum_{\bv\in\mathcal{V}}\bigg[{\rm Var}^*\big(Y_{i_{j_1},i_{j_1}'}^\dag\big)\prod_{\ell\neq j_1}\varphi_\ell\big(Y_{i_\ell,i_\ell'}\big)\sum_{\tilde{\bv}\in\mathcal{V}_{j_1,j_2}(\bv)}\prod_{\ell\neq j_2}\varphi_\ell\big(Y_{\tilde{i}_\ell,\tilde{i}_\ell'}\big)\bigg]\\
&~~~~~~~~~-\frac{1}{|\mathcal{V}|^2}\sum_{\bv\in\mathcal{V}}\bigg[{\rm Var}\big(Y_{i_{j_1},i_{j_1}'}\big)\prod_{\ell\neq j_1}\mathbb{E}\big\{\varphi_\ell\big(Y_{i_\ell,i_\ell'}\big)\big\}\sum_{\tilde{\bv}\in\mathcal{V}_{j_1,j_2}(\bv)}\prod_{\ell\neq j_2}\mathbb{E}\big\{\varphi_\ell\big(Y_{\tilde{i}_\ell,\tilde{i}_\ell'}\big)\big\}\bigg]\\
=&~o_p(N^{-1})
\end{split}
\]
for any $j_1,j_2=1,\ldots,k$. Notice that ${\rm Var}^*(Y_{i_j,i_j'}^\dag)=Y_{i_j,i_j'}(\beta-\alpha)+\alpha(1-\beta)$ and $\mathbb{E}\{{\rm Var}^*(Y_{i_j,i_j'}^\dag)\}={\rm Var}(Y_{i_j,i_j'})$. Given $j_1$, define $\tilde{\varphi}_\ell(Y_{i_\ell,i_\ell'})=\varphi_\ell(Y_{i_\ell,i_\ell'})$ for any $\ell\neq j_1$, and $\tilde{\varphi}_{j_1}(Y_{i_{j_1},i_{j_1}'})=Y_{i_{j_1},i_{j_1}'}(\beta-\alpha)+\alpha(1-\beta)$. Then
\[
\begin{split}
\Delta_{j_1,j_2}=&~\frac{1}{|\mathcal{V}|^2}\sum_{\bv\in\mathcal{V}}\bigg[\prod_{\ell=1}^k\tilde{\varphi}_\ell\big(Y_{i_\ell,i_\ell'}\big)-\prod_{\ell=1}^k\mathbb{E}\big\{\tilde{\varphi}_\ell\big(Y_{i_\ell,i_\ell'}\big)\big\}\bigg]\sum_{\tilde{\bv}\in\mathcal{V}_{j_1,j_2}(\bv)}\prod_{\ell\neq j_2}\mathbb{E}\big\{\varphi_\ell\big(Y_{\tilde{i}_\ell,\tilde{i}_\ell'}\big)\big\}\\
&+\frac{1}{|\mathcal{V}|^2}\sum_{\bv\in\mathcal{V}}\bigg[\prod_{\ell=1}^k\mathbb{E}\big\{\tilde{\varphi}_\ell\big(Y_{i_\ell,i_\ell'}\big)\big\}\bigg]\sum_{\tilde{\bv}\in\mathcal{V}_{j_1,j_2}(\bv)}\bigg[\prod_{\ell\neq j_2}\varphi_\ell\big(Y_{\tilde{i}_\ell,\tilde{i}_\ell'}\big)-\prod_{\ell\neq j_2}\mathbb{E}\big\{\varphi_\ell\big(Y_{\tilde{i}_\ell,\tilde{i}_\ell'}\big)\big\}\bigg]\\
&+\frac{1}{|\mathcal{V}|^2}\sum_{\bv\in\mathcal{V}}\bigg[\prod_{\ell=1}^k\tilde{\varphi}_\ell\big(Y_{i_\ell,i_\ell'}\big)-\prod_{\ell=1}^k\mathbb{E}\big\{\tilde{\varphi}_\ell\big(Y_{i_\ell,i_\ell'}\big)\big\}\bigg]\\
&~~~~~~~~~~~~~~~~~~~\times\sum_{\tilde{\bv}\in\mathcal{V}_{j_1,j_2}(\bv)}\bigg[\prod_{\ell\neq j_2}\varphi_\ell\big(Y_{\tilde{i}_\ell,\tilde{i}_\ell'}\big)-\prod_{\ell\neq j_2}\mathbb{E}\big\{\varphi_\ell\big(Y_{\tilde{i}_\ell,\tilde{i}_\ell'}\big)\big\}\bigg]\\
:=&~\Delta_{j_1,j_2}(1)+\Delta_{j_1,j_2}(2)+\Delta_{j_1,j_2}(3)\,.
\end{split}
\]
We will show $|\Delta_{j_1,j_2}(1)|=o_p(N^{-1})$, $|\Delta_{j_1,j_2}(2)|=o_p(N^{-1})$ and $|\Delta_{j_1,j_2}(3)|=o_p(N^{-1})$.

For $\Delta_{j_1,j_2}(1)$, it holds that
\[
\Delta_{j_1,j_2}(1)=\sum_{\xi_1+\cdots+\xi_k=1\atop \xi_1,\ldots,\xi_k\in\{0,1\}}^k\frac{1}{|\mathcal{V}|^2}\sum_{\bv\in\mathcal{V}}\prod_{\ell=1}^k\mathring{\tilde{\varphi}}_\ell(Y_{i_\ell,i_\ell'})^{\xi_\ell}\big[\mathbb{E}\big\{\tilde{\varphi}_\ell(Y_{i_\ell,i_\ell'})\big\}\big]^{1-\xi_\ell}B_1(\bv)
\]
where $\mathring{\tilde{\varphi}}_\ell(Y_{i_\ell,i_\ell'})={\tilde{\varphi}}_\ell(Y_{i_\ell,i_\ell'})-\mathbb{E}\{{\tilde{\varphi}}_\ell(Y_{i_\ell,i_\ell'})\}$ and
\[
B_1(\bv)=\sum_{\tilde{\bv}\in\mathcal{V}_{j_1,j_2}(\bv)}\prod_{\ell\neq j_2}\mathbb{E}\big\{\varphi_\ell\big(Y_{\tilde{i}_\ell,\tilde{i}_\ell'}\big)\big\}\,.
\]
Same as (\ref{eq:bb1}) and (\ref{eq:b1}), we have
\[
\mathbb{E}\big\{|\Delta_{j_1,j_2}(1)|^2\big\}\leq \frac{C\aleph_{\mathcal{V}}^3}{|\mathcal{V}|^3}=O(N^{-3})\,,
\]
which implies $|\Delta_{j_1,j_2}(1)|=O_p(N^{-3/2})=o_p(N^{-1})$. Notice that if $\tilde{\bv}\in\mathcal{V}_{j_1,j_2}(\bv)$, then $\bv\in\mathcal{V}_{j_2,j_1}(\tilde{\bv})$. We can reformulate $\Delta_{j_1,j_2}(2)$ as
\[
\Delta_{j_1,j_2}(2)=\frac{1}{|\mathcal{V}|^2}\sum_{\tilde{\bv}\in\mathcal{V}}\bigg[\prod_{\ell\neq j_2}\varphi_\ell\big(Y_{\tilde{i}_\ell,\tilde{i}_\ell'}\big)-\prod_{\ell\neq j_2}\mathbb{E}\big\{\varphi_\ell\big(Y_{\tilde{i}_\ell,\tilde{i}_\ell'}\big)\big\}\bigg]\sum_{{\bv}\in\mathcal{V}_{j_2,j_1}(\tilde{\bv})}\bigg[\prod_{\ell=1}^k\mathbb{E}\big\{\tilde{\varphi}_\ell\big(Y_{i_\ell,i_\ell'}\big)\big\}\bigg]\,.
\]
Following the same arguments to bound $\mathbb{E}\{|\Delta_{j_1,j_2}(1)|^2\}$, we have $|\Delta_{j_1,j_2}(2)|=o_p(N^{-1})$. For $\Delta_{j_1,j_2}(3)$, we can reformulate it as
\begin{align*}
\Delta_{j_1,j_2}(3)=&~\sum_{\xi_1+\cdots+\xi_k=1\atop \xi_1,\ldots,\xi_k\in\{0,1\}}^k\sum_{\tilde{\xi}_1+\cdots+\tilde{\xi}_{j_2-1}+\tilde{\xi}_{j_2+1}+\cdots+\tilde{\xi}_k=1\atop \tilde{\xi}_1,\ldots,\tilde{\xi}_{j_2-1},\tilde{\xi}_{j_2+1},\ldots,\tilde{\xi}_k\in\{0,1\}}^{k-1}\notag\\
&~~~~~~~~~\frac{1}{|\mathcal{V}|^2}\sum_{\bv\in\mathcal{V}}\bigg(\prod_{\ell=1}^k\mathring{\tilde{\varphi}}_\ell\big(Y_{i_\ell,i_\ell'}\big)^{\xi_\ell}\big[\mathbb{E}\big\{\tilde{\varphi}_\ell\big(Y_{i_\ell,i_\ell'}\big)\big\}\big]^{1-\xi_\ell}\bigg)\notag\\
&~~~~~~~~~~~~~~~~~~~~~~~~~~~~~~\times\sum_{\tilde{\bv}\in\mathcal{V}_{j_1,j_2}(\bv)}\bigg(\prod_{\ell\neq j_2}\mathring{{\varphi}}_\ell\big(Y_{\tilde{i}_\ell,\tilde{i}_\ell'}\big)^{\tilde{\xi}_\ell}\big[\mathbb{E}\big\{{\varphi}_\ell\big(Y_{\tilde{i}_\ell,\tilde{i}_\ell'}\big)\big\}\big]^{1-\tilde{\xi}_\ell}\bigg)\,.\notag
\end{align*}
By Cauchy-Schwarz inequality, we have
\begin{align*}
&\mathbb{E}\bigg\{\bigg|\frac{1}{|\mathcal{V}|^2}\sum_{\bv\in\mathcal{V}}\bigg(\prod_{\ell=1}^k\mathring{\tilde{\varphi}}_\ell\big(Y_{i_\ell,i_\ell'}\big)^{\xi_\ell}\big[\mathbb{E}\big\{\tilde{\varphi}_\ell\big(Y_{i_\ell,i_\ell'}\big)\big\}\big]^{1-\xi_\ell}\bigg)\\
&~~~~~~~~~~~~~~~~~~~~~~~~~~~~~~\times\sum_{\tilde{\bv}\in\mathcal{V}_{j_1,j_2}(\bv)}\bigg(\prod_{\ell\neq j_2}\mathring{{\varphi}}_\ell\big(Y_{\tilde{i}_\ell,\tilde{i}_\ell'}\big)^{\tilde{\xi}_\ell}\big[\mathbb{E}\big\{{\varphi}_\ell\big(Y_{\tilde{i}_\ell,\tilde{i}_\ell'}\big)\big\}\big]^{1-\tilde{\xi}_\ell}\bigg)\bigg|^2\bigg\}\\
&~~~~~~~~~\leq \frac{1}{|\mathcal{V}|^4}\mathbb{E}\bigg\{\sum_{\bv\in\mathcal{V}}\bigg(\prod_{\ell=1}^k\mathring{\tilde{\varphi}}_\ell\big(Y_{i_\ell,i_\ell'}\big)^{\xi_\ell}\big[\mathbb{E}\big\{\tilde{\varphi}_\ell\big(Y_{i_\ell,i_\ell'}\big)\big\}\big]^{1-\xi_\ell}\bigg)^2\\
&~~~~~~~~~~~~~~~~~~~~~~~~~~~~~~\times\sum_{\bv\in\mathcal{V}}\bigg(\sum_{\tilde{\bv}\in\mathcal{V}_{j_1,j_2}(\bv)}\prod_{\ell\neq j_2}\mathring{{\varphi}}_\ell\big(Y_{\tilde{i}_\ell,\tilde{i}_\ell'}\big)^{\tilde{\xi}_\ell}\big[\mathbb{E}\big\{{\varphi}_\ell\big(Y_{\tilde{i}_\ell,\tilde{i}_\ell'}\big)\big\}\big]^{1-\tilde{\xi}_\ell}\bigg)^2\bigg\}\,.
\end{align*}
Notice that $\tilde{\varphi}_\ell\big(Y_{i_\ell,i_\ell'}\big)$ is bounded, then
\begin{align*}
&\mathbb{E}\bigg\{\bigg|\frac{1}{|\mathcal{V}|^2}\sum_{\bv\in\mathcal{V}}\bigg(\prod_{\ell=1}^k\mathring{\tilde{\varphi}}_\ell\big(Y_{i_\ell,i_\ell'}\big)^{\xi_\ell}\big[\mathbb{E}\big\{\tilde{\varphi}_\ell\big(Y_{i_\ell,i_\ell'}\big)\big\}\big]^{1-\xi_\ell}\bigg)\\
&~~~~~~~~~~~~~~~~~~~~~~~~~~~~~~\times\sum_{\tilde{\bv}\in\mathcal{V}_{j_1,j_2}(\bv)}\bigg(\prod_{\ell\neq j_2}\mathring{{\varphi}}_\ell\big(Y_{\tilde{i}_\ell,\tilde{i}_\ell'}\big)^{\tilde{\xi}_\ell}\big[\mathbb{E}\big\{{\varphi}_\ell\big(Y_{\tilde{i}_\ell,\tilde{i}_\ell'}\big)\big\}\big]^{1-\tilde{\xi}_\ell}\bigg)\bigg|^2\bigg\}\\
&~~~~~~~~~\leq \frac{C}{|\mathcal{V}|^3}\sum_{\bv\in\mathcal{V}}\mathbb{E}\bigg\{\bigg(\sum_{\tilde{\bv}\in\mathcal{V}_{j_1,j_2}(\bv)}\prod_{\ell\neq j_2}\mathring{{\varphi}}_\ell\big(Y_{\tilde{i}_\ell,\tilde{i}_\ell'}\big)^{\tilde{\xi}_\ell}\big[\mathbb{E}\big\{{\varphi}_\ell\big(Y_{\tilde{i}_\ell,\tilde{i}_\ell'}\big)\big\}\big]^{1-\tilde{\xi}_\ell}\bigg)^2\bigg\}\,.
\end{align*}
Same as (\ref{eq:b2}), we have
\begin{align*}
\mathbb{E}\bigg\{\bigg(\sum_{\tilde{\bv}\in\mathcal{V}_{j_1,j_2}(\bv)}\prod_{\ell\neq j_2}\mathring{{\varphi}}_\ell\big(Y_{\tilde{i}_\ell,\tilde{i}_\ell'}\big)^{\tilde{\xi}_\ell}\big[\mathbb{E}\big\{{\varphi}_\ell\big(Y_{\tilde{i}_\ell,\tilde{i}_\ell'}\big)\big\}\big]^{1-\tilde{\xi}_\ell}\bigg)^2\bigg\}\leq C\aleph_{\mathcal{V}}\aleph_{\mathcal{V}}(k-2)\,,
\end{align*}
which implies
\begin{align*}
&\mathbb{E}\bigg\{\bigg|\frac{1}{|\mathcal{V}|^2}\sum_{\bv\in\mathcal{V}}\bigg(\prod_{\ell=1}^k\mathring{\tilde{\varphi}}_\ell\big(Y_{i_\ell,i_\ell'}\big)^{\xi_\ell}\big[\mathbb{E}\big\{\tilde{\varphi}_\ell\big(Y_{i_\ell,i_\ell'}\big)\big\}\big]^{1-\xi_\ell}\bigg)\\
&~~~~~~~~~~~~~~~~~~~~~~~~~~~~~~\times\sum_{\tilde{\bv}\in\mathcal{V}_{j_1,j_2}(\bv)}\bigg(\prod_{\ell\neq j_2}\mathring{{\varphi}}_\ell\big(Y_{\tilde{i}_\ell,\tilde{i}_\ell'}\big)^{\tilde{\xi}_\ell}\big[\mathbb{E}\big\{{\varphi}_\ell\big(Y_{\tilde{i}_\ell,\tilde{i}_\ell'}\big)\big\}\big]^{1-\tilde{\xi}_\ell}\bigg)\bigg|^2\bigg\}\\
&~~~~~~~~~\leq \frac{C\aleph_{\mathcal{V}}\aleph_{\mathcal{V}}(k-2)}{|\mathcal{V}|^2}=o\bigg(\frac{\aleph_{\mathcal{V}}^2}{|\mathcal{V}|^2}\bigg)=o(N^{-2})\,.
\end{align*}
Then $|\Delta_{j_1,j_2}(3)|=o_p(N^{-1})$. We complete the proof of Theorem \ref{tm:bootstrap}. $\hfill\Box$

\subsection*{Proof of Proposition \ref{pn2}}
To simplify the notation, we write $\hat{T}_{\mathcal{V}}(\tau_1,\ldots,\tau_k)$, $\tilde{T}_{\mathcal{V}}(\tau_1,\ldots,\tau_k)$ and $T_{\mathcal{V}}(\tau_1,\ldots,\tau_k)$ as $\hat{T}_{\mathcal{V}}$, $\tilde{T}_{\mathcal{V}}$ and $T_{\mathcal{V}}$, respectively. For given $\tau_1,\ldots,\tau_k\in\{0,1\}$, we define $\hat{\varphi}_\ell(x)=(x-\tilde{\alpha})^{\tau_\ell}(1-\tilde{\beta}-x)^{1-\tau_\ell}$ for $x\in\{0,1\}$. Recall that
\[
\tilde{T}_{\mathcal{V}}=\frac{1}{|\mathcal{V}|}\sum_{\bv\in\mathcal{V}}\prod_{\ell=1}^k\varphi_\ell\big(Y_{i_\ell,i_\ell'}\big)\,.
\]
As we have shown in Proposition \ref{tm:pre} that $|\tilde{T}_{\mathcal{V}}-T_{\mathcal{V}}|=O_p(N^{-1/2})$. To show $|\hat{T}_{\mathcal{V}}-T_{\mathcal{V}}|=O_p(N^{-1/2})$, we only need to prove $|\hat{T}_{\mathcal{V}}-\tilde{T}_{\mathcal{V}}|=O_p(N^{-1/2})$.

For each $\bv\in\mathcal{V}$, we have the following identity
\[
\begin{split}
&\prod_{\ell=1}^k\hat{\varphi}_\ell\big(Y_{i_\ell,i_\ell'}\big)-\prod_{\ell=1}^k{\varphi}_\ell\big(Y_{i_\ell,i_\ell'}\big)=\sum_{\xi_1+\cdots+\xi_k=1\atop \xi_1,\ldots,\xi_k\in\{0,1\}}^k \prod_{\ell=1}^k\big\{\hat{\varphi}_\ell\big(Y_{i_\ell,i_\ell'}\big)-{\varphi}_\ell\big(Y_{i_\ell,i_\ell'}\big)\big\}^{\xi_\ell}\big\{{\varphi}_\ell\big(Y_{i_\ell,i_\ell'}\big)\big\}^{1-\xi_\ell}\,.
\end{split}
\]
Recall that
$
\hat{\varphi}_\ell(Y_{i_\ell,i_\ell'})-{\varphi}_\ell(Y_{i_\ell,i_\ell'})=(\alpha-\tilde{\alpha})^{\tau_\ell}(\beta-\tilde{\beta})^{1-\tau_\ell}
$ and $Y_{i_\ell,i_\ell'}\in\{0,1\}$. Let $r_{\max}=\max\{|\tilde{\alpha}-\alpha|,|\tilde{\beta}-\beta|\}$. Notice that $r_{\max}=O_p(N^{-1/2})$. Then
\[
\begin{split}
\bigg|\prod_{\ell=1}^k\hat{\varphi}_\ell\big(Y_{i_\ell,i_\ell'}\big)-\prod_{\ell=1}^k{\varphi}_\ell\big(Y_{i_\ell,i_\ell'}\big)\bigg|\leq&~ \sum_{\xi_1+\cdots+\xi_k=1\atop \xi_1,\ldots,\xi_k\in\{0,1\}}^k \prod_{\ell=1}^k \big(|\hat{\alpha}-{\alpha}|^{\tau_\ell}|\hat{\beta}-{\beta}|^{1-\tau_\ell}\big)^{\xi_\ell}\\
\leq&~\sum_{\xi_1+\cdots+\xi_k=1\atop \xi_1,\ldots,\xi_k\in\{0,1\}}^k r_{\max}^{\xi_1+\cdots+\xi_k}=\sum_{\ell=1}^kC_k^\ell r_{\max}^\ell\,,
\end{split}
\]
which implies that
$
|\hat{T}_{\mathcal{V}}-\tilde{T}_{\mathcal{V}}|\leq \sum_{\ell=1}^kC_k^\ell r_{\max}^\ell=O_p(N^{-1/2})
$.

Recall that $\tilde{\alpha}-\alpha=O_p(N^{-1/2})$, $\tilde{\beta}-\beta=O_p(N^{-1/2})$ and $\hat{T}_{\mathcal{V}}-T_{\mathcal{V}}=O_p(N^{-1/2})$. It holds that
\begin{equation}\label{eq:Cv}
\begin{split}
\sqrt{N}\big(\hat{C}_{\mathcal{V}}-C_{\mathcal{V}}\big)=&~\frac{\sqrt{N}\hat{T}_{\mathcal{V}}}{(1-\tilde{\alpha}-\tilde{\beta})^k}-\frac{\sqrt{N}{T}_{\mathcal{V}}}{(1-{\alpha}-{\beta})^k}\\
=&~\frac{\sqrt{N}(\hat{T}_{\mathcal{V}}-T_{\mathcal{V}})}{(1-\alpha-\beta)^k}+\frac{kT_{\mathcal{V}}\sqrt{N}(\tilde{\alpha}-\alpha)}{(1-\alpha-\beta)^{k+1}}+\frac{kT_{\mathcal{V}}\sqrt{N}(\tilde{\beta}-\beta)}{(1-\alpha-\beta)^{k+1}}+O_p(N^{-1/2})\\
=&~\frac{\sqrt{N}(\hat{T}_{\mathcal{V}}-T_{\mathcal{V}})}{(1-\alpha-\beta)^k}+\frac{kC_{\mathcal{V}}\sqrt{N}(\tilde{\alpha}-\alpha)}{1-\alpha-\beta}+\frac{kC_{\mathcal{V}}\sqrt{N}(\tilde{\beta}-\beta)}{1-\alpha-\beta}+O_p(N^{-1/2})\,.
\end{split}
\end{equation}
In the sequel, we will specify the leading term of $\sqrt{N}(\hat{T}_{\mathcal{V}}-T_{\mathcal{V}})$. Notice that $\sqrt{N}(\hat{T}_{\mathcal{V}}-T_{\mathcal{V}})=\sqrt{N}(\hat{T}_{\mathcal{V}}-\tilde{T}_{\mathcal{V}})+\sqrt{N}(\tilde{T}_{\mathcal{V}}-T_{\mathcal{V}})$. Recall that
\[
\begin{split}
\hat{T}_{\mathcal{V}}-\tilde{T}_{\mathcal{V}}=&~\frac{1}{|\mathcal{V}|}\sum_{\bv\in\mathcal{V}}\prod_{\ell=1}^k\hat{\varphi}_\ell\big(Y_{i_\ell,i_\ell'}\big)-\frac{1}{|\mathcal{V}|}\sum_{\bv\in\mathcal{V}}\prod_{\ell=1}^k{\varphi}_\ell\big(Y_{i_\ell,i_\ell'}\big)\\
=&~\frac{1}{|\mathcal{V}|}\sum_{\bv\in\mathcal{V}}\sum_{\xi_1+\cdots+\xi_k=1\atop \xi_1,\ldots,\xi_k\in\{0,1\}}^k\prod_{\ell=1}^k\big\{\hat{\varphi}_\ell\big(Y_{i_\ell,i_\ell'}\big)-{\varphi}_\ell\big(Y_{i_\ell,i_\ell'}\big)\big\}^{\xi_\ell}\big\{{\varphi}_\ell\big(Y_{i_\ell,i_\ell'}\big)\big\}^{1-\xi_\ell}\,.
\end{split}
\]
Since $\hat{\varphi}_\ell(Y_{i_\ell,i_\ell'})-{\varphi}_\ell(Y_{i_\ell,i_\ell'})=(\alpha-\tilde{\alpha})^{\tau_\ell}(\beta-\tilde{\beta})^{1-\tau_\ell}$, we have that
\[
\hat{T}_{\mathcal{V}}-\tilde{T}_{\mathcal{V}}=\sum_{\xi_1+\cdots+\xi_k=1\atop \xi_1,\ldots,\xi_k\in\{0,1\}}^k(\alpha-\tilde{\alpha})^{\sum_{\ell=1}^k\tau_\ell\xi_\ell}(\beta-\tilde{\beta})^{\sum_{\ell=1}^k(1-\tau_\ell)\xi_\ell}\frac{1}{|\mathcal{V}|}\sum_{\bv\in\mathcal{V}}\prod_{\ell=1}^k\big\{{\varphi}_\ell\big(Y_{i_\ell,i_\ell'}\big)\big\}^{1-\xi_\ell}\,.
\]
If $\sum_{\ell=1}^k\xi_\ell\geq2$, then
\[
(\alpha-\tilde{\alpha})^{\sum_{\ell=1}^k\tau_\ell\xi_\ell}(\beta-\tilde{\beta})^{\sum_{\ell=1}^k(1-\tau_\ell)\xi_\ell}=O_p(N^{-1})\,.
\]
for any $\xi_1,\ldots,\xi_k,\tau_1,\ldots,\tau_k\in\{0,1\}$. Due to $|{\varphi}_\ell(Y_{i_\ell,i_\ell'})|\leq \max\{1-\alpha,\alpha,1-\beta,\beta\}$, then
\[
\begin{split}
&\hat{T}_{\mathcal{V}}-\tilde{T}_{\mathcal{V}}\\
&~~~~~=\sum_{\xi_1+\cdots+\xi_k=1\atop \xi_1,\ldots,\xi_k\in\{0,1\}}(\alpha-\tilde{\alpha})^{\sum_{\ell=1}^k\tau_\ell\xi_\ell}(\beta-\tilde{\beta})^{\sum_{\ell=1}^k(1-\tau_\ell)\xi_\ell}\frac{1}{|\mathcal{V}|}\sum_{\bv\in\mathcal{V}}\prod_{\ell=1}^k\big\{{\varphi}_\ell\big(Y_{i_\ell,i_\ell'}\big)\big\}^{1-\xi_\ell}+O_p(N^{-1})\\
&~~~~~=\sum_{j=1}^k(\alpha-\tilde{\alpha})^{\tau_j}(\beta-\tilde{\beta})^{1-\tau_j}\frac{1}{|\mathcal{V}|}\sum_{\bv\in\mathcal{V}}\prod_{\ell\neq j}{\varphi}_\ell\big(Y_{i_\ell,i_\ell'}\big)+O_p(N^{-1})\,.
\end{split}
\]
Similar to (\ref{eq:b1}), we have
\[
\bigg|\frac{1}{|\mathcal{V}|}\sum_{\bv\in\mathcal{V}}\prod_{\ell\neq j}\varphi_\ell \big(Y_{i_\ell,i_\ell'}\big)-\frac{1}{|\mathcal{V}|}\sum_{\bv\in\mathcal{V}}\prod_{\ell\neq j}\mathbb{E}\big\{\varphi_\ell\big(Y_{i_\ell,i_\ell'}\big)\big\}\bigg|=O_p\bigg(\sqrt{\frac{\aleph_{\mathcal{V}}}{|\mathcal{V}|}}\bigg)
\]
for any $j=1,\ldots,k$. Since $\aleph_{\mathcal{V}}/|\mathcal{V}|\asymp N^{-1}$, it holds that
\[
\begin{split}
\hat{T}_{\mathcal{V}}-\tilde{T}_{\mathcal{V}}=&~\sum_{j=1}^k(\alpha-\tilde{\alpha})^{\tau_j}(\beta-\tilde{\beta})^{1-\tau_j}\frac{1}{|\mathcal{V}|}\sum_{\bv\in\mathcal{V}}\prod_{\ell\neq j}\mathbb{E}\big\{\varphi_\ell\big(Y_{i_\ell,i_\ell'}\big)\big\}+O_p(N^{-1})\,.
\end{split}
\]
As we have shown in (\ref{eq:expandor}),
\[
\begin{split}
\tilde{T}_{\mathcal{V}}-T_{\mathcal{V}}
=&~\sum_{j=1}^k\frac{(-1)^{1-\tau_j}}{|\mathcal{V}|}\sum_{\bv\in\mathcal{V}}\bigg[\mathring{Y}_{i_j,i_j'}\prod_{\ell\neq j}\mathbb{E}\big\{\varphi_\ell\big(Y_{i_\ell,i_\ell'}\big)\big\}\bigg]+o_p(N^{-1/2})\,.
\end{split}
\]
Thus, it follows from (\ref{eq:Cv}) that
\[
\begin{split}
\sqrt{N}\big(\hat{C}_{\mathcal{V}}-C_{\mathcal{V}}\big)=&~\frac{\sqrt{N}}{(1-\alpha-\beta)^k}\sum_{j=1}^k\frac{(-1)^{1-\tau_j}}{|\mathcal{V}|}\sum_{\bv\in\mathcal{V}}\bigg[\mathring{Y}_{i_j,i_j'}\prod_{\ell\neq j}\mathbb{E}\big\{\varphi_\ell\big(Y_{i_\ell,i_\ell'}\big)\big\}\bigg]\\
&-\frac{1}{(1-\alpha-\beta)^k}\sum_{j=1}^k\sqrt{N}(\tilde{\alpha}-{\alpha})^{\tau_j}(\tilde{\beta}-{\beta})^{1-\tau_j}\frac{1}{|\mathcal{V}|}\sum_{\bv\in\mathcal{V}}\prod_{\ell\neq j}\mathbb{E}\big\{\varphi_\ell\big(Y_{i_\ell,i_\ell'}\big)\big\}\\
&+\frac{kC_{\mathcal{V}}\sqrt{N}(\tilde{\alpha}-\alpha)}{1-\alpha-\beta}+\frac{kC_{\mathcal{V}}\sqrt{N}(\tilde{\beta}-\beta)}{1-\alpha-\beta}+o_p(1)\,.
\end{split}
\]
We complete the proof of Proposition \ref{pn2}. $\hfill\Box$

\subsection*{Proof of Theorem \ref{tm:5}}

Recall that $\hat{u}_1-u_1=(2N)^{-1}\sum_{i\neq j}\mathring{Y}_{i,j}$, $\hat{u}_2-u_2=(4N)^{-1}\sum_{i\neq j}\{\eta_{i,j}-\mathbb{E}(\eta_{i,j})\}$ and $\hat{u}_3-u_3=(6N)^{-1}\sum_{i\neq j}\{\xi_{i,j}-\mathbb{E}(\xi_{i,j})\}$ with $\eta_{i,j}=|Y_{i,j,*}-Y_{i,j}|$ and $\xi_{i,j}=I(Y_{i,j,**}-2Y_{i,j,*}+Y_{i,j}=1~\textrm{or}-2)$. Let $\mathring{\eta}_{i,j}=\eta_{i,j}-\mathbb{E}(\eta_{i,j})$ and $\mathring{\xi}_{i,j}=\xi_{i,j}-\mathbb{E}(\xi_{i,j})$. Define $\kappa_1=\alpha(1-\alpha)$ and $\kappa_2=\beta(1-\beta)$. Due to $\{(Y_{i,j},Y_{i,j,*},Y_{i,j,**})\}_{i<j}$ are independent, and $Y_{i,j}=Y_{j,i}$, $Y_{i,j,*}=Y_{j,i,*}$ and $Y_{i,j,**}=Y_{j,i,**}$, thus
$
\mathbb{E}(\mathring{Y}_{s_1,t_1}\mathring{Y}_{s_2,t_2})=A_{s_1,t_1}\kappa_2+(1-A_{s_1,t_1})\kappa_1
$ if $\{s_1,t_1\}=\{s_2,t_2\}$, $\mathbb{E}(\mathring{Y}_{s_1,t_1}\mathring{Y}_{s_2,t_2})=0
$ if $\{s_1,t_1\}\neq\{s_2,t_2\}$, $
\mathbb{E}(\mathring{Y}_{s_1,t_1}\mathring{\eta}_{s_2,t_2})=A_{s_1,t_1}\kappa_2(2\beta-1)+(1-A_{s_1,t_1})\kappa_1(1-2\alpha)
$ if $\{s_1,t_1\}=\{s_2,t_2\}$, $\mathbb{E}(\mathring{Y}_{s_1,t_1}\mathring{\eta}_{s_2,t_2})=0
$ if $\{s_1,t_1\}\neq\{s_2,t_2\}$, $
\mathbb{E}(\mathring{Y}_{s_1,t_1}\mathring{\xi}_{s_2,t_2})=A_{s_1,t_1}\kappa_2(\beta^2-2\kappa_2)+(1-A_{s_1,t_1})\kappa_1\{(1-\alpha)^2-2\kappa_1\}
$ if $\{s_1,t_1\}=\{s_2,t_2\}$ and $\mathbb{E}(\mathring{Y}_{s_1,t_1}\mathring{\xi}_{s_2,t_2})=0
$ if $\{s_1,t_1\}\neq\{s_2,t_2\}$. Notice that $\sqrt{N}\{\hat{C}_{\mathcal{V}}(\tau_1,\ldots,\tau_k)-C_{\mathcal{V}}(\tau_1,\ldots,\tau_k)\}=S_{\mathcal{V}}(\tau_1,\ldots,\tau_k)+\Xi_{\mathcal{V}}(\tau_1,\ldots,\tau_k)$ with $\Xi_{\mathcal{V}}(\tau_1,\ldots,\tau_k)=\Delta_{\alpha,\mathcal{V}}(\tau_1,\ldots,\tau_k)\sqrt{N}(\tilde{\alpha}-\alpha)+\Delta_{\beta,\mathcal{V}}(\tau_1,\ldots,\tau_k)\sqrt{N}(\tilde{\beta}-\beta)
$. The asymptotic variances of $S_{\mathcal{V}}(\tau_1,\ldots,\tau_k)$ has been specified in \eqref{eq:asyv} and the asymptotic variance of $\Xi_{\mathcal{V}}(\tau_1,\ldots,\tau_k)$ can be obtained via Theorems \ref{tm:twounknown} and \ref{tm:3}. Here we only need to specify ${\rm Cov}\{S_{\mathcal{V}}(\tau_1,\ldots,\tau_k),\Xi_{\mathcal{V}}(\tau_1,\ldots,\tau_k)\}$. Due to $\Xi_{\mathcal{V}}(\tau_1,\ldots,\tau_k)$ is a linear combination of $\sqrt{N}(\tilde{\alpha}-\alpha)$ and $\sqrt{N}(\tilde{\beta}-\beta)$, and the leading terms of $\tilde{\alpha}-\alpha$ and $\tilde{\beta}-\beta$ are both linear combinations of $\hat{u}_1-u_1$, $\hat{u}_2-u_2$ and $\hat{u}_3-u_3$, then the leading term of $\Xi_{\mathcal{V}}(\tau_1,\ldots,\tau_k)$ is also a linear combination of $\hat{u}_1-u_1$, $\hat{u}_2-u_2$ and $\hat{u}_3-u_3$. We first calculate a more general result ${\rm Cov}\{S_{\mathcal{V}}(\tau_1,\ldots,\tau_k),x_{1}\sqrt{N}(\hat{u}_1-u_1)+x_{2}\sqrt{N}(\hat{u}_2-u_2)+x_3\sqrt{N}(\hat{u}_3-u_3)\}$ for any $(x_1,x_2,x_3)\in\mathbb{R}^3$.

Notice that
\begin{equation}\label{eq:cov}
\begin{split}
&{\rm Cov}\big\{S_{\mathcal{V}}(\tau_1,\ldots,\tau_k),x_{1}\sqrt{N}(\hat{u}_1-u_1)+x_{2}\sqrt{N}(\hat{u}_2-u_2)+x_3\sqrt{N}(\hat{u}_3-u_3)\big\}\\
&~~~~~~=\frac{x_{1}}{2(1-\alpha-\beta)^k}\sum_{j=1}^k\frac{(-1)^{1-\tau_j}}{|\mathcal{V}|}\sum_{\bv\in\mathcal{V}}\sum_{s\neq t}\mathbb{E}\big(\mathring{Y}_{i_j,i_j'}\mathring{Y}_{s,t}\big)\prod_{\ell\neq j}\mathbb{E}\big\{\varphi_\ell\big(Y_{i_\ell,i_\ell'}\big)\big\}\\
&~~~~~~~~~+\frac{x_{2}}{4(1-\alpha-\beta)^k}\sum_{j=1}^k\frac{(-1)^{1-\tau_j}}{|\mathcal{V}|}\sum_{\bv\in\mathcal{V}}\sum_{s\neq t}\mathbb{E}\big(\mathring{Y}_{i_j,i_j'}\mathring{\eta}_{s,t}\big)\prod_{\ell\neq j}\mathbb{E}\big\{\varphi_\ell\big(Y_{i_\ell,i_\ell'}\big)\big\}\\
&~~~~~~~~~+\frac{x_{3}}{6(1-\alpha-\beta)^k}\sum_{j=1}^k\frac{(-1)^{1-\tau_j}}{|\mathcal{V}|}\sum_{\bv\in\mathcal{V}}\sum_{s\neq t}\mathbb{E}\big(\mathring{Y}_{i_j,i_j'}\mathring{\xi}_{s,t}\big)\prod_{\ell\neq j}\mathbb{E}\big\{\varphi_\ell\big(Y_{i_\ell,i_\ell'}\big)\big\}\\
%&~~~~~~=\frac{x_{1}}{(1-\alpha-\beta)^k}\sum_{j=1}^k\frac{(-1)^{1-\tau_j}}{|\mathcal{V}|}\sum_{\bv\in\mathcal{V}}\{A_{i_j,i_j'}(\kappa_2-\kappa_1)+\kappa_1\}\prod_{\ell\neq j}\mathbb{E}\big\{\varphi_\ell\big(Y_{i_\ell,i_\ell'}\big)\big\}\\
%&~~~~~~~~~+\frac{x_{2}}{2(1-\alpha-\beta)^k}\sum_{j=1}^k\frac{(-1)^{1-\tau_j}}{|\mathcal{V}|}\sum_{\bv\in\mathcal{V}}[A_{i_j,i_j'}\{\kappa_2(2\beta-1)-\kappa_1(1-2\alpha)\}\\
%&~~~~~~~~~~~~~~~~~~~~~~~~~~~~~~~~~~~~~~~~~~~~~~~~~~~~~~~+\kappa_1(1-2\alpha)]\prod_{\ell\neq j}\mathbb{E}\big\{\varphi_\ell\big(Y_{i_\ell,i_\ell'}\big)\big\}\\
%&~~~~~~~~~+\frac{x_{3}}{3(1-\alpha-\beta)^k}\sum_{j=1}^k\frac{(-1)^{1-\tau_j}}{|\mathcal{V}|}\sum_{\bv\in\mathcal{V}}[A_{i_j,i_j'}\kappa_2(\beta^2-2\kappa_2)\\
%&~~~~~~~~~~~~~~~~~~~~~~~~~~~~~~~~~~~~~~~~~~~~~~~~~~~~~~~+(1-A_{i_j,i_j'})\kappa_1\{(1-\alpha)^2-2\kappa_1\}]\prod_{\ell\neq j}\mathbb{E}\big\{\varphi_\ell\big(Y_{i_\ell,i_\ell'}\big)\big\}\\
&~~~~~~=\bigg\{\frac{x_{1}(\kappa_2-\kappa_1)}{1-\alpha-\beta}+\frac{x_{2}\{\kappa_2(2\beta-1)-\kappa_1(1-2\alpha)\}}{2(1-\alpha-\beta)}\\
&~~~~~~~~~~~~~~~~~~~~~~~~~~~~+\frac{x_3[\kappa_2(\beta^2-2\kappa_2)-\kappa_1\{(1-\alpha)^2-2\kappa_1\}]}{3(1-\alpha-\beta)}\bigg\}\\
&~~~~~~~~~~~~~~~~~~~~~~~~~~~~~~~~\times\sum_{j=1}^k(-1)^{1-\tau_j}C_{\mathcal{V}}(\tau_1,\ldots,\tau_{j-1},1,\tau_{j+1},\ldots,\tau_k)\\
&~~~~~~~~~+\bigg[\frac{x_{1}\kappa_1}{(1-\alpha-\beta)^k}+\frac{x_{2}\kappa_1(1-2\alpha)}{2(1-\alpha-\beta)^k}+\frac{x_{3}\kappa_1\{(1-\alpha)^2-2\kappa_1\}}{3(1-\alpha-\beta)^k}\bigg]\\
&~~~~~~~~~~~~~~~~~~~~~~~~~~~~~~~~\times\sum_{j=1}^k\frac{(-1)^{1-\tau_j}}{|\mathcal{V}|}\sum_{\bv\in\mathcal{V}}\prod_{\ell\neq j}\mathbb{E}\big\{\varphi_\ell\big(Y_{i_\ell,i_\ell'}\big)\big\}\,.
\end{split}
\end{equation}

If $\alpha$ is known, we have $\tilde{\alpha}=\alpha$ and $\tilde{\beta}=\hat{\beta}$. Then
$
\Xi_{\mathcal{V}}(\tau_1,\ldots,\tau_k)
=\Delta_{\beta,\mathcal{V}}(\tau_1,\ldots,\tau_k)\sqrt{N}(\hat{\beta}-\beta)
$.
As we have shown in the proof of Theorem \ref{tm:twounknown}, $\hat{\beta}-\beta=g_{\beta,1}(\hat{u}_1-u_1)+g_{\beta,2}(\hat{u}_2-u_2)+o_p(N^{-1/2})$. With selecting $x_1=g_{\beta,1}\Delta_{\beta,\mathcal{V}}(\tau_1,\ldots,\tau_k)$, $x_2=g_{\beta,2}\Delta_{\beta,\mathcal{V}}(\tau_1,\ldots,\tau_k)$ and $x_3=0$ in (\ref{eq:cov}), we then have part (i).

If $\beta$ is known, we have $\tilde{\alpha}=\hat{\alpha}$ and $\tilde{\beta}=\beta$. Then
$
\Xi_{\mathcal{V}}(\tau_1,\ldots,\tau_k)
=\Delta_{\alpha,\mathcal{V}}(\tau_1,\ldots,\tau_k)\sqrt{N}(\hat{\alpha}-\alpha)$.
As we have shown in the proof of Theorem \ref{tm:twounknown}, $\hat{\alpha}-\alpha=g_{\alpha,1}(\hat{u}_1-u_1)+g_{\alpha,2}(\hat{u}_2-u_2)+o_p(N^{-1/2})$.
With selecting $x_1=g_{\alpha,1}\Delta_{\alpha,\mathcal{V}}(\tau_1,\ldots,\tau_k)$, $x_2=g_{\alpha,2}\Delta_{\alpha,\mathcal{V}}(\tau_1,\ldots,\tau_k)$ and $x_3=0$ in (\ref{eq:cov}), we then have part (ii).

If $\alpha$ and $\beta$ are unknown, we have $\tilde{\alpha}=\hat{\alpha}$ and $\tilde{\beta}=\hat{\beta}$. Then
$
\Xi_{\mathcal{V}}(\tau_1,\ldots,\tau_k)
=\Delta_{\alpha,\mathcal{V}}(\tau_1,\ldots,\tau_k)\sqrt{N}(\hat{\alpha}-\alpha)+\Delta_{\beta,\mathcal{V}}(\tau_1,\ldots,\tau_k)\sqrt{N}(\hat{\beta}-\beta)
$.
As we have shown in the proof of Theorem \ref{tm:3}, $\hat{\alpha}-\alpha=g_{\alpha,1}(\hat{u}_1-u_1)+g_{\alpha,2}(\hat{u}_2-u_2)+g_{\alpha,3}(\hat{u}_3-u_3)+o_p(N^{-1/2})$ and $\hat{\beta}-\beta=g_{\beta,1}(\hat{u}_1-u_1)+g_{\beta,2}(\hat{u}_2-u_2)+g_{\beta,3}(\hat{u}_3-u_3)+o_p(N^{-1/2})$. With selecting $x_1=g_{\alpha,1}\Delta_{\alpha,\mathcal{V}}(\tau_1,\ldots,\tau_k)+g_{\beta,1}\Delta_{\beta,\mathcal{V}}(\tau_1,\ldots,\tau_k)$, $x_2=g_{\alpha,2}\Delta_{\alpha,\mathcal{V}}(\tau_1,\ldots,\tau_k)+g_{\beta,2}\Delta_{\beta,\mathcal{V}}(\tau_1,\ldots,\tau_k)$ and $x_3=g_{\alpha,3}\Delta_{\alpha,\mathcal{V}}(\tau_1,\ldots,\tau_k)+g_{\beta,3}\Delta_{\beta,\mathcal{V}}(\tau_1,\ldots,\tau_k)$ in (\ref{eq:cov}), we then have part (iii). $\hfill\Box$

\end{document}